\documentclass{aastex631}
\usepackage{amsmath, ntheorem}
\usepackage{algorithm2e}
\usepackage{color}

\newcommand{\B}{\mathbf{B}}
\newcommand{\R}{\mathbb{R}}
\newcommand{\bb}{\mathbf{b}}
\newcommand{\kk}{\mathbf{k}}
\newcommand{\M}{\mathbf{M}}
\newcommand{\HH}{\mathbf{H}}
\newcommand{\rr}{\mathbf{r}}
\DeclareMathOperator*{\argmin}{argmin}
\DeclareMathOperator*{\argmax}{argmax}

\received{March 9, 2023}
\accepted{March 21, 2023}

\submitjournal{ApJS}

\shorttitle{UQ of the Wave-Telescope}
\shortauthors{Broeren and Klein}

\begin{document}

\title{Data-Driven Uncertainty Quantification of the Wave-Telescope Technique:\\
General Equations and Application to HelioSwarm}

\correspondingauthor{\href{https://sites.google.com/math.arizona.edu/broeren}{Theodore Broeren}}
\email{broeren@arizona.edu}

\author[0000-0002-2649-020X]{T. Broeren}
\affiliation{Department of Applied Mathematics, University of Arizona, Tucson, AZ, USA}

\author[0000-0001-6038-1923]{K. G. Klein}
\affiliation{Lunar and Planetary Laboratory, University of Arizona, Tucson, AZ, USA}

\begin{abstract}
The upcoming NASA mission HelioSwarm will use nine spacecraft to make the first simultaneous multi-point measurements of space plasmas spanning multiple scales. 
Using the wave-telescope technique, HelioSwarm’s measurements will allow for both the calculation of the power in wavevector-and-frequency space and the characterization of the associated dispersion relations of waves present in the plasma at MHD and ion-kinetic scales. 
This technique has been applied to the four-spacecraft missions of CLUSTER and MMS and its effectiveness has previously been characterized in a handful of case studies. 
We expand this uncertainty quantification analysis to arbitrary configurations of four through nine spacecraft for three-dimensional plane waves. 
We use Bayesian inference to learn equations that approximate the error in reconstructing the wavevector as a function of relative wavevector magnitude, spacecraft configuration shape, and number of spacecraft. 
We demonstrate the application of these equations to data drawn from a nine-spacecraft configuration to both improve the accuracy of the technique, as well as expand the magnitudes of wavevectors that can be characterized.
\end{abstract}


\section{Introduction}
\label{sec:intro}
Plasmas are composed of charged particles, the movement of which creates and responds to electromagnetic fields. Electromagnetic waves are a fundamental collective response of the self-consistent coupling between the fields and charged particles. These electromagnetic waves propagate through plasmas carrying energy and transferring that energy to objects in their path. To better understand plasma systems, we wish to study the generation and dissipation of these waves. However, plasmas are difficult to create and contain on Earth. Therefore, space is often used as a natural laboratory for the study of plasmas \citep{Verscharen:2019}. 

The wave-telescope technique (sometimes referred to as k-filtering) was developed to extract wave information out of in situ magnetic field measurements made by multi-spacecraft missions \citep{Pincon1988}. This technique, which computes the Spectral Energy Density as a function of frequency $\omega$ and wavevector $\mathbf{k}$, allows us to estimate the direction, velocity, and frequency of plasma waves. This technique has been used to analyze data from the European Space Agency mission CLUSTER II \citep{Glassmeier:2001} as well as the NASA mission MMS \citep{Narita:2016}. Because this technique utilizes in situ data from a configuration of spacecraft, the reliability of the technique is dependent upon the size of the spacecraft configuration, as well as its shape \citep{Sahraoui:2010b}. A recent overview of applications of the wave-telescope technique can be found in \cite{Narita:2022}.

Numerous many-spacecraft missions (including the nine-spacecraft NASA mission HelioSwarm) have been proposed that hope to utilize the wave-telescope technique of wavevector identification \citep{Plice:2020,Maruca:2021,Retino:2019}. We wish to quantify the accuracy of this wavevector identification technique so that we can verify the wave-telescope's applicability in future multi-spacecraft missions. As the instrument common to all of these mission concepts is a magnetometer, we focus our analysis on the version of the wave-telescope that only requires in situ magnetic field measurements. Specific questions that we wish to answer include:
\begin{enumerate}
    \item Can we formulate an equation that gives the expected error in identifying the wavevector $\mathbf{k}$ for an arbitrary spacecraft configuration?
    \item How does the expected error vary for different numbers of spacecraft?
    \item How does the expected error vary for differently shaped spacecraft configurations?
    \item What is the level of variance from this mean error value that typically occurs?
\end{enumerate}

Quantitative answers to these questions are provided in the following sections. The methodology used in this study is described in \S \ref{sec:meth} and resulting learned equations in \S \ref{sec:learn_eqns}. A posteriori analysis of the equations is discussed in \S \ref{sec:Results}, with an application to nine-spacecraft configurations for the HelioSwarm design reference mission in \S \ref{ssec:app_HS}. We show extra details of the wave-telescope derivation in Appendix \S \ref{sec:appendix.lagrange}, outline the algorithm that we used to efficiently scan wavevector space in Appendix \S \ref{sec:appendix.scan_alg}, analyze the geometry of the example HelioSwarm spacecraft configuration used in Appendix \S \ref{sec:appendix.hour_205}, and give an additional verification of the fitted equations in Appendix \S \ref{sec:appendix.verify}. 

\section{Methodology}
\label{sec:meth}
In this section we first work through the computation of the classic wave-telescope in \S \ref{ssec:WT}. We next discuss the limitations of this technique derived from theory and previous numerical studies in \S \ref{ssec:limits}. Then, we introduce the concept of Bayesian inference, which we use to determine the coefficients in our learned equations in \S \ref{ssec:Bayes_inf}. Finally, we detail the spacecraft configurations and magnetic field waves that were used in our simulations to create our dataset in \S  \ref{ssec:data}.

\subsection{Wave-Telescope Technique}
\label{ssec:WT}
The wave-telescope technique identifies the strongest waves present in a plasma by searching for maxima in the field energy density $P(\omega,\kk)$ \citep{Pincon1988,Motschmann:1998,Narita:2022}. We start with the assumption that the magnetic field can be expressed as a superposition of plane waves
\begin{equation}
    \bb(t, \rr_n) = \sum_{\omega} \sum_{\kk} \bb(\omega, \kk) e^{i (\kk \cdot \rr_n - \omega t)}.
\end{equation}
The left-hand side, $\bb(t, \rr_n)$, is the measured value of the magnetic field at time $t$ and at spacecraft $n$'s position $\rr_n$. We let $N$ be the total number of spacecraft in a configuration. To modify this technique to other applications, one can replace the plane wave basis function by another basis function, such as spherical waves \citep{Constantinescu:2006}. 

The quantity that we wish to solve for, the field energy density $P(\omega, \kk)$, is the trace of the $3\times 3$ field energy matrix $\mathbf{P}(\omega, \kk)$, defined as
\begin{equation}
    \mathbf{P}(\omega, \kk) = \mathbb{E}\left[ \bb(\omega, \kk) \bb^{\dagger}(\omega, \kk) \right]. \label{eqn:power_def}
\end{equation}
We use lowercase bold letters to denote vector quantities, capital bold letters to denote matrix (or higher dimensional) quantities, and non-bold letters to denote scalar values. We use the symbol $\mathbb{E}$ to denote the computation of an expected value. The dagger symbol $\dagger$ denotes taking the Hermitian (complex conjugate) transpose of a vector/matrix quantity.

As spacecraft magnetometers take measurements at a cadence fast in time (10-100Hz) compared to their spatial sampling (100-1000km), we can transform a time series of our magnetic field measurements at each spacecraft $n \leq N$ into the frequency domain using a discrete Fourier transformation

\begin{equation}
    \bb(\omega, \rr_n) = \frac{1}{2\pi} \sum_t \bb(t, \rr_n)e^{i \omega t}.
\end{equation}
We cannot perform an analogous transform between the $\rr$ and $\kk$ spaces, as the $N$ measurement locations are not sufficient to approximate the required sum. Therefore, we take a minimum variance estimator approach to describe the spatial structure.

The frequency transformed magnetic field measurements, $\bb(\omega, \rr_n)$, can be expressed as a single sum of plane waves
\begin{equation}
     \bb(\omega, \rr_n) = \sum_{\kk} \bb(\omega, \kk) e^{i \kk \cdot \rr_n }. \label{eqn:w_wave}
\end{equation}
Equation \ref{eqn:w_wave} can be written in matrix form by combining measurements from all $N$ spacecraft. The resulting equation is 
\begin{equation}
    \tilde{\bb}(\omega) = \sum_{\kk} \HH(\kk) \bb(\omega,\kk) \label{eqn:B}
\end{equation}
where the measured field quantities (transformed into the frequency domain) are captured in the matrix
\begin{equation}
    \tilde{\bb}(\omega) = \begin{pmatrix} \bb(\omega, \rr_1) \\ \vdots \\ \bb(\omega, \rr_N) \end{pmatrix}_{3N\times 1}.
\end{equation}
Letting $\mathbf{I}$ be the $3\times 3$ identity matrix, we define the propagation matrix as
\begin{equation}
    \HH(\kk) = \begin{pmatrix} \mathbf{I} e^{i\kk \cdot \rr_1} \\ \vdots \\ \mathbf{I} e^{i\kk \cdot \rr_N} \end{pmatrix}_{3N\times 3}.
\end{equation}
We also define the $3N\times 3N$ spatial correlation matrix as the following expectation
\begin{equation}
    \M(\omega) = \mathbb{E} \left[ \tilde{\bb}(\omega) \tilde{\bb}^{\dagger}(\omega) \right].
\end{equation}

To approximate the expectation $\M$, we divide our time series of data into a partition of $Q$ subintervals, and computing $\tilde{\bb}_q(\omega)$ for each. Then, we average over the different intervals 
\begin{equation}
    \M(\omega) = \mathbb{E} \left[ \tilde{\bb}(\omega) \tilde{\bb}^{\dagger}(\omega) \right] \approx \frac{1}{Q}\sum_{q=1}^Q \tilde{\bb}_q(\omega) \tilde{\bb}_q^{\dagger}(\omega). \label{eqn:M_calc}
\end{equation}
To relate $\M(\omega)$ to our known parameters we take the definition of $\tilde{\bb}(\omega)$ (eqn \ref{eqn:B}), multiply it on the right with its Hermitian adjoint, and take the expectation. We then invoke the definition of $\mathbf{P}$ (eqn \ref{eqn:power_def}) to find that it is related to $\M$ through the expression
\begin{equation}
     \M(\omega) = \sum_{\kk} \HH(\kk) \mathbf{P}(\omega, \kk) \HH^{\dagger}(\kk). \label{eqn:M}
\end{equation}
The objective now is to find the matrix $\mathbf{P}$ which satisfies eqn \ref{eqn:M} for all computed values of $\M(\omega)$ and $\HH(\kk)$. We cannot directly invert eqn \ref{eqn:M} to solve for $\mathbf{P}$ because of the summation over wavevectors, $\kk$. We will instead use a filter-bank approach to find a solution. 

Each filter $\mathbf{W}(\omega, \kk)$ is a matrix that relates the desired $\bb(\omega, \kk)$ (needed for computation of $P$ via eqn \ref{eqn:power_def}) to the computed value of $\tilde{\bb}(\omega)$ via the equation
\begin{equation}
    \bb(\omega, \kk) = \mathbf{W}^\dagger(\omega, \kk) \tilde{\bb}(\omega). \label{eqn:filter}
\end{equation}
This definition implies that our filter $\mathbf{W}$ is a $3N\times 3$ matrix. We can multiply eqn \ref{eqn:filter} with its Hermitian adjoint and take the expectation (again invoking eqn \ref{eqn:power_def}) to find that 
\begin{equation}
    \mathbf{P}(\omega, \kk) = \mathbf{W}^\dagger(\omega, \kk) \M(\omega) \mathbf{W}(\omega, \kk). \label{eqn:Pmatrix}
\end{equation}

We now uniquely determine each filter $\mathbf{W}(\omega, \kk)$ by requiring that the filter will absorb all energy which does not have frequency and wavevector exactly equal to $\omega$ and $\kk$. Stated as a minimization problem, we determine the filters by minimizing the total power
\begin{equation}
   \mathbf{W}(\omega,\kk) = \argmin_{\mathbf{W}(\omega,\kk)} \text{Tr}\left[\mathbf{W}^\dagger(\omega, \kk) \M(\omega) \mathbf{W}(\omega, \kk) \right]
\end{equation}
subject to the constraint that a wave with frequency $\omega$ and wavevector $\kk$ remains unchanged after multiplication by the filter matrix. Using equations \ref{eqn:B} and \ref{eqn:filter} we see that this constraint is equivalent to
\begin{equation}
   \mathbf{W}^\dagger(\omega, \kk) \HH(\kk) \bb(\omega, \kk) = \bb(\omega, \kk) .
\end{equation}
However, to properly account for imaginary components of these matrices we must ensure that the Hermitian adjoint of this constraint is also satisfied. This means that the power can be found via the formulation
\begin{align}
    P(\omega, \kk) &= \min_{\mathbf{W}(\omega,\kk)} \text{Tr}\left[\mathbf{W}^\dagger(\omega, \kk) \M(\omega) \mathbf{W}(\omega, \kk) \right] \label{eqn:min}\\
   \text{  s.t. \hspace{.2cm} }    &\mathbf{W}^\dagger(\omega, \kk) \HH(\kk)  = \mathbf{I} \hspace{0.2cm}\text{and} \hspace{0.2cm} \HH^\dagger(\kk) \mathbf{W}(\omega, \kk)  = \mathbf{I}. \nonumber
\end{align}
The system \ref{eqn:min} can be solved using the method of Lagrange multipliers. The exact solution to this system (derived in Appendix \S \ref{sec:appendix.lagrange}) is
\begin{equation}
    P(\omega, \kk) = \text{Tr} \left[ \HH^\dagger(\kk) \M^{-1}(\omega) \HH(\kk) \right]^{-1}.  \label{eqn:power}
\end{equation}

To determine the $\omega$ and $\kk$ of the magnetic waves present, we scan all possible values of $\omega$ and $\kk$ and use eqn \ref{eqn:power} to compute the quantity $P(\omega,\kk)$ at each. Once this scan is completed, we find that $P(\omega,\kk)$ will have local maxima at the values of $\omega$ and $\kk$ corresponding to the frequency and wavevectors of the waves present in the plasma (see Figure \ref{fig:k_search} for an example).

\subsection{Limitations of the Wave-Telescope}
\label{ssec:limits}
The domain in $(\omega, \kk)$ space which can be scanned using this technique is not unbounded. If $T$ is the number of time samples of data that we are processing and $1/\Delta t$ is the sampling frequency, then the (positive) frequency domain using a standard Fourier transform is limited to 
\begin{equation}
    \omega \in \left\{0, \frac{1}{T\Delta t}, \frac{2}{T\Delta t},...,\frac{1}{2 \Delta t} - \frac{1}{T \Delta t}\right\} \label{eqn:w_valid}.
\end{equation}

Let us express our wavevector $\kk$ in spherical form $\left[ k, k_\theta, k_\phi\right]$. If the configuration of spacecraft used is not co-linear or co-planar, then we can reconstruct all $k_\theta \in [0, 2\pi)$ and $k_\phi \in [0, \pi)$. However, the magnitude of the wavevectors, $k$, has an upper bound due to the Nyquist sampling theorem. If we let $d_{max}$ be the maximum inter-spacecraft distance, then the magnitudes of wavevector that we can theoretically reconstruct are $k < k_{max}$ \citep{Sahraoui:2010b,Constantinescu:2006} where 
\begin{equation}
    k_{max} = \pi/d_{max}. \label{eqn:k_max}
\end{equation} 

\begin{figure}[ht]
\centering
\includegraphics[width=0.55\textwidth]{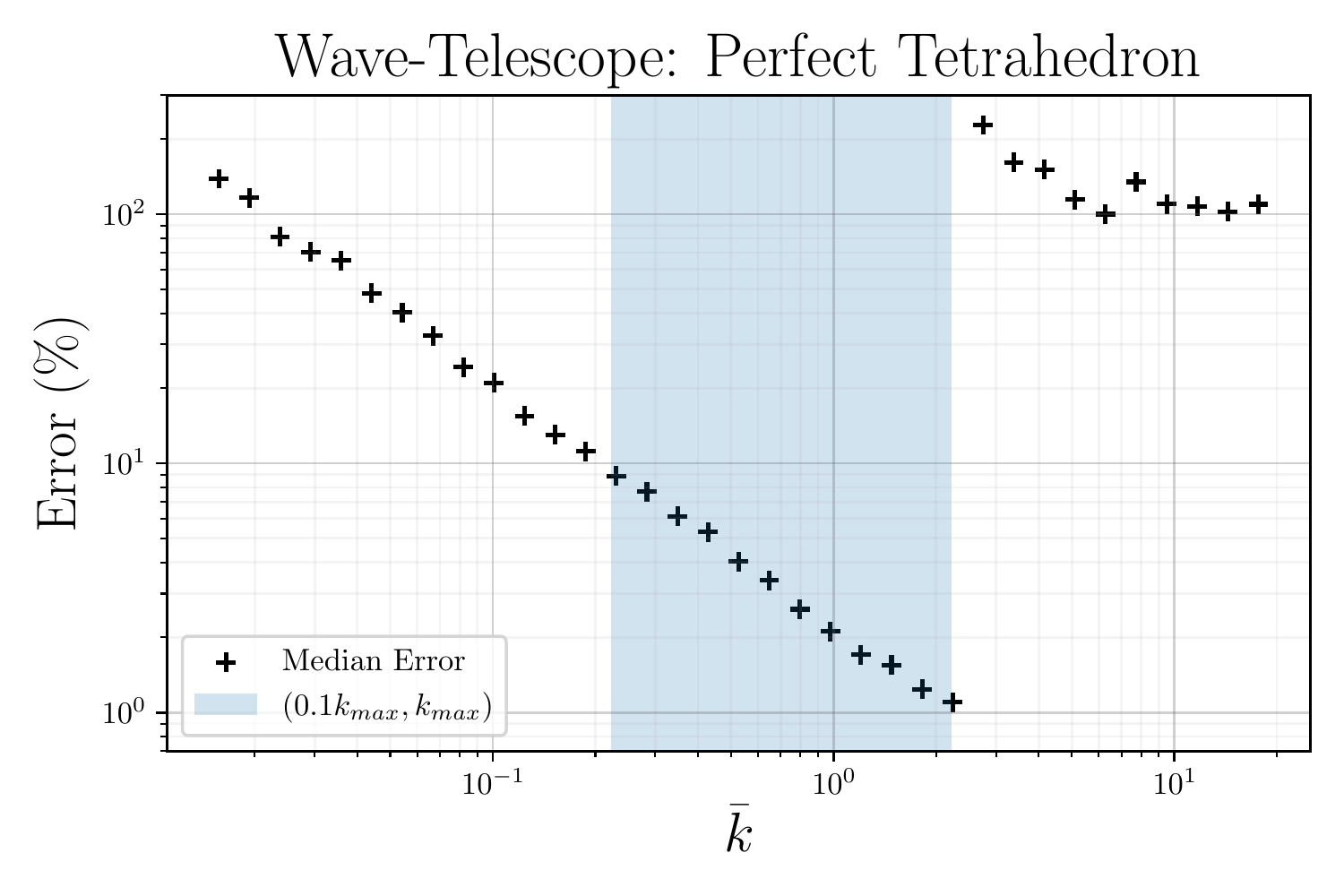}
\caption{{\small A numerical experiment using the wave-telescope technique to identify wavevectors using a single perfectly shaped four-spacecraft configuration ($\chi=0$, $L=1$, $d_{max}=\sqrt{2}$). We plot the median value of error in wavevector identification as a function of wavevector magnitude. }}
\label{fig:Sahrauri} 
\end{figure}

Previous studies \citep{Sahraoui:2010b} have numerically computed the relative error in the determination of the wavevector derived from the wave-telescope technique for a simulated plane wave using one four-spacecraft configuration. We perform a numerical experiment with a single perfectly shaped tetrahedral four-spacecraft configuration. We simulated plane-waves with 35 magnitudes and 50 directions (see \S \ref{sssec:data_waves}). Because there is a random shift of the computed wavevector using the wave-telescope, we plot the median value of error at each wavevector magnitude in Figure \ref{fig:Sahrauri}. We see that the error decreases log-linearly with magnitude of wavevector for values less than $k_{max}$. This leads us to the conclusion, in agreement with \cite{Sahraoui:2010b}, that the wave-telescope technique is accurate (had $\leq 10\%$ error) for about one order of magnitude of wavevector, where $k \in (0.1 k_{max}, k_{max})$.

Because the wavevector magnitude that a configuration of spacecraft is sensitive to changes with the size of the spacecraft configuration, we define the relative wavevector magnitude, $\bar{k}$, as
\begin{equation}
    \bar{k} = k L \label{eqn:k_unitless}.
\end{equation}
In this definition, $k$ is the absolute magnitude of the wavevector (in inverse distance units) and $L$ is the characteristic size of the spacecraft configuration (in distance units) that is used for the wave-telescope computation (see eqn \ref{eqn:L}). Therefore, we can translate the unit-less relative wavevector magnitude $\bar{k}$ (used in Figures \ref{fig:Sahrauri},\ref{fig:mu_data},\ref{fig:data_slice}, and \ref{fig:PPC}) into real inverse length units for any spacecraft configuration via eqn \ref{eqn:k_unitless}.

\subsection{Bayesian Inference}
\label{ssec:Bayes_inf}
In this work we make use of a statistical method called Bayesian inference to estimate unknown model parameters. This method utilizes Bayes theorem to update a user-defined prior distribution for each unknown parameter by sequentially incorporating data. Once all data has been incorporated, we are left with a posterior distribution for each quantity. These posterior distributions represent the probability density function of each parameter given our data and informed prior \citep{McElreath2016}.

We have chosen to implement Bayesian inference using the package PyMC3 \citep{Salvatier:2016}. This tool uses Markov Chain Monte Carlo (MCMC) sampling to construct the posterior distributions. MCMC does not directly compute or approximate the posterior distribution, but it instead directly draws samples from the posterior distributions of the parameters. We use histograms to create an image of these posterior distributions. Finally, we compute statistics of these posterior samples, such as mean and standard deviation, to obtain an expected value and error bars for each unknown model parameter that we are estimating.

\subsection{Dataset Generation}
\label{ssec:data}
To extract equations that describe the errors in reconstructing wavevectors from differently sized and shaped configurations of spacecraft, we must first define general parameters describing size and shape that can be applied to an arbitrary observatory. Because we are taking a data-driven approach to learning these equations, we also must ensure that our dataset uniformly samples all possible combinations of spacecraft configurations and wavevectors as defined by these general shape and size parameters.



\figsetgrpstart
\figsetgrpnum{2.1}
\figsetgrptitle{Four-Spacecraft Configurations}
\figsetplot{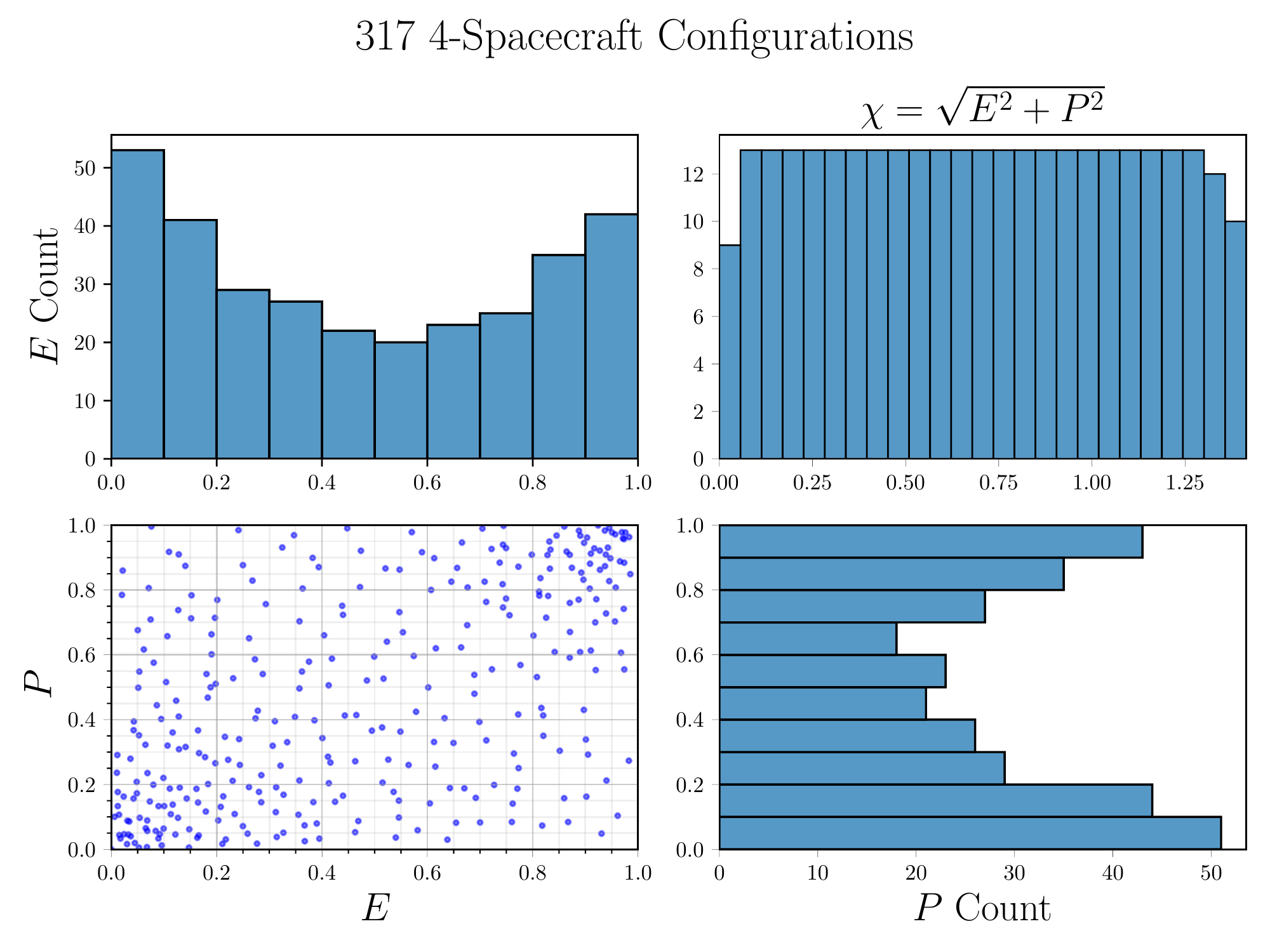}
\figsetgrpnote{A histogram showing the distribution of shapes of the four-spacecraft configurations used in our wave-telescope simulations.}
\figsetgrpend

\figsetgrpstart
\figsetgrpnum{2.2}
\figsetgrptitle{Five-Spacecraft Configurations}
\figsetplot{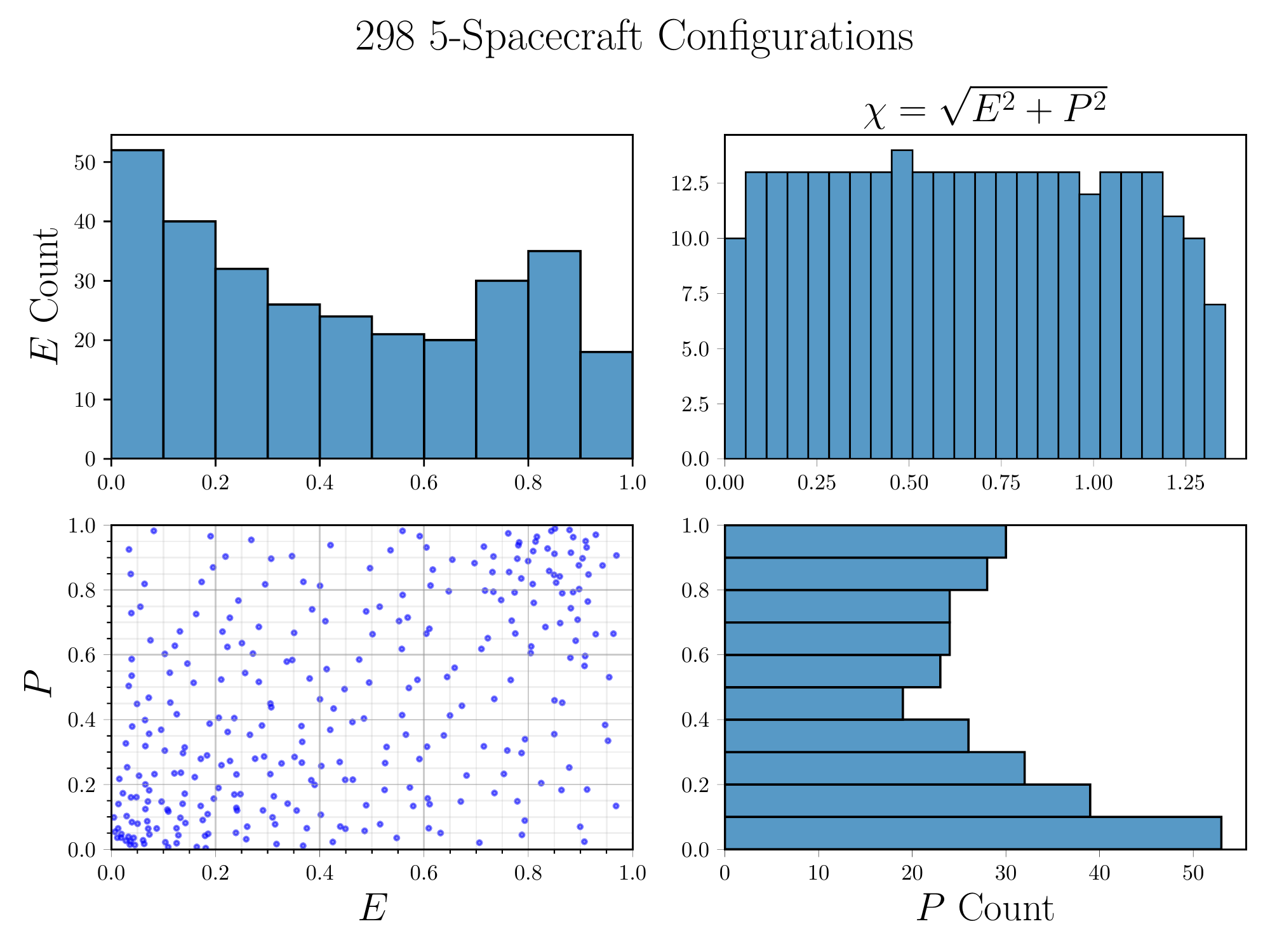}
\figsetgrpnote{A histogram showing the distribution of shapes of the five-spacecraft configurations used in our wave-telescope simulations.}
\figsetgrpend

\figsetgrpstart
\figsetgrpnum{2.3}
\figsetgrptitle{Six-Spacecraft Configurations}
\figsetplot{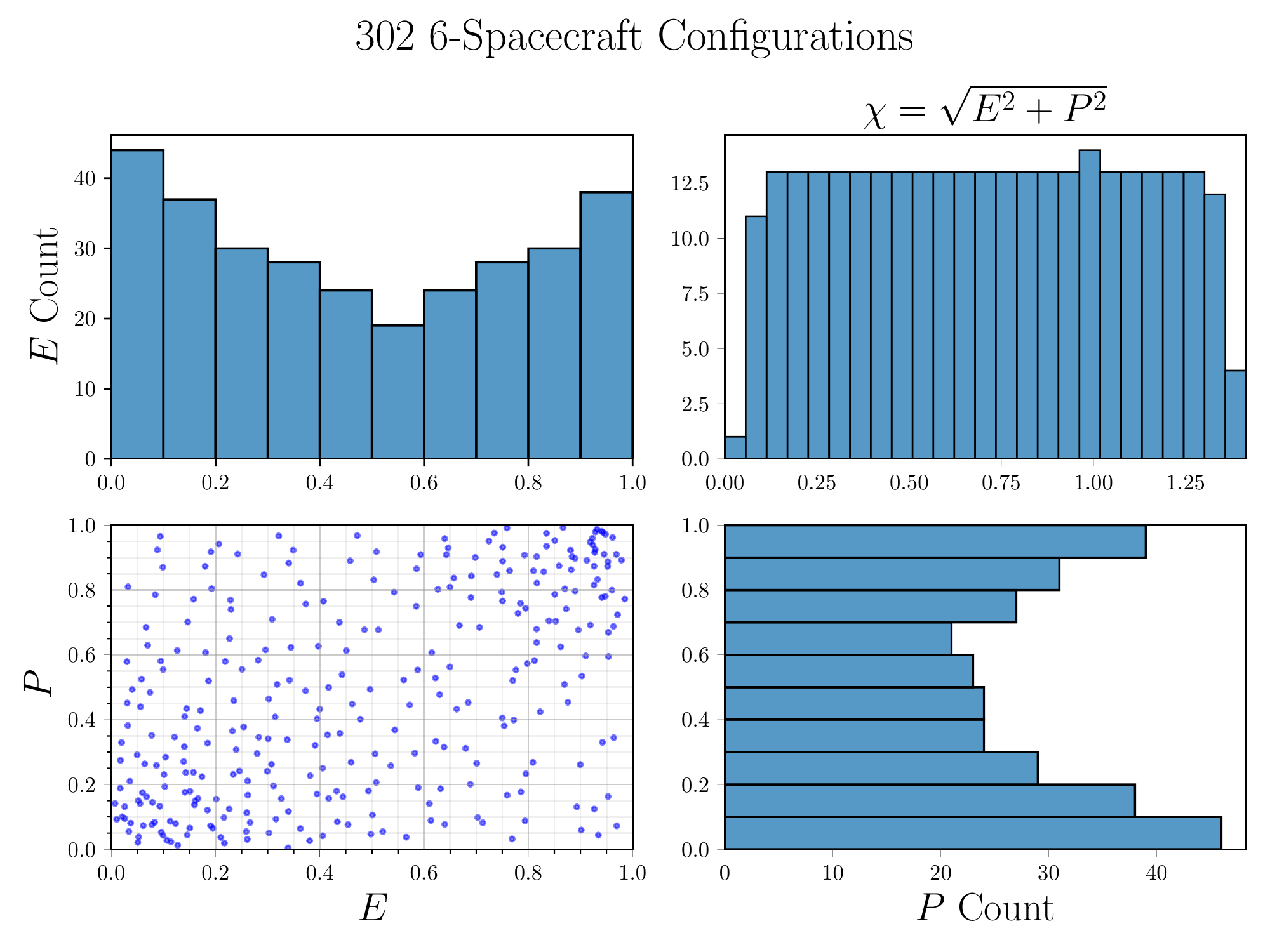}
\figsetgrpnote{A histogram showing the distribution of shapes of the six-spacecraft configurations used in our wave-telescope simulations.}
\figsetgrpend

\figsetgrpstart
\figsetgrpnum{2.4}
\figsetgrptitle{Seven-Spacecraft Configurations}
\figsetplot{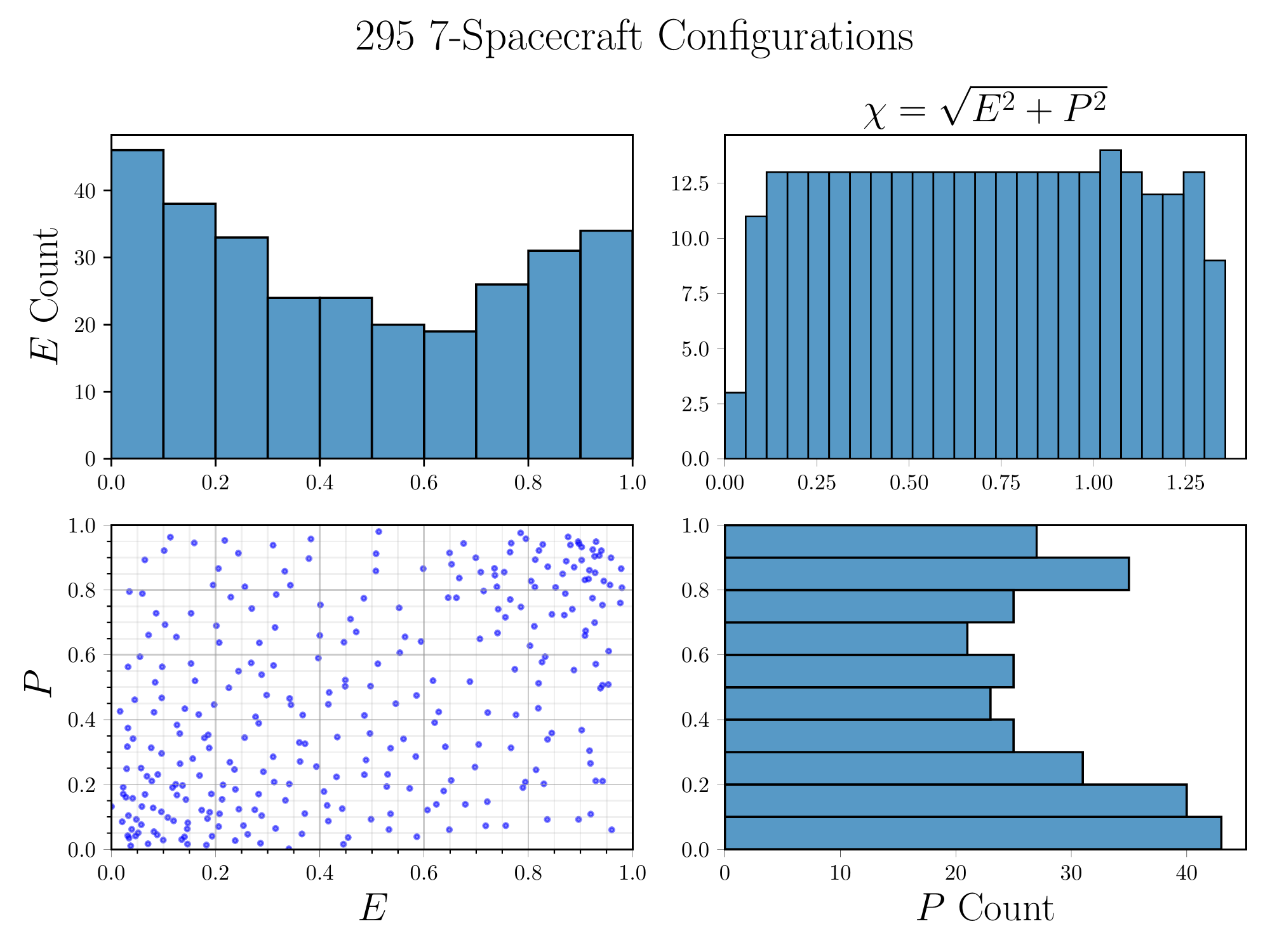}
\figsetgrpnote{A histogram showing the distribution of shapes of the seven-spacecraft configurations used in our wave-telescope simulations.}
\figsetgrpend

\figsetgrpstart
\figsetgrpnum{2.5}
\figsetgrptitle{Eight-Spacecraft Configurations}
\figsetplot{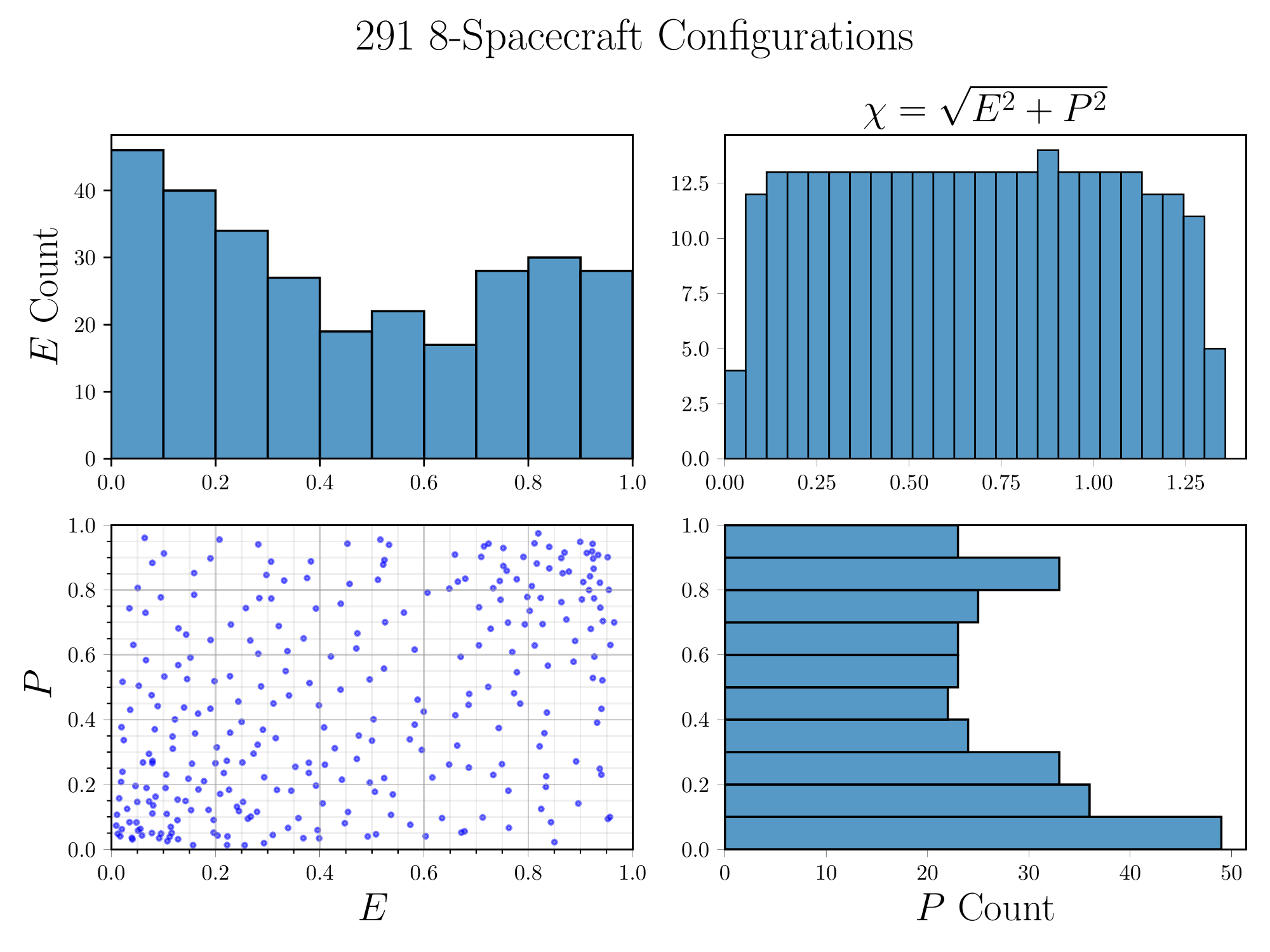}
\figsetgrpnote{A histogram showing the distribution of shapes of the eight-spacecraft configurations used in our wave-telescope simulations.}
\figsetgrpend

\figsetgrpstart
\figsetgrpnum{2.6}
\figsetgrptitle{Nine-Spacecraft Configurations}
\figsetplot{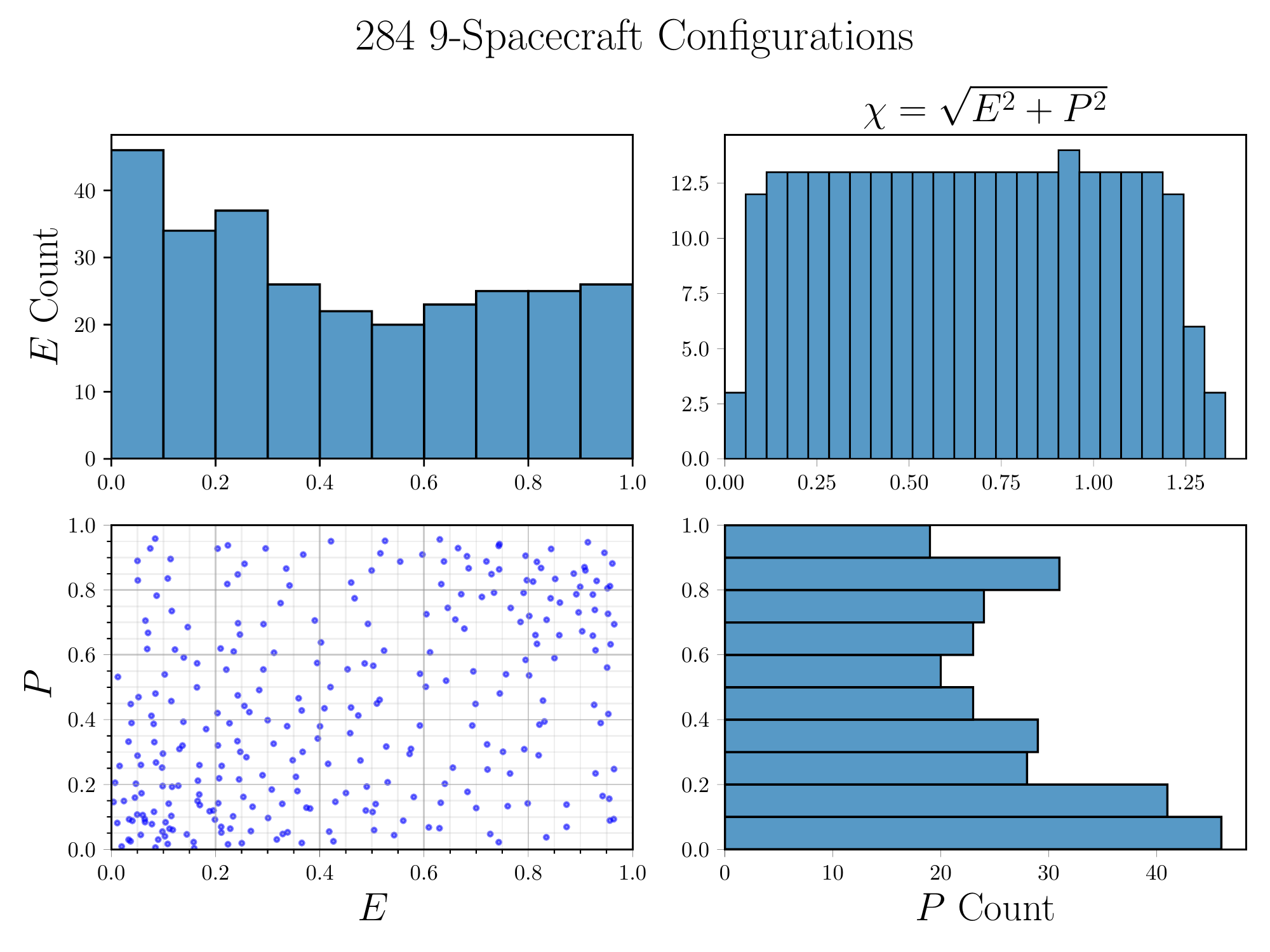}
\figsetgrpnote{A histogram showing the distribution of shapes of the nine-spacecraft configurations used in our wave-telescope simulations.}
\figsetgrpend

\figsetend

\begin{figure}[ht]
\centering
\includegraphics[width=0.75\textwidth]{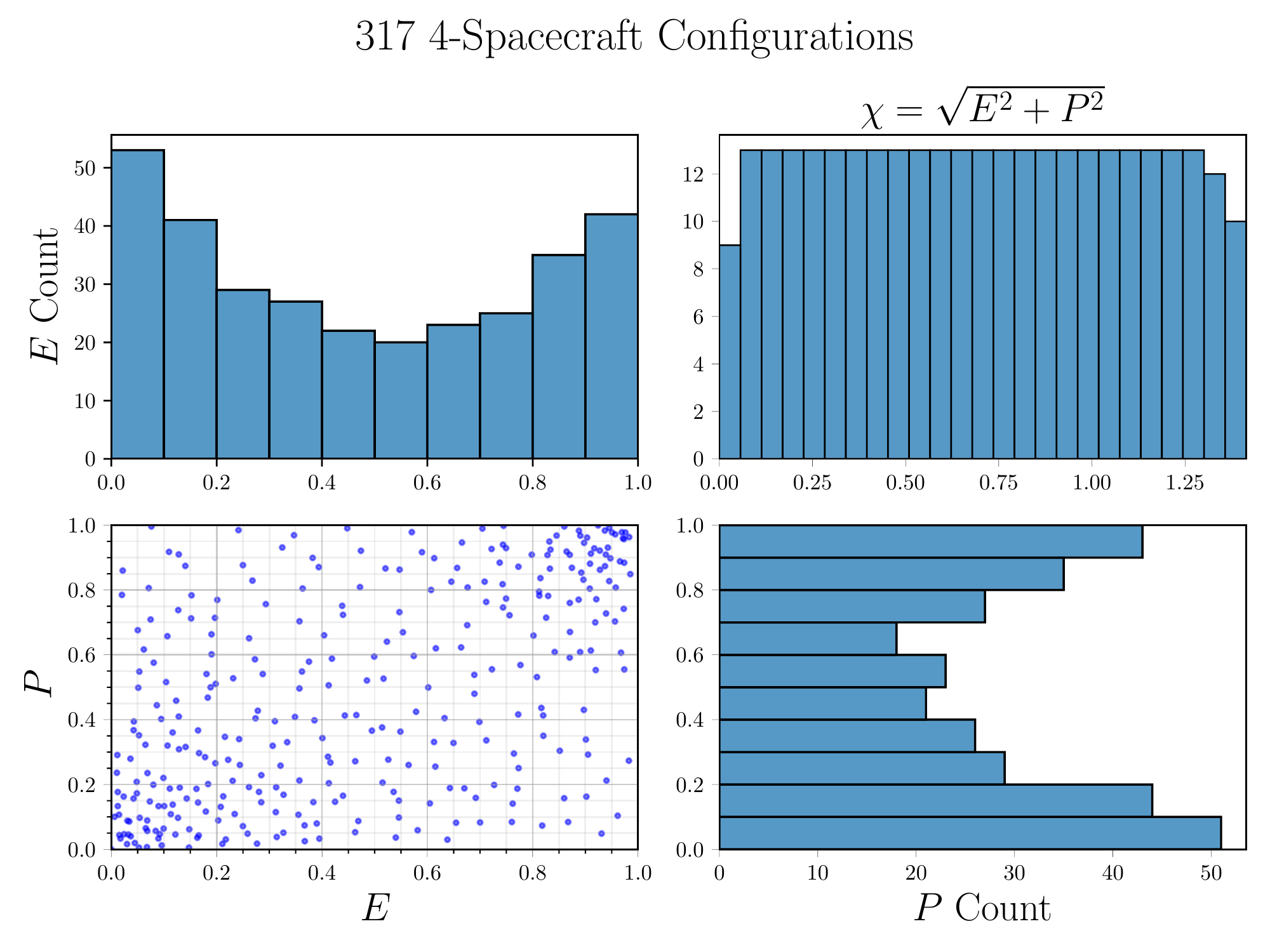}
\caption{{\small A histogram showing the distribution of shapes of the four-spacecraft configurations. The distribution of shape parameter $\chi$, defined in eqn \ref{eqn:chi}, is chosen to be approximately uniform over the range $(0,\sqrt{2})$ for each number of spacecraft. A complete figure set (6 images) for four through nine spacecraft configurations is available in the online journal.} }
\label{fig:EP_chi} 
\end{figure}

\subsubsection{Spacecraft Configurations}
\label{sssec:data_sc_config}
To quantify the shape and size of a configuration of spacecraft, we use the definitions of planarity, elongation, and characteristic size described in chapter 12 of \cite{Paschmann:1998}. From the positions of all $N$ spacecraft, we compute the barycenter of the configuration as the average spacecraft position
\begin{equation}
    \overline{\mathbf{r}} = \frac{1}{N} \sum_{n=1}^N \mathbf{r}_n .
\end{equation}
We then use the barycenter to define the $3\times3$ volumetric tensor matrix
\begin{equation}
    R_{jk} = \frac{1}{N} \sum_{n=1}^N \left[ r_n^{(j)} - (\overline{r})^{(j)} \right]  \left[ r_n^{(k)} - (\overline{r})^{(k)} \right] ,
\end{equation}
where $r_n^{(j)}$ is the $j^{th}$ component of the $n^{th}$ spacecraft position. We take the square-roots of the volumetric tensor's eigenvalues to find its singular values $a \geq b \geq c$. These singular values describe the 3 semi-axis of an ellipsoid that approximates the size and orientation of the spacecraft configuration. We use these singular values to define planarity ($P$), elongation ($E$), and characteristic size ($L$). Elongation describes the ratio of the longest two semi-axes of the ellipsoid
\begin{equation}
    E = 1 - b/a \in [0,1], \label{eqn:E}
\end{equation}
and approaches its maximal value for configurations that are nearly co-linear. Planarity describes the ratio of the smallest two semi-axes of the ellipsoid
\begin{equation}
    P = 1 - c/b \in [0,1], \label{eqn:P}
\end{equation}
and approaches its maximal value for configurations that are nearly co-planar. Characteristic size describes the overall size of the ellipsoid along its semi-major axis
\begin{align}
    L &= 2a \in [0,\infty). \label{eqn:L}
\end{align}

Previous studies of these quantities have shown that the well-shapedness of a configuration is related to elongation and planarity symmetrically (p343 \cite{Paschmann:1998},\cite{Broeren:2021}). Therefore we define a new shape parameter $\chi$ that combines elongation and planarity
\begin{equation}
    \chi = \sqrt{E^2 + P^2} . \label{eqn:chi}
\end{equation}
This new parameter is near its minimal value of 0 for well-shaped 'spherical' configurations of spacecraft and near its maximal value of $\sqrt{2}$ for poorly-shaped (co-linear and co-planar) configurations.

To test the accuracy of the wave-telescope technique on a variety of spacecraft configurations, we define ensembles of spacecraft configurations that contain $N = 4, 5, 6, 7, 8$, and $9$ spacecraft. For each number of spacecraft, we randomly generate approximately 300 configurations that uniformly span the shape parameter $\chi$ regime. The generated four-spacecraft configurations are shown in Figure \ref{fig:EP_chi}, and the $N\in \{5,6,7,8,9\}$ configurations can be found in the corresponding figure set. Each spacecraft configuration is scaled such that it has unit size ($L=1$ in eqn \ref{eqn:L}).

\subsubsection{Plane-Wave Distribution}
\label{sssec:data_waves}
Using the Fibonacci sphere algorithm \citep{Keinert:2015}, we define 50 unit directions that plane-waves in our simulations will travel. This number of unit directions gives us an average minimum angular separation of $28^{\circ}$. For each of these 50 unit directions, we simulate plane-waves of 35 logarithmically-spaced wavevector magnitudes (i.e. values of $\bar{k}$) in the interval $[0.005 \pi, 5.62 \pi]$. This gives us $1,750$ plane-waves simulated for every configuration of spacecraft.

Because we have normalized each spacecraft configuration to unit characteristic size ($L=1$) we know that the maximum inter-spacecraft distance, $d_{max}$, will typically be less than 1. From eqn \ref{eqn:k_max} we know that this gives us a corresponding value of $k_{max}>\pi$. The maximum relative wavevector magnitude in our simulations was therefore chosen to be a greater than $\pi$ so that aliasing is likely to occur for all spacecraft configurations.

In our simulations of the wave-telescope technique, we first generate synthetic magnetic field measurements at each spacecraft 
\begin{equation}
    \mathbf{b}(\mathbf{r}_n,t) = \text{Re} \left[ e^{ i\left(\mathbf{r}_n\cdot \mathbf{k} - \omega t\right)} \right]. \label{eqn:B_field}
\end{equation}
These field measurements represent a single wave that has unit amplitude, zero phase, wavevector $\mathbf{k}$, and frequency $\omega$. The frequency value $\omega$ is randomly drawn from a uniform distribution over the interval of detectable frequencies $\omega \sim U\left[0, 1/2\Delta t \right)$. We sample 64 points in time for each spacecraft and partition the data into $Q=4$ subintervals to obtain our expected value quantities via ensemble averaging. We then perform a discrete Fourier transform to move from the time to the frequency domain. Finally, we perform the wave-telescope technique outlined in \S \ref{ssec:WT} for each spacecraft configuration and wavevector combination to estimate the wavevector $\mathbf{k}_{calc}$ from the simulated spacecraft data. We use the scanning procedure outlined in Appendix \S \ref{sec:appendix.scan_alg} to speed up the computations of the wave-telescope technique. This yields a dataset of approximately 525,000 samples for each number of spacecraft $N$. For each of these simulated waves, we compute the error in its wavevector identification as
\begin{equation}
    Error = 100 \frac{\|\mathbf{k}_{calc} - \mathbf{k} \|_2}{\|\mathbf{k}\|_2}. \label{eqn:Error}
\end{equation}

\section{Learning Equations}
\label{sec:learn_eqns}
In section \S \ref{ssec:explore} we visualize the complete results of the wave-telescope simulations. We then describe our method of filtering the data and selecting the subset of data that we will use in the rest of the analysis in \S \ref{ssec:data_filter}. We define equation forms that fit these results in \S \ref{ssec:eqn_forms}. We then verify that Bayesian inference is capable of learning coefficients in these equations using our dataset in \S \ref{ssec:verify_bayes} before we apply Bayesian inference to the filtered simulation data to gain posterior distributions for all 12 coefficients in \S \ref{ssec:coefs}. Finally, we verify that the found equations correctly classify the errors occurring from the wave-telescope technique in \S \ref{ssec:PPC}.

\subsection{Data Exploration}
\label{ssec:explore}
\begin{figure}[ht]
\centering
\begin{tabular}{ccc}
\includegraphics[width=0.31\textwidth]{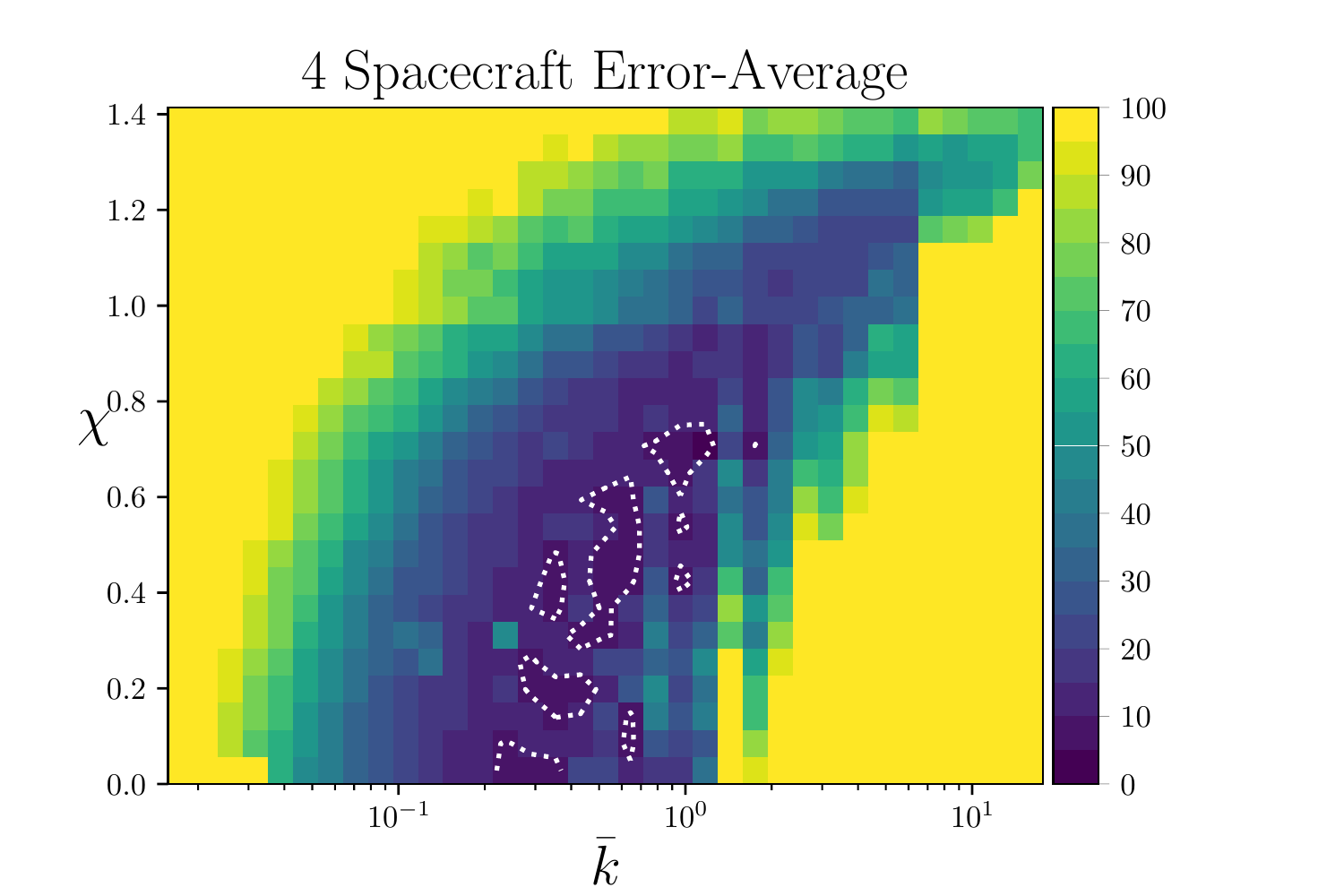} &
\includegraphics[width=0.31\textwidth]{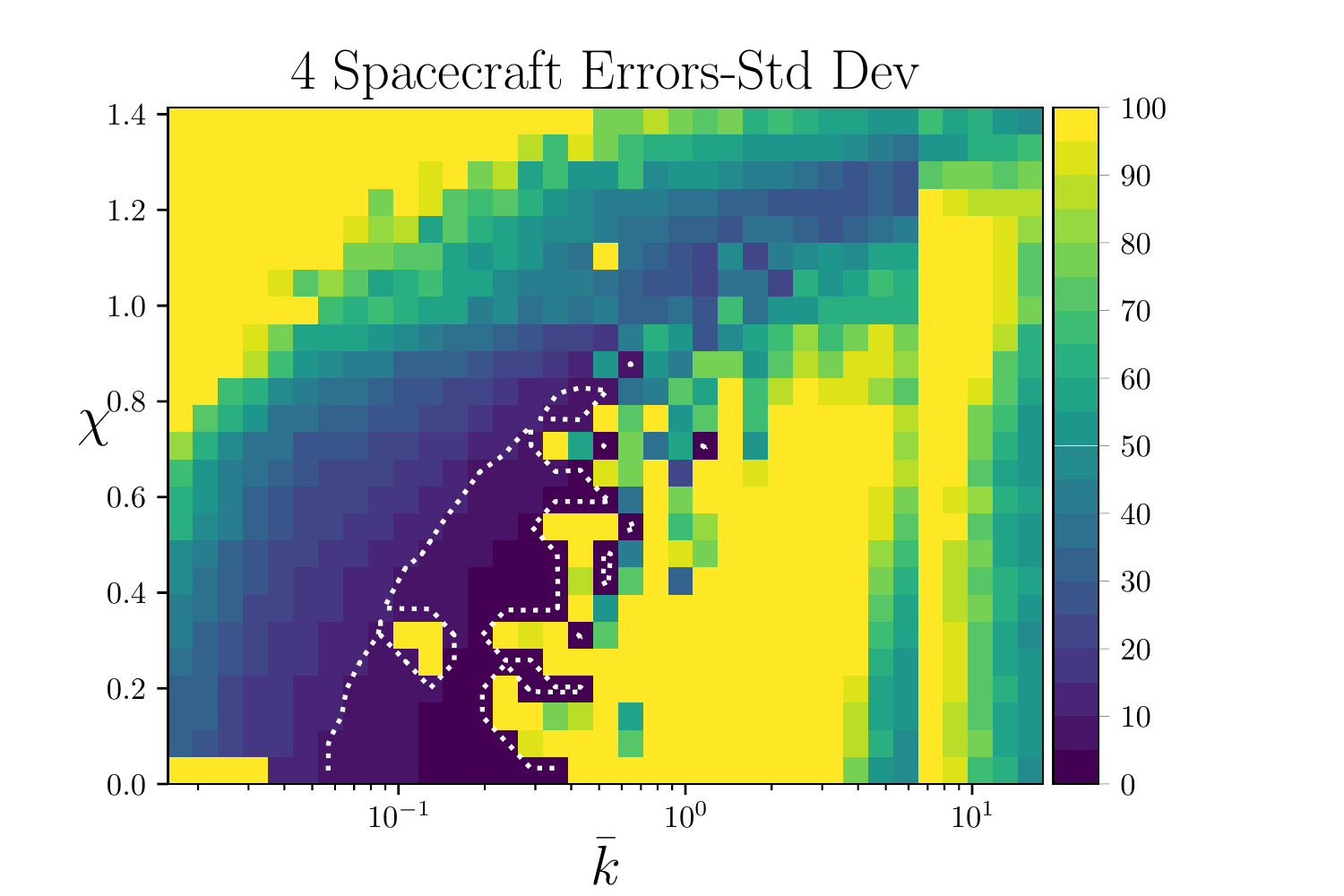} &
\includegraphics[width=0.31\textwidth]{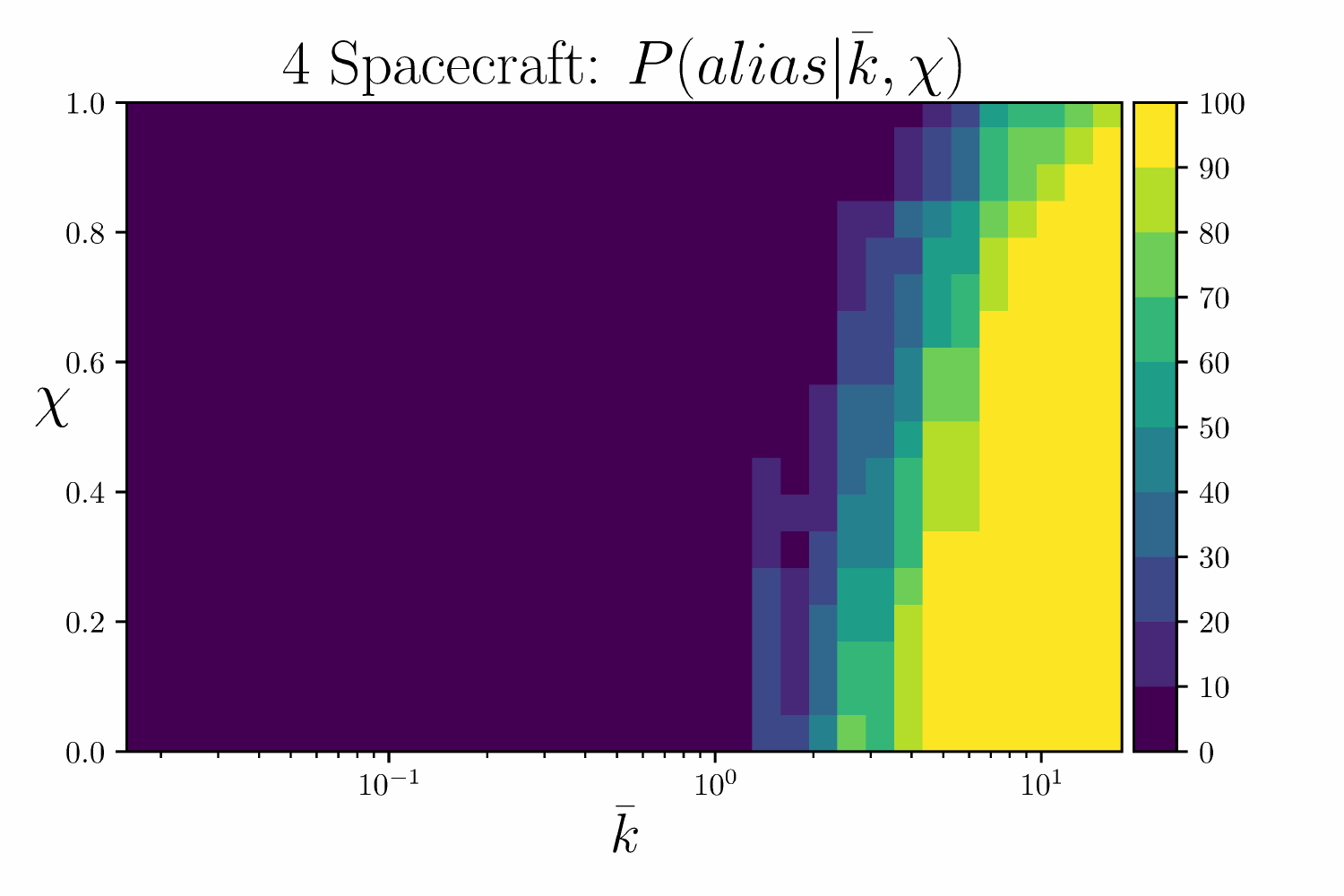} \\
\includegraphics[width=0.31\textwidth]{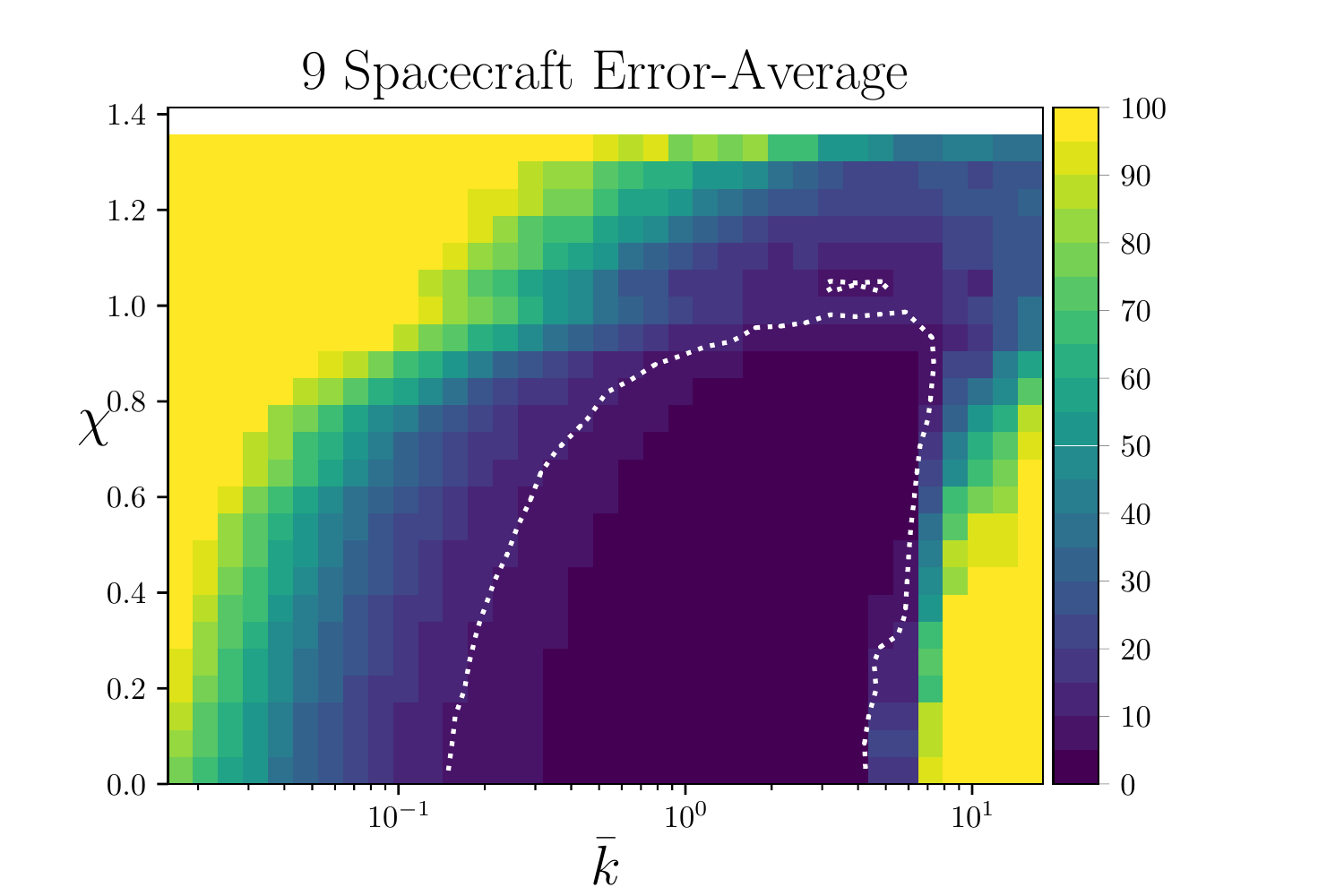} &
\includegraphics[width=0.31\textwidth]{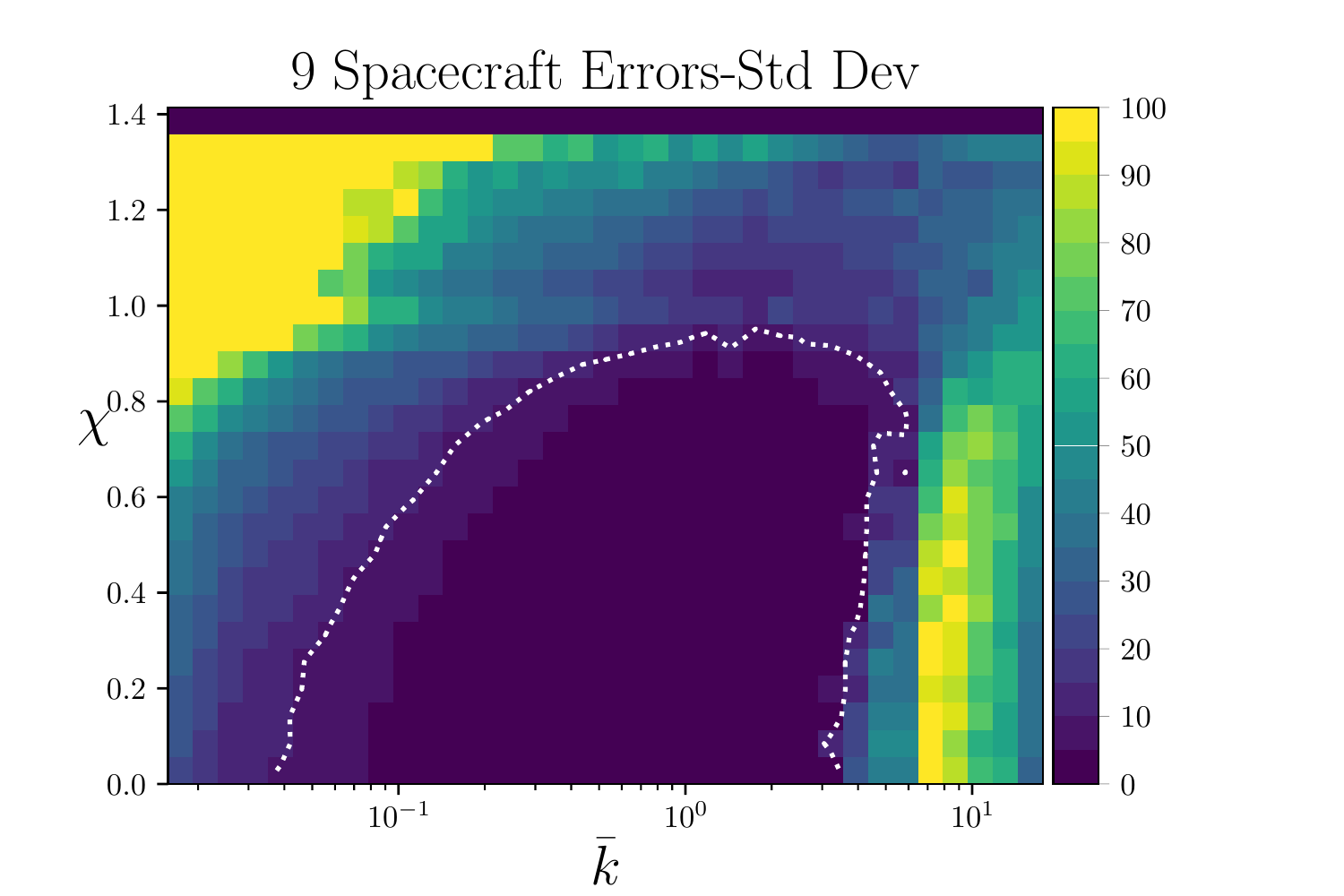} &
\includegraphics[width=0.31\textwidth]{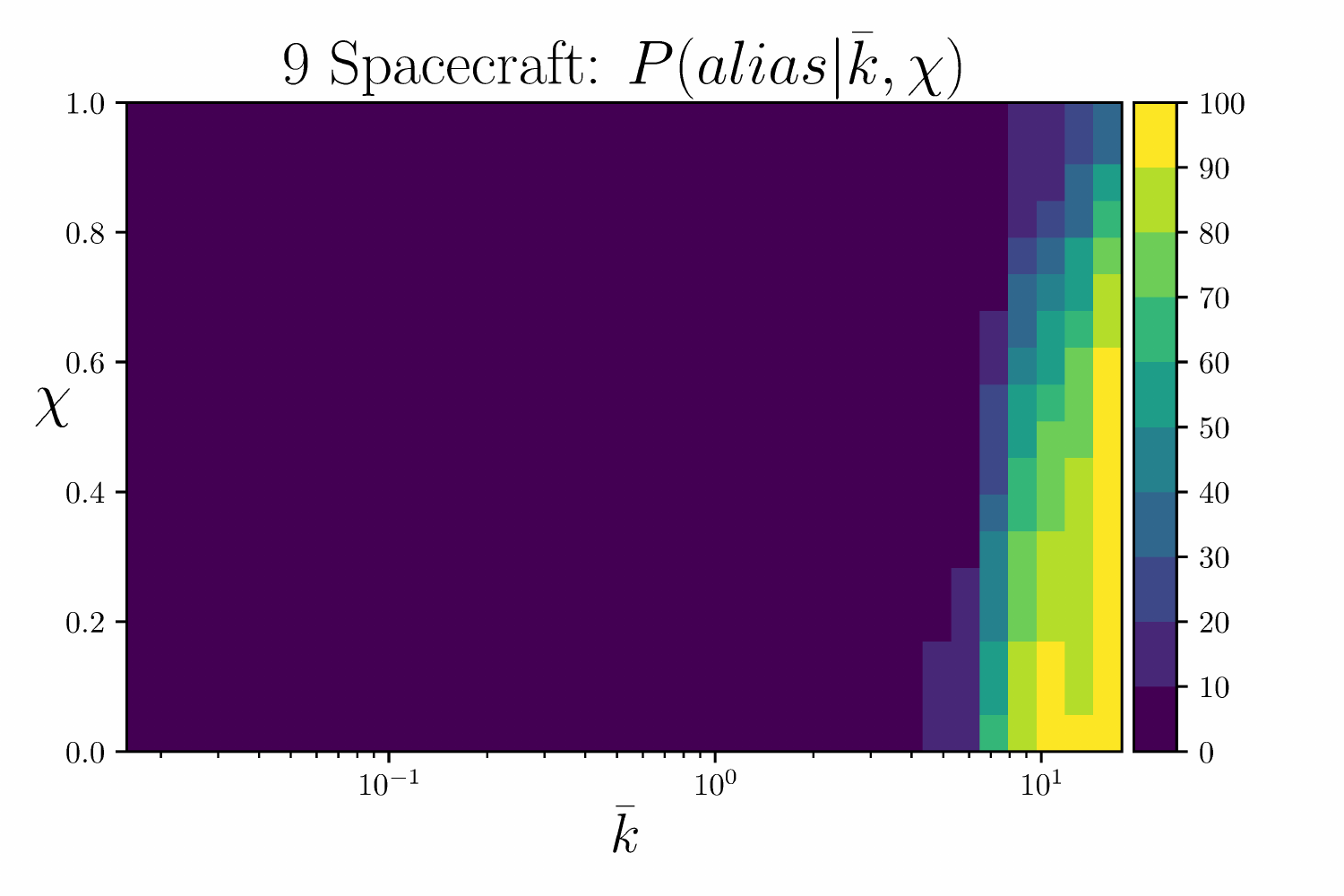} 
\end{tabular}
\caption{{\small Using data from many numerical simulations, we show the average error (in color, defined by eqn \ref{eqn:Error}) associated with detecting a wave with wavevector relative magnitude $\bar{k}$ using a four or nine-spacecraft configuration with shape parameter $\chi$ in the left panels. In the center panels we show the standard deviation of the data at each wavevector magnitude and shape parameter value. In the right panels we show the observed probability that a wavevector signal was aliased. The white dotted lines show the regions where average error and standard deviation are less than 10\%. A complete figure set (18 images) for four through nine spacecraft configurations is available in the online journal. }}
\label{fig:mu_data} 
\end{figure}

In order to extend our display of errors found in Figure \ref{fig:Sahrauri}, we plot the average error as a function of the relative magnitude of the true wavevector $\bar{k}$ and the shape parameters $\chi$ in the left panels of Figure \ref{fig:mu_data}. We plot the standard deviation as a function of $\bar{k}$ and $\chi$ in the central panels of Figure \ref{fig:mu_data} and plot the observed probability that a wavevector signal was aliased in the right panels of Figure \ref{fig:mu_data}. The average errors, standard deviations, and aliasing probabilities are calculated over elements located within linearly binned values for $\chi$. Results for all numbers of spacecraft $N$ can be found in the figure set corresponding to Figure \ref{fig:mu_data}.



\figsetgrpstart
\figsetgrpnum{3.1}
\figsetgrptitle{Four-Spacecraft Average Error}
\figsetplot{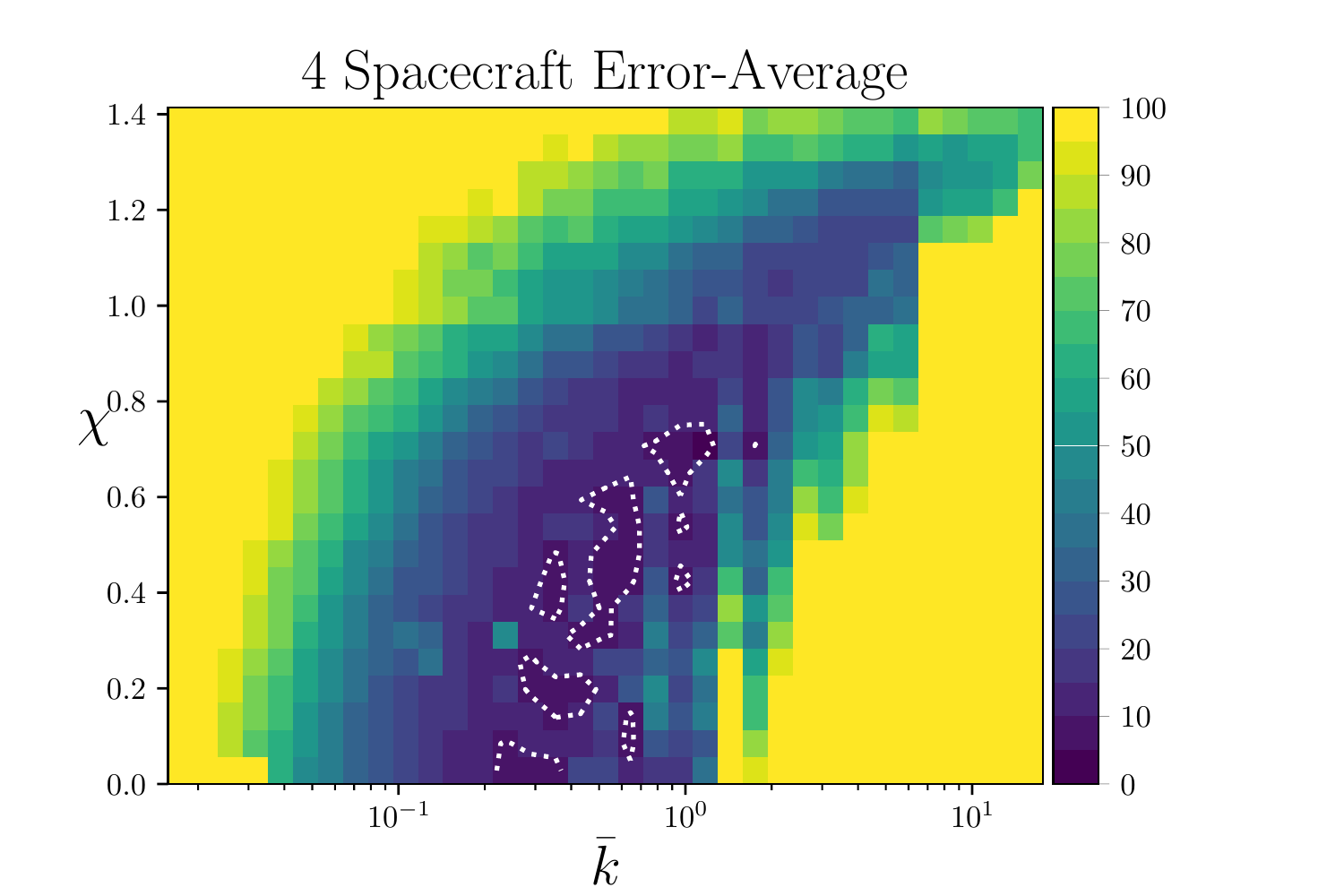}
\figsetgrpnote{Using data from many numerical simulations, we show the average error (in color, defined by eqn \ref{eqn:Error}) associated with detecting a wave with wavevector relative magnitude $\bar{k}$ using a four-spacecraft configuration with shape parameter $\chi$. The white dotted lines show the regions where average error is less than 10\%.}
\figsetgrpend

\figsetgrpstart
\figsetgrpnum{3.2}
\figsetgrptitle{Five-Spacecraft Average Error}
\figsetplot{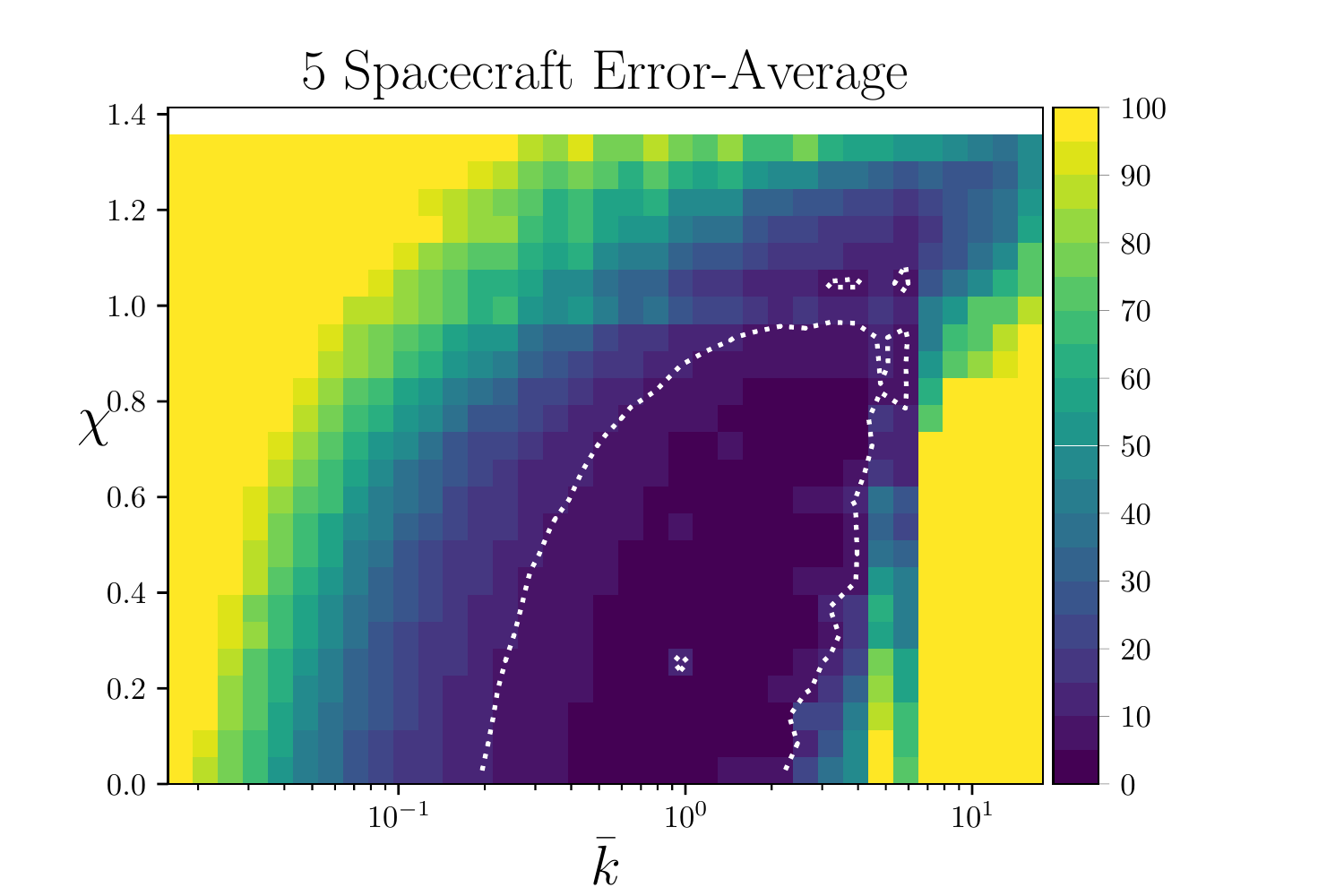}
\figsetgrpnote{Using data from many numerical simulations, we show the average error (in color, defined by eqn \ref{eqn:Error}) associated with detecting a wave with wavevector relative magnitude $\bar{k}$ using a five-spacecraft configuration with shape parameter $\chi$. The white dotted lines show the regions where average error is less than 10\%.}
\figsetgrpend

\figsetgrpstart
\figsetgrpnum{3.3}
\figsetgrptitle{Six-Spacecraft Average Error}
\figsetplot{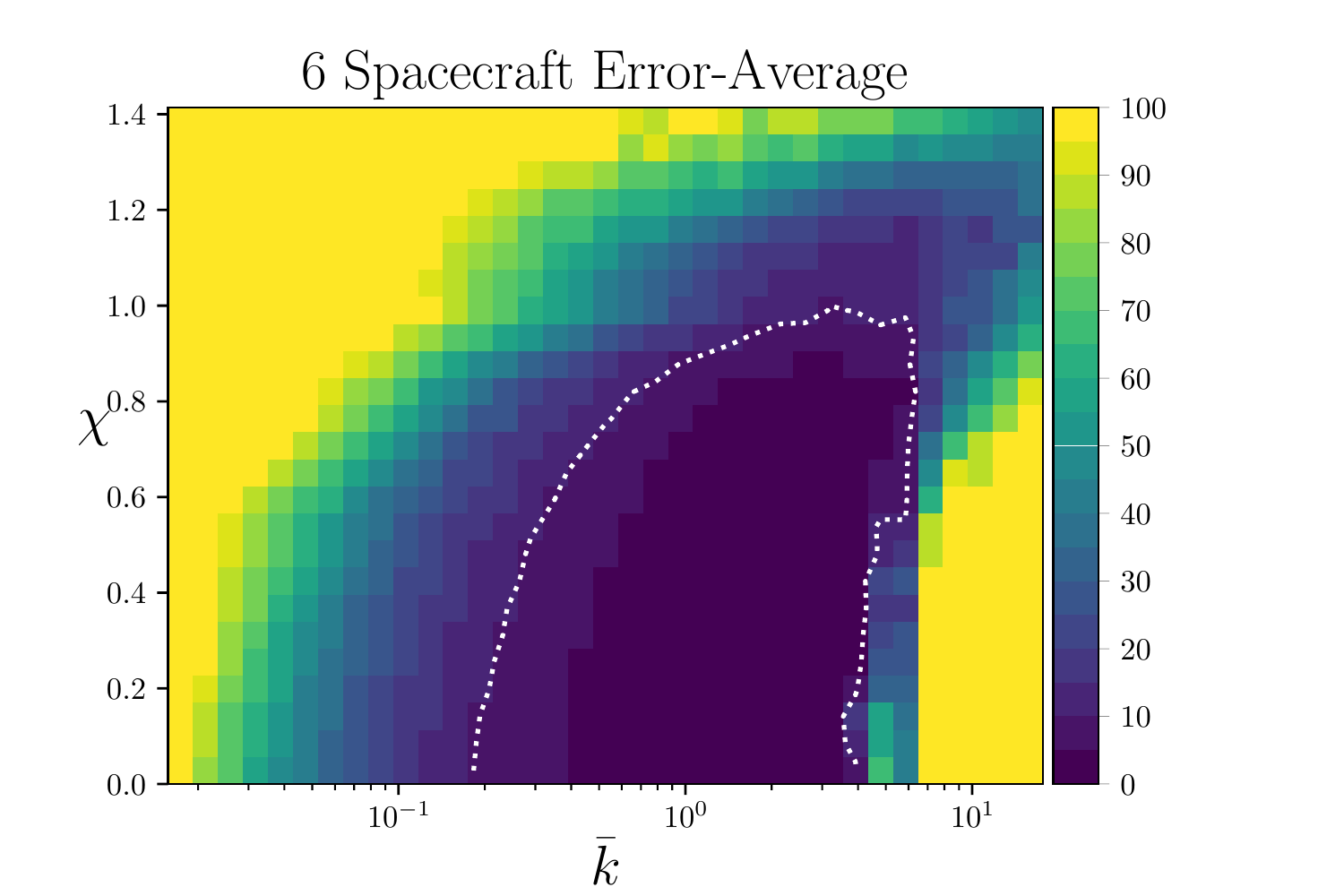}
\figsetgrpnote{Using data from many numerical simulations, we show the average error (in color, defined by eqn \ref{eqn:Error}) associated with detecting a wave with wavevector relative magnitude $\bar{k}$ using a six-spacecraft configuration with shape parameter $\chi$. The white dotted lines show the regions where average error is less than 10\%.}
\figsetgrpend

\figsetgrpstart
\figsetgrpnum{3.4}
\figsetgrptitle{Seven-Spacecraft Average Error}
\figsetplot{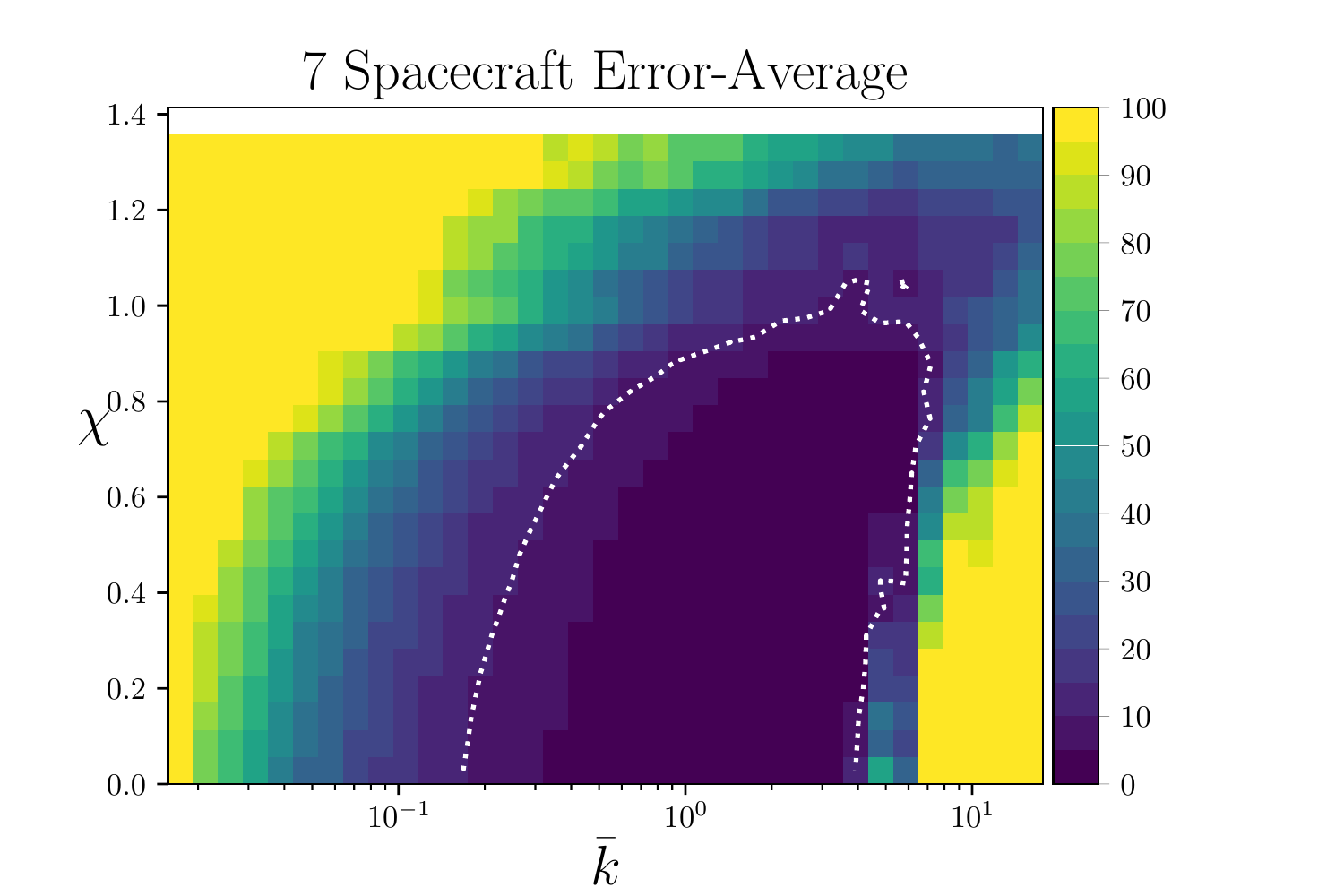}
\figsetgrpnote{Using data from many numerical simulations, we show the average error (in color, defined by eqn \ref{eqn:Error}) associated with detecting a wave with wavevector relative magnitude $\bar{k}$ using a seven-spacecraft configuration with shape parameter $\chi$. The white dotted lines show the regions where average error is less than 10\%.}
\figsetgrpend

\figsetgrpstart
\figsetgrpnum{3.5}
\figsetgrptitle{Eight-Spacecraft Average Error}
\figsetplot{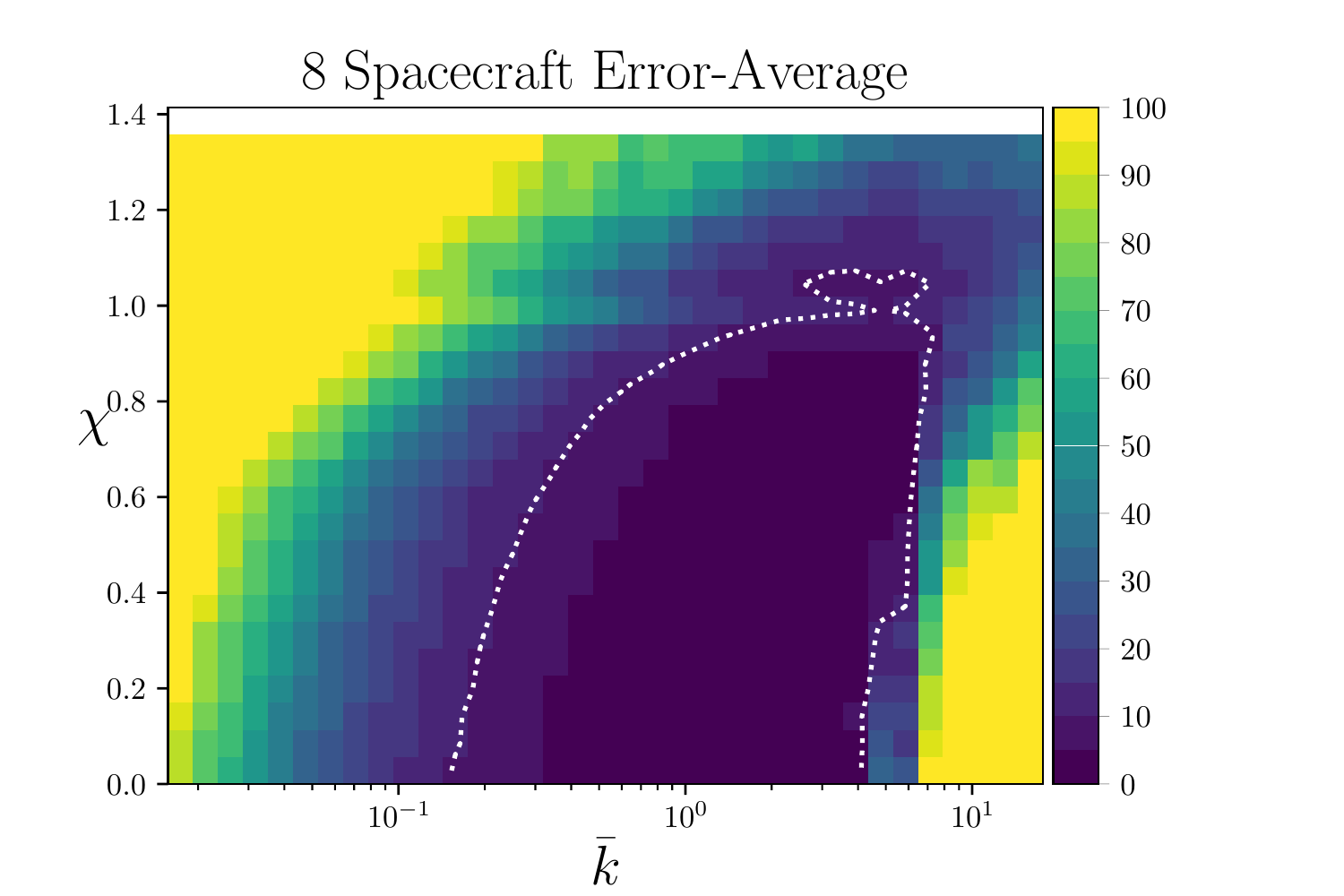}
\figsetgrpnote{Using data from many numerical simulations, we show the average error (in color, defined by eqn \ref{eqn:Error}) associated with detecting a wave with wavevector relative magnitude $\bar{k}$ using a eight-spacecraft configuration with shape parameter $\chi$. The white dotted lines show the regions where average error is less than 10\%.}
\figsetgrpend

\figsetgrpstart
\figsetgrpnum{3.6}
\figsetgrptitle{Nine-Spacecraft Average Error}
\figsetplot{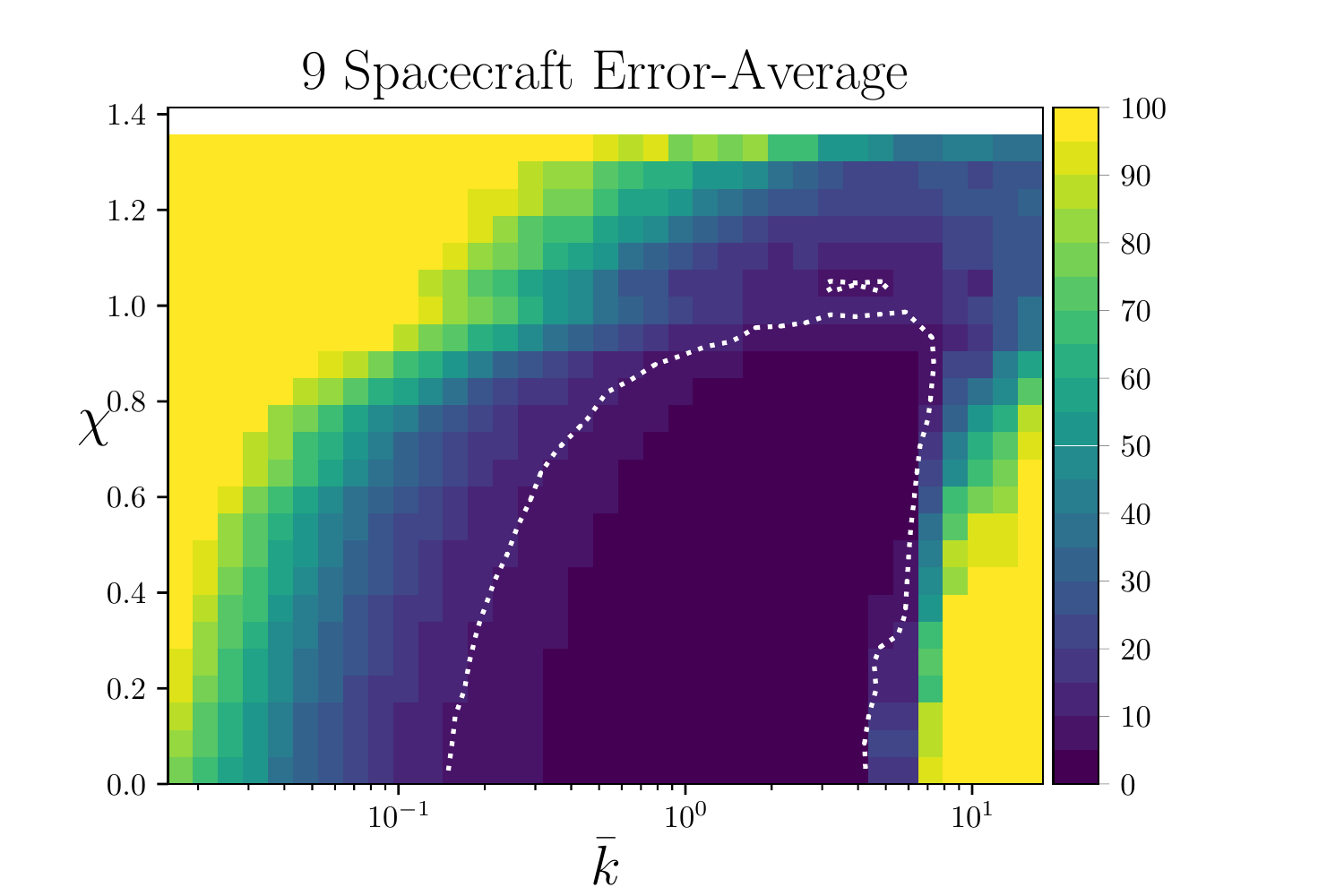}
\figsetgrpnote{Using data from many numerical simulations, we show the average error (in color, defined by eqn \ref{eqn:Error}) associated with detecting a wave with wavevector relative magnitude $\bar{k}$ using a nine-spacecraft configuration with shape parameter $\chi$. The white dotted lines show the regions where average error is less than 10\%.}
\figsetgrpend

\figsetgrpstart
\figsetgrpnum{3.7}
\figsetgrptitle{Four-Spacecraft Standard Deviation of Error}
\figsetplot{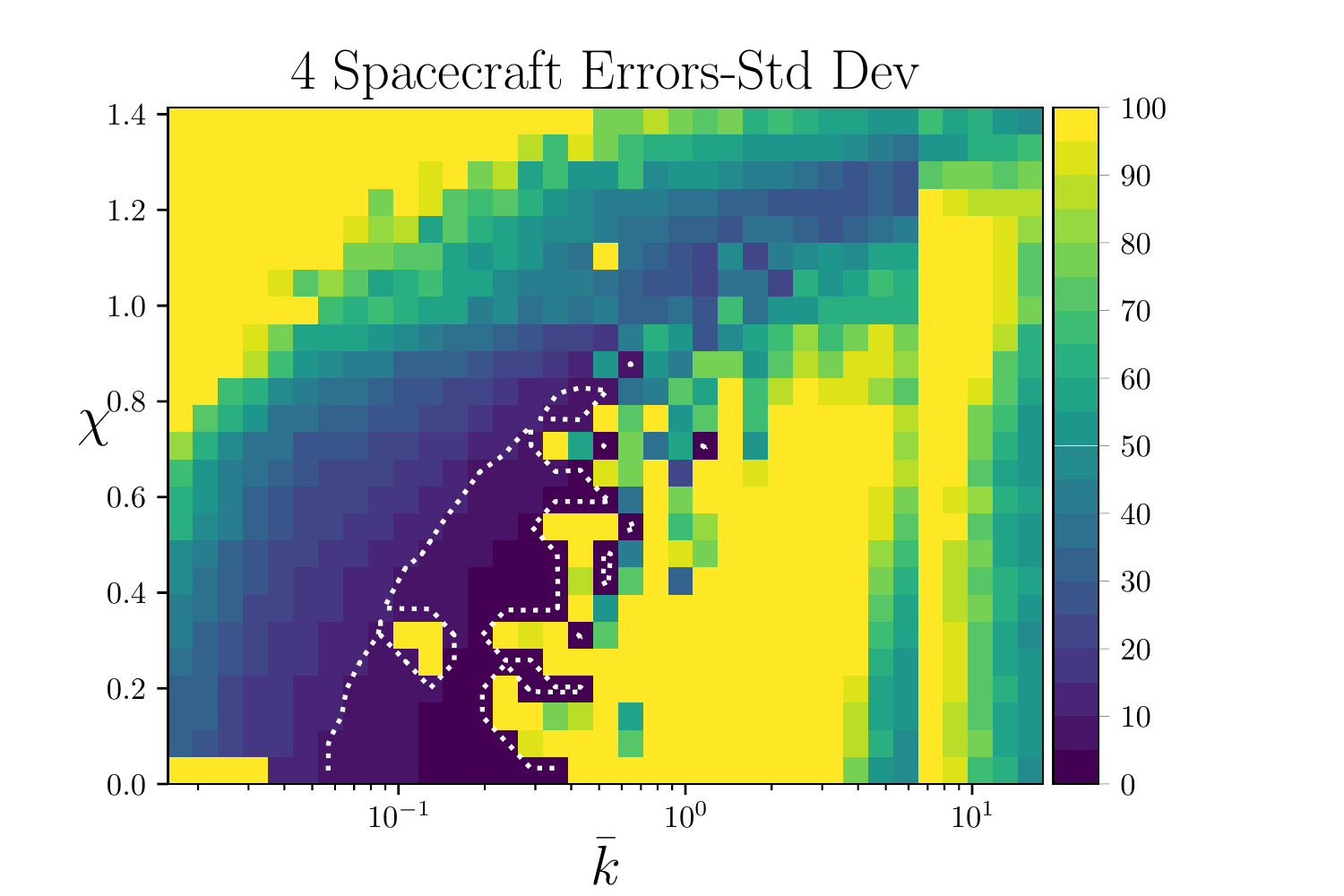}
\figsetgrpnote{Using data from many numerical simulations, we show the standard deviation of the error (in color, defined by eqn \ref{eqn:Error}) associated with detecting a wave with wavevector relative magnitude $\bar{k}$ using a four-spacecraft configuration with shape parameter $\chi$. The white dotted lines show the regions where the standard deviation of the error is less than 10\%.}
\figsetgrpend

\figsetgrpstart
\figsetgrpnum{3.8}
\figsetgrptitle{Five-Spacecraft Standard Deviation of Error}
\figsetplot{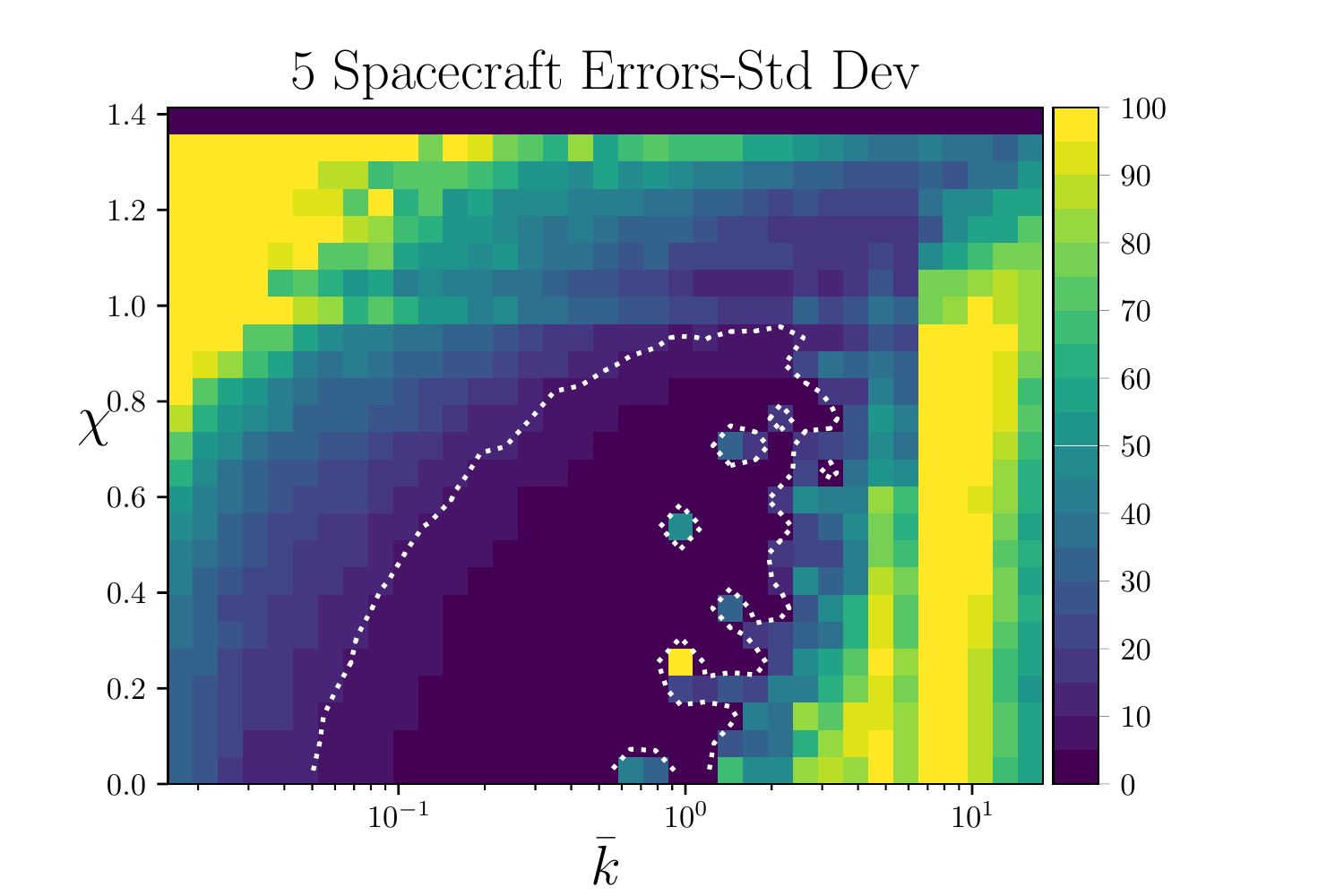}
\figsetgrpnote{Using data from many numerical simulations, we show the standard deviation of the error (in color, defined by eqn \ref{eqn:Error}) associated with detecting a wave with wavevector relative magnitude $\bar{k}$ using a five-spacecraft configuration with shape parameter $\chi$. The white dotted lines show the regions where the standard deviation of the error is less than 10\%.}
\figsetgrpend

\figsetgrpstart
\figsetgrpnum{3.9}
\figsetgrptitle{Six-Spacecraft Standard Deviation of Error}
\figsetplot{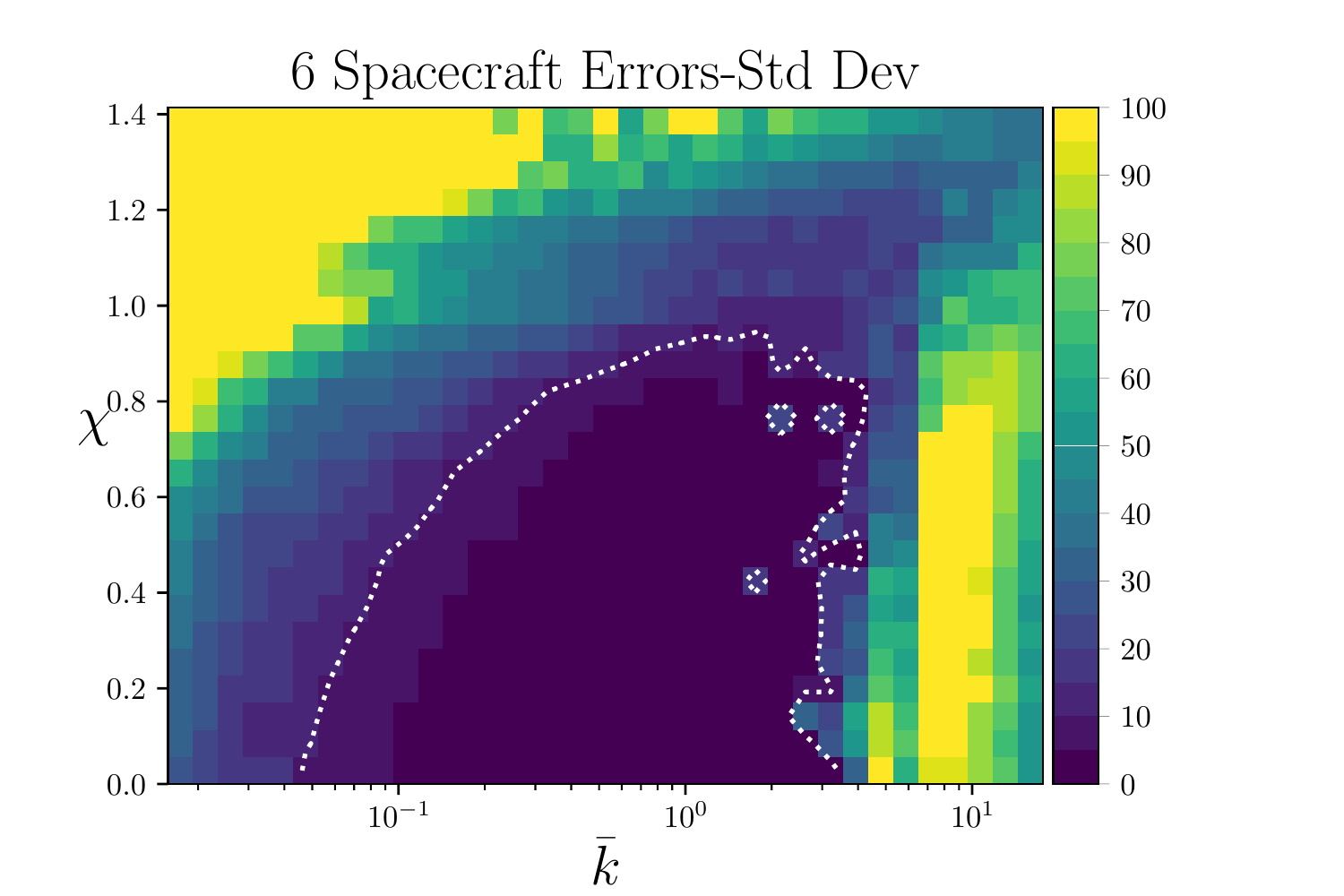}
\figsetgrpnote{Using data from many numerical simulations, we show the standard deviation of the error (in color, defined by eqn \ref{eqn:Error}) associated with detecting a wave with wavevector relative magnitude $\bar{k}$ using a six-spacecraft configuration with shape parameter $\chi$. The white dotted lines show the regions where the standard deviation of the error is less than 10\%.}
\figsetgrpend

\figsetgrpstart
\figsetgrpnum{3.10}
\figsetgrptitle{Seven-Spacecraft Standard Deviation of Error}
\figsetplot{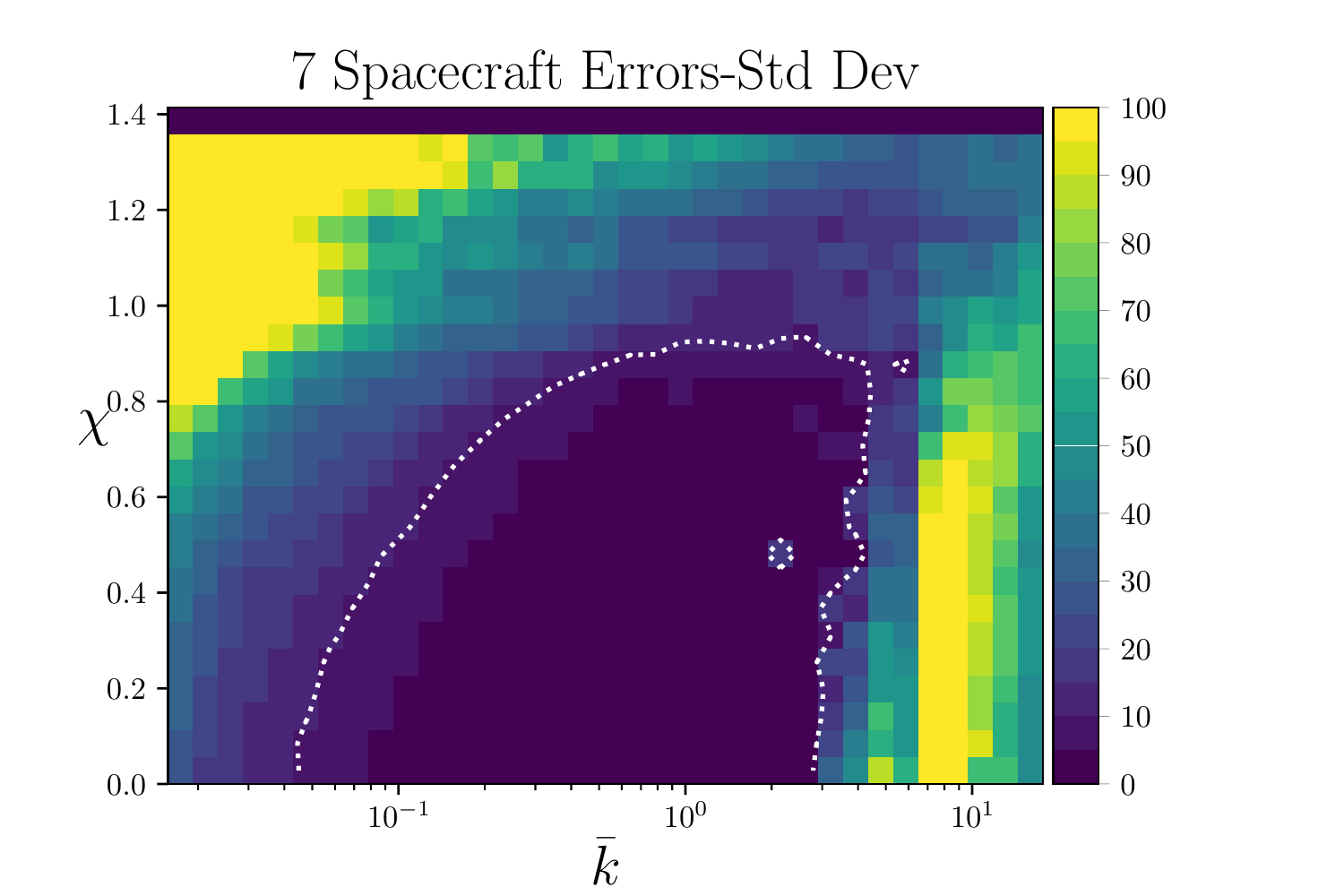}
\figsetgrpnote{Using data from many numerical simulations, we show the standard deviation of the error (in color, defined by eqn \ref{eqn:Error}) associated with detecting a wave with wavevector relative magnitude $\bar{k}$ using a seven-spacecraft configuration with shape parameter $\chi$. The white dotted lines show the regions where the standard deviation of the error is less than 10\%.}
\figsetgrpend

\figsetgrpstart
\figsetgrpnum{3.11}
\figsetgrptitle{Eight-Spacecraft Standard Deviation of Error}
\figsetplot{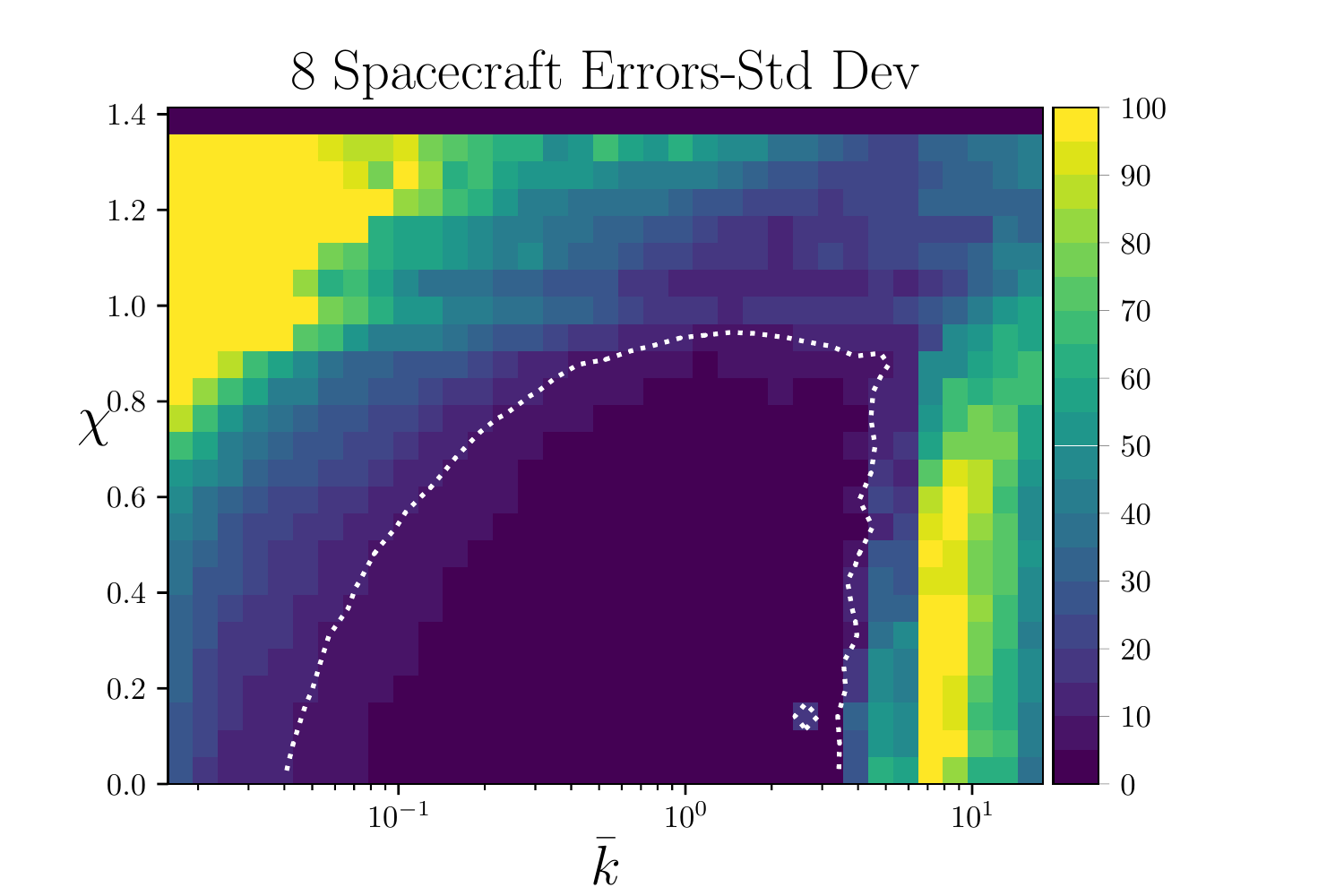}
\figsetgrpnote{Using data from many numerical simulations, we show the standard deviation of the error (in color, defined by eqn \ref{eqn:Error}) associated with detecting a wave with wavevector relative magnitude $\bar{k}$ using a eight-spacecraft configuration with shape parameter $\chi$. The white dotted lines show the regions where the standard deviation of the error is less than 10\%.}
\figsetgrpend

\figsetgrpstart
\figsetgrpnum{3.12}
\figsetgrptitle{Nine-Spacecraft Standard Deviation of Error}
\figsetplot{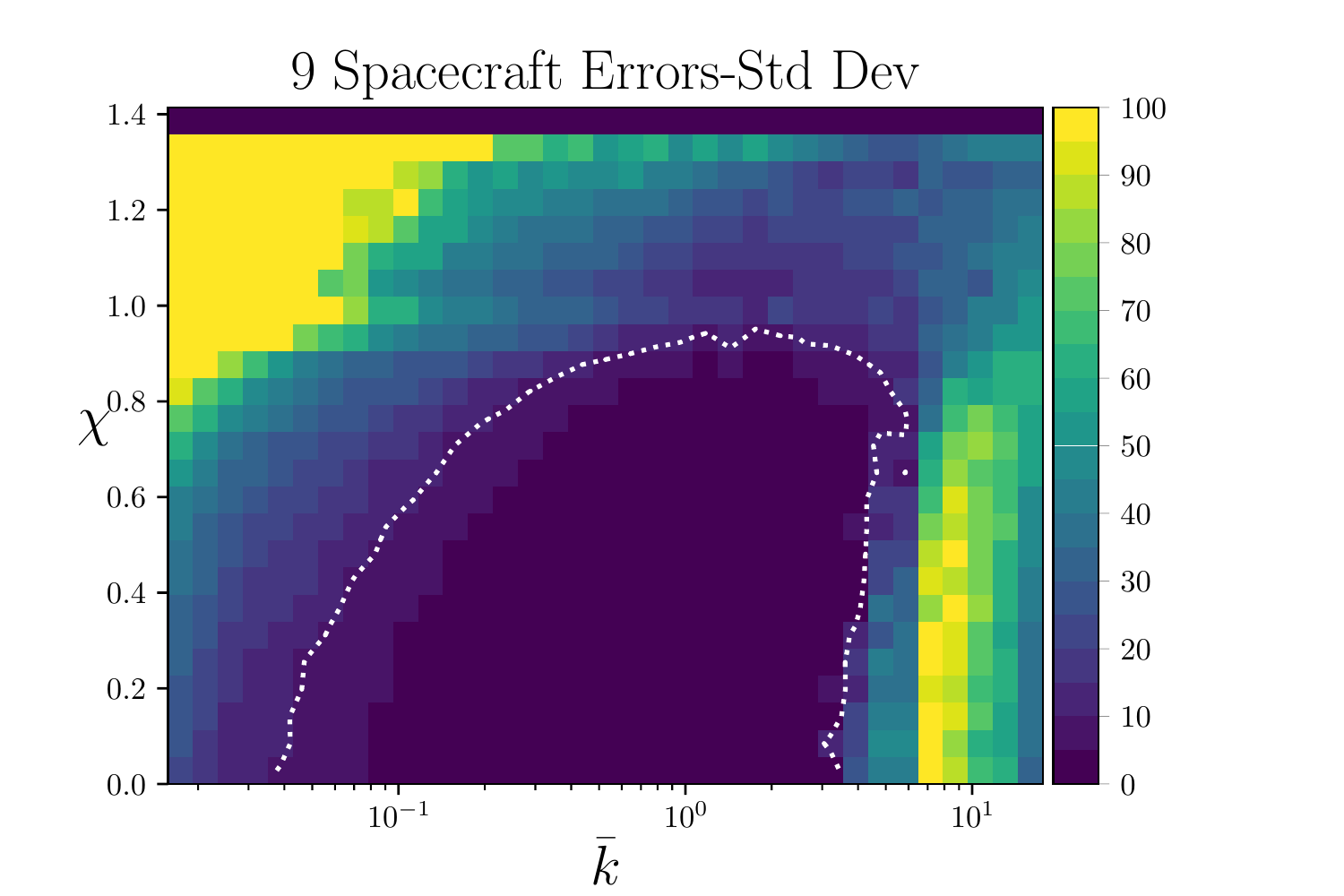}
\figsetgrpnote{Using data from many numerical simulations, we show the standard deviation of the error (in color, defined by eqn \ref{eqn:Error}) associated with detecting a wave with wavevector relative magnitude $\bar{k}$ using a nine-spacecraft configuration with shape parameter $\chi$. The white dotted lines show the regions where the standard deviation of the error is less than 10\%.}
\figsetgrpend

\figsetgrpnum{3.13}
\figsetgrptitle{Four-Spacecraft Probability of Aliasing}
\figsetplot{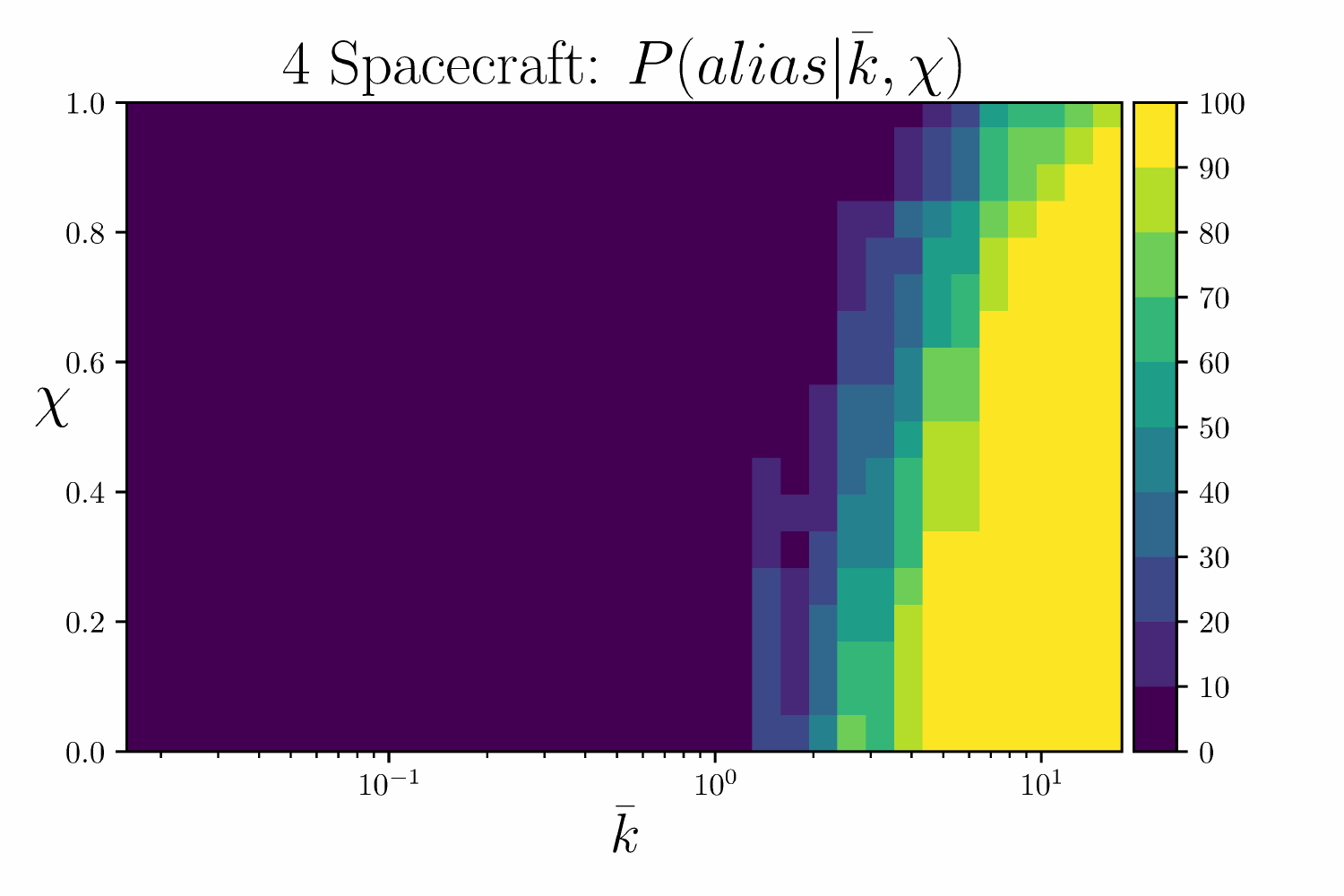}
\figsetgrpnote{Using data from many numerical simulations, we show the Probability of aliasing (in color, defined by eqn \ref{eqn:Error}) associated with detecting a wave with wavevector relative magnitude $\bar{k}$ using a four-spacecraft configuration with shape parameter $\chi$.}
\figsetgrpend

\figsetgrpstart
\figsetgrpnum{3.14}
\figsetgrptitle{Five-Spacecraft Probability of Aliasing}
\figsetplot{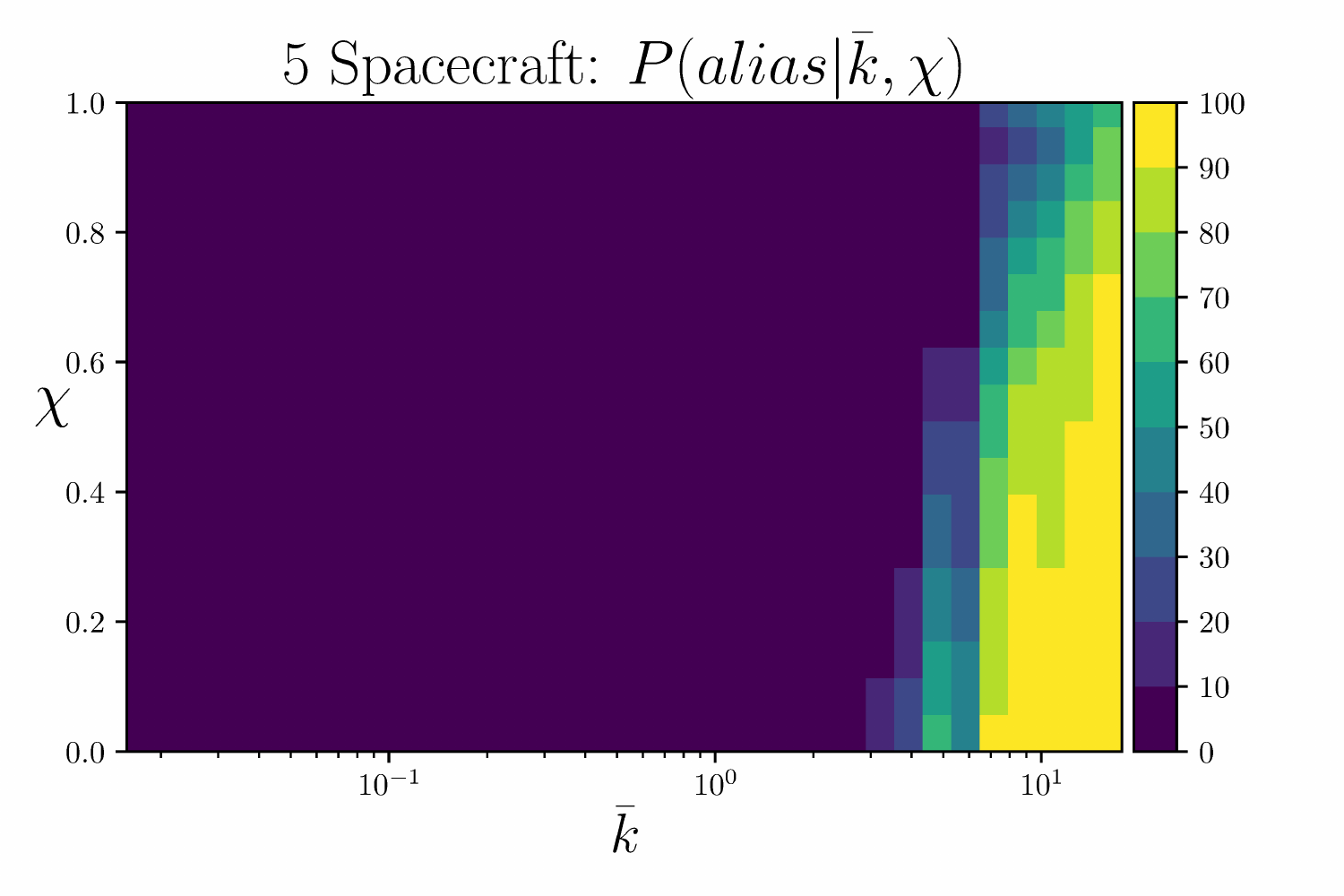}
\figsetgrpnote{Using data from many numerical simulations, we show the Probability of aliasing (in color, defined by eqn \ref{eqn:Error}) associated with detecting a wave with wavevector relative magnitude $\bar{k}$ using a five-spacecraft configuration with shape parameter $\chi$.}
\figsetgrpend

\figsetgrpstart
\figsetgrpnum{3.15}
\figsetgrptitle{Six-Spacecraft Probability of Aliasing}
\figsetplot{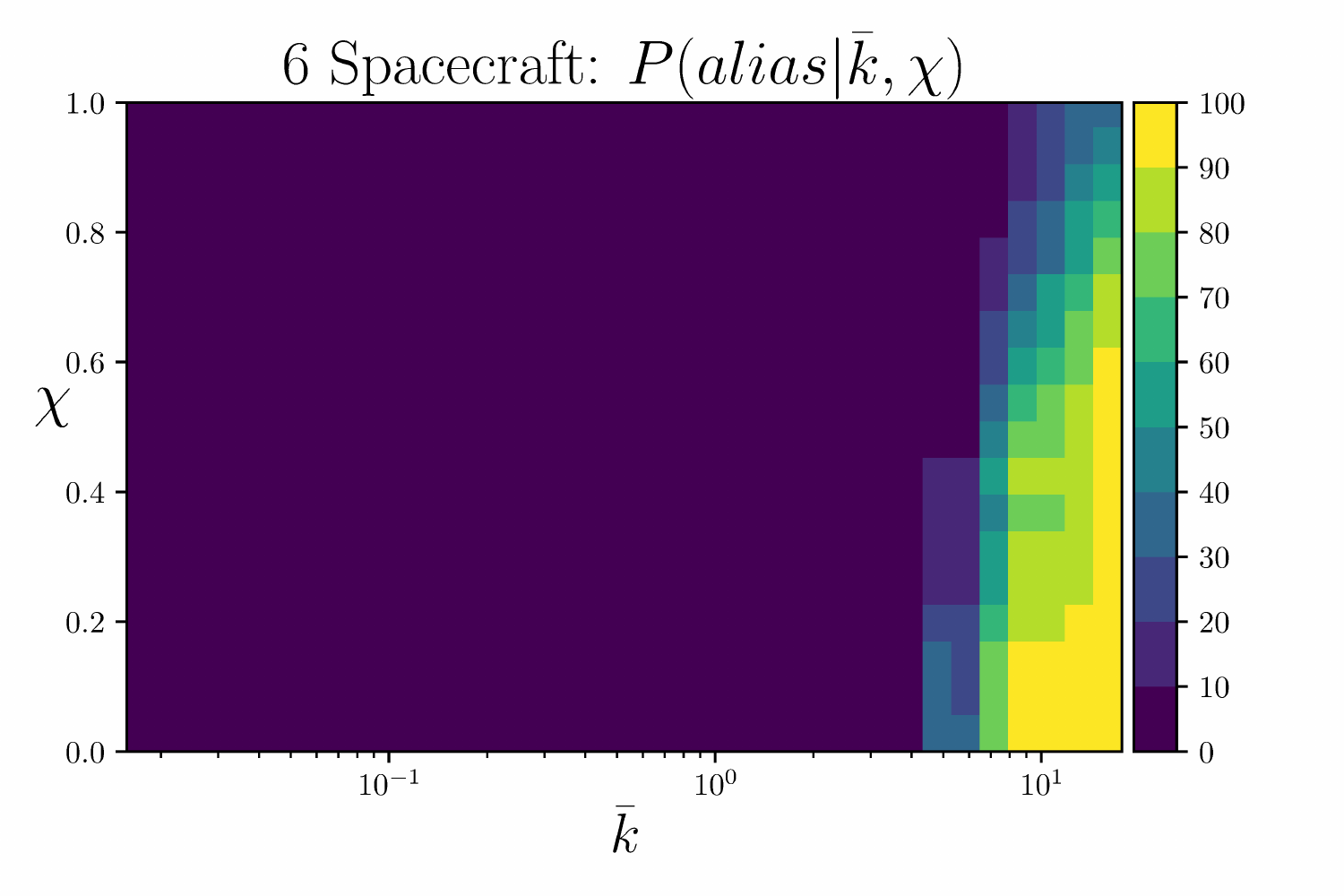}
\figsetgrpnote{Using data from many numerical simulations, we show the Probability of aliasing (in color, defined by eqn \ref{eqn:Error}) associated with detecting a wave with wavevector relative magnitude $\bar{k}$ using a six-spacecraft configuration with shape parameter $\chi$.}
\figsetgrpend

\figsetgrpstart
\figsetgrpnum{3.16}
\figsetgrptitle{Seven-Spacecraft Probability of Aliasing}
\figsetplot{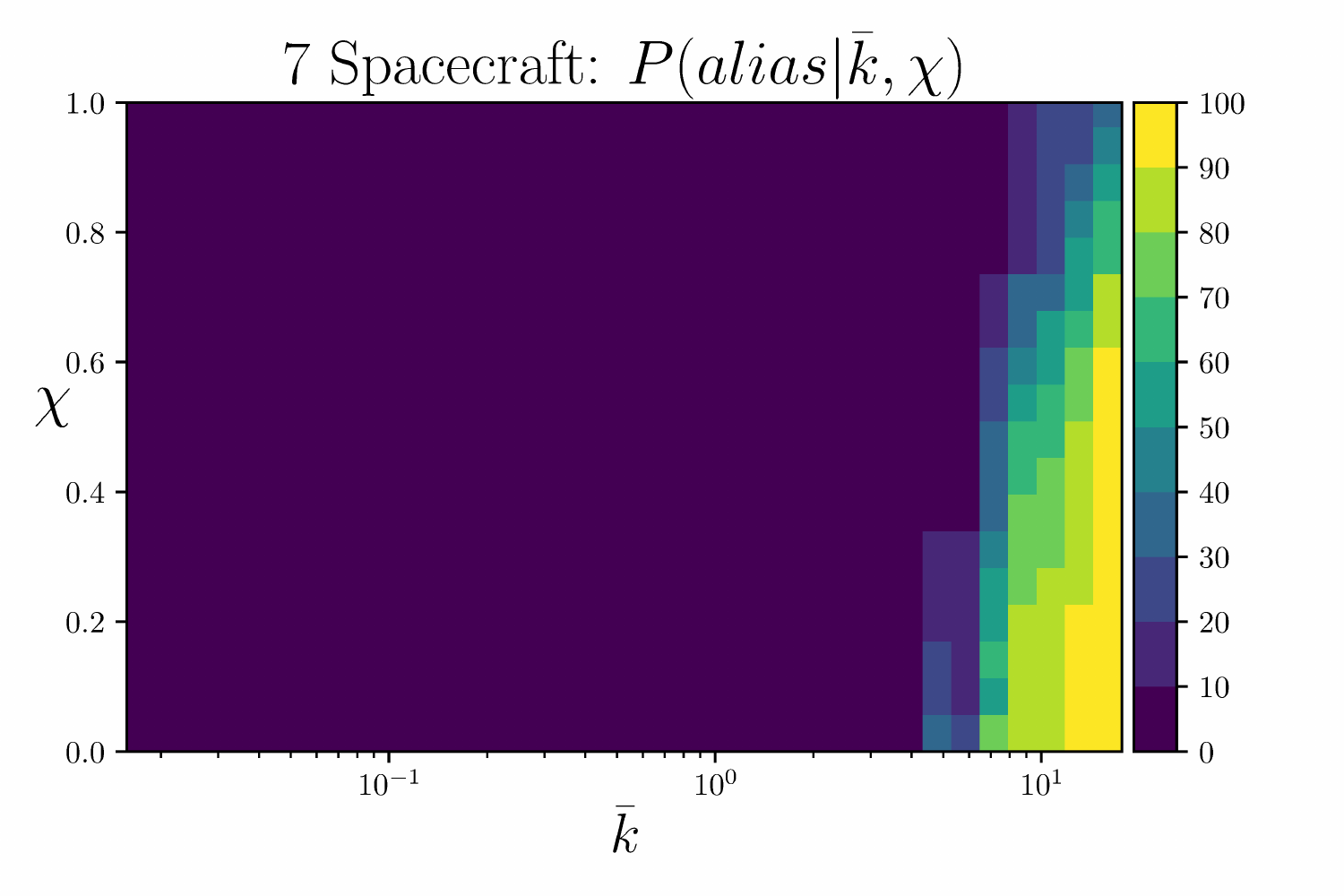}
\figsetgrpnote{Using data from many numerical simulations, we show the Probability of aliasing (in color, defined by eqn \ref{eqn:Error}) associated with detecting a wave with wavevector relative magnitude $\bar{k}$ using a seven-spacecraft configuration with shape parameter $\chi$.}
\figsetgrpend

\figsetgrpstart
\figsetgrpnum{3.17}
\figsetgrptitle{Eight-Spacecraft Probability of Aliasing}
\figsetplot{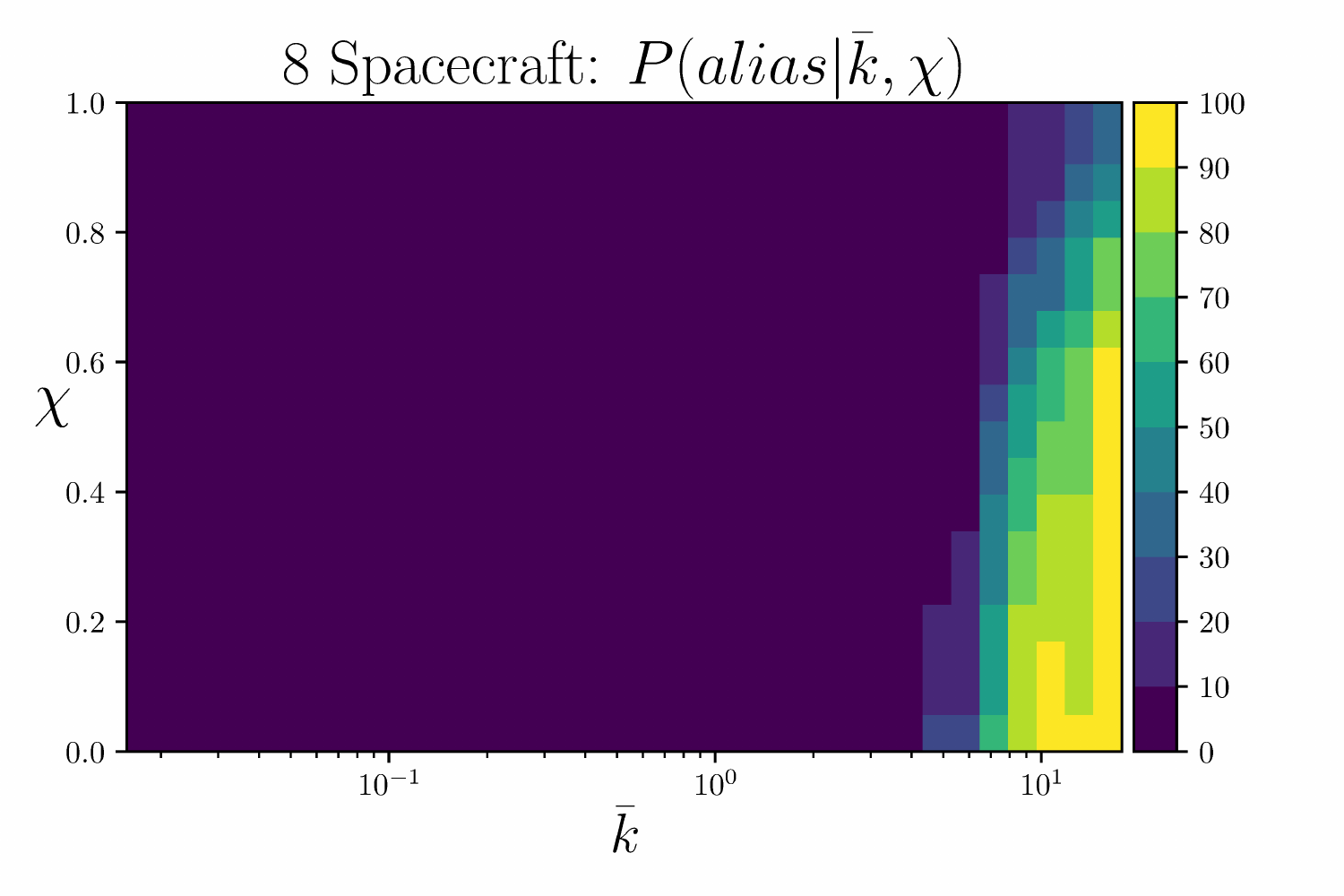}
\figsetgrpnote{Using data from many numerical simulations, we show the Probability of aliasing (in color, defined by eqn \ref{eqn:Error}) associated with detecting a wave with wavevector relative magnitude $\bar{k}$ using a eight-spacecraft configuration with shape parameter $\chi$.}
\figsetgrpend

\figsetgrpstart
\figsetgrpnum{3.18}
\figsetgrptitle{Nine-Spacecraft Probability of Aliasing}
\figsetplot{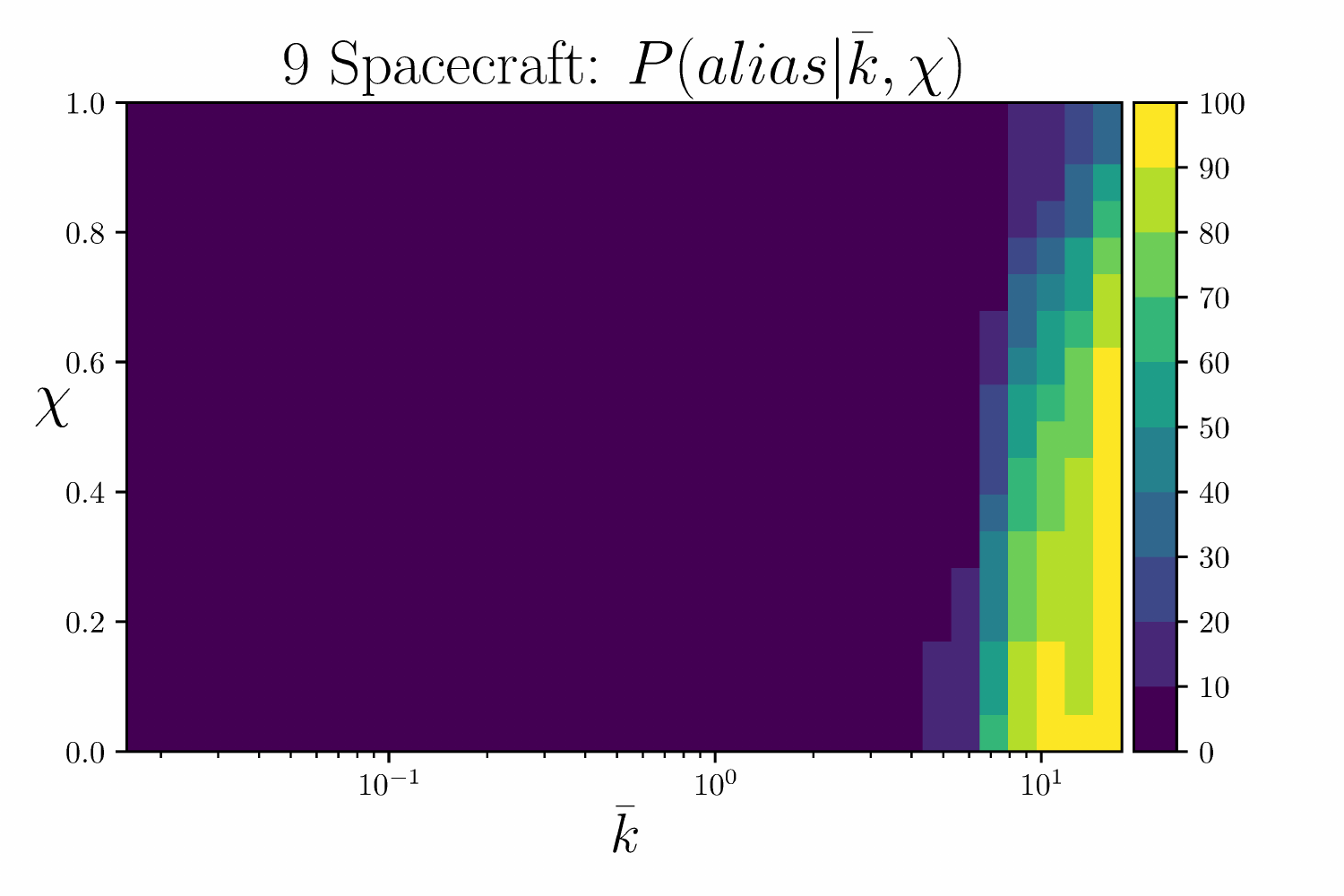}
\figsetgrpnote{Using data from many numerical simulations, we show the Probability of aliasing (in color, defined by eqn \ref{eqn:Error}) associated with detecting a wave with wavevector relative magnitude $\bar{k}$ using a nine-spacecraft configuration with shape parameter $\chi$.}
\figsetgrpend
\figsetend

Figure \ref{fig:mu_data} shows that using a four-spacecraft configuration that is well-shaped ($\chi \approx 0$) does not achieve the same minimum average error as a similarly well-shaped nine-spacecraft configuration. For the same configurations shape value $\chi$, we also note that the nine-spacecraft configuration reconstructs about 1.5 orders of magnitude of wavevectors with high accuracy, while the four-spacecraft configuration can only reconstruct 1 order of magnitude of wavevectors with similar accuracy.

\begin{figure}[ht]
\centering
\begin{tabular}{cc}
\includegraphics[width=0.46\textwidth]{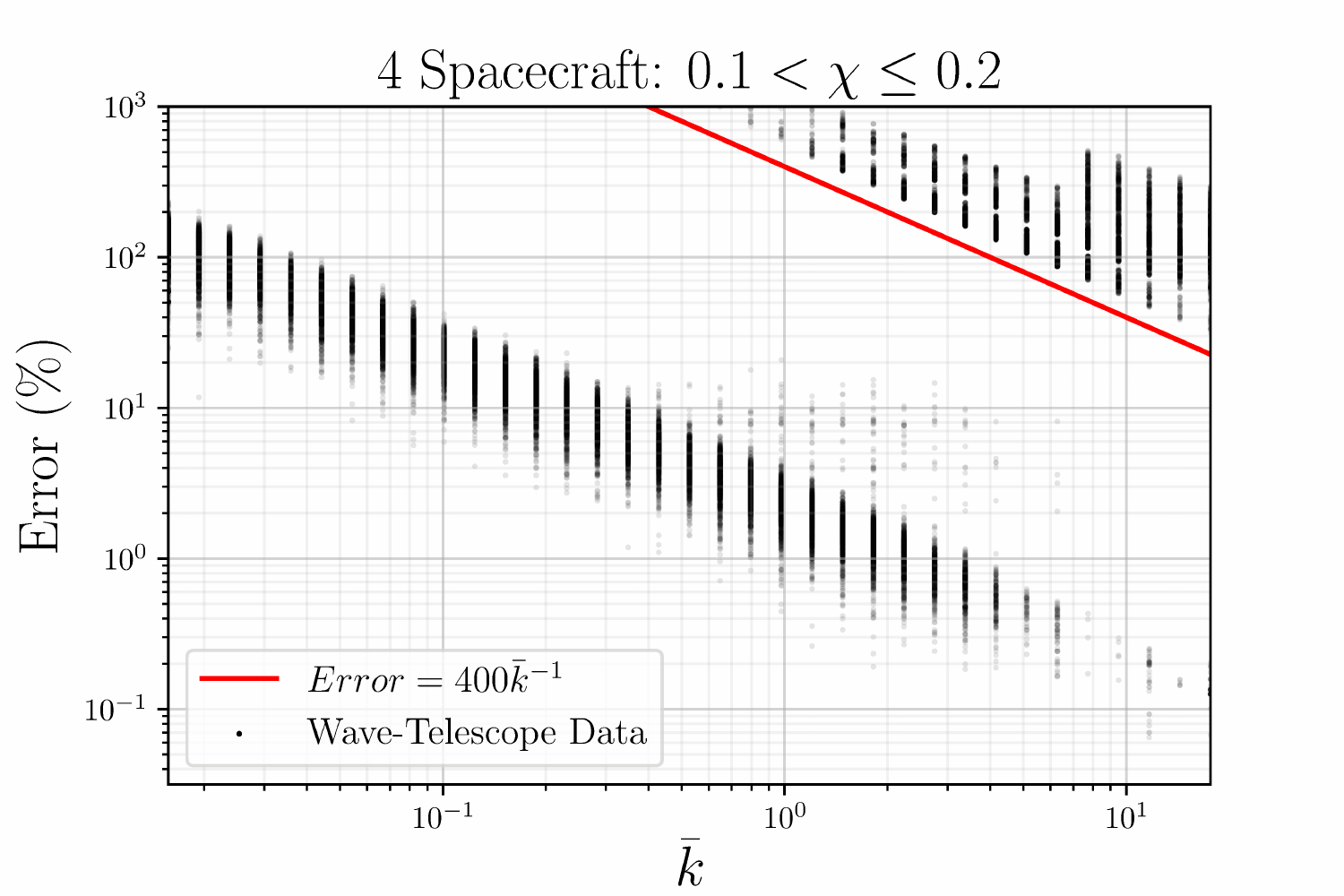} &
\includegraphics[width=0.46\textwidth]{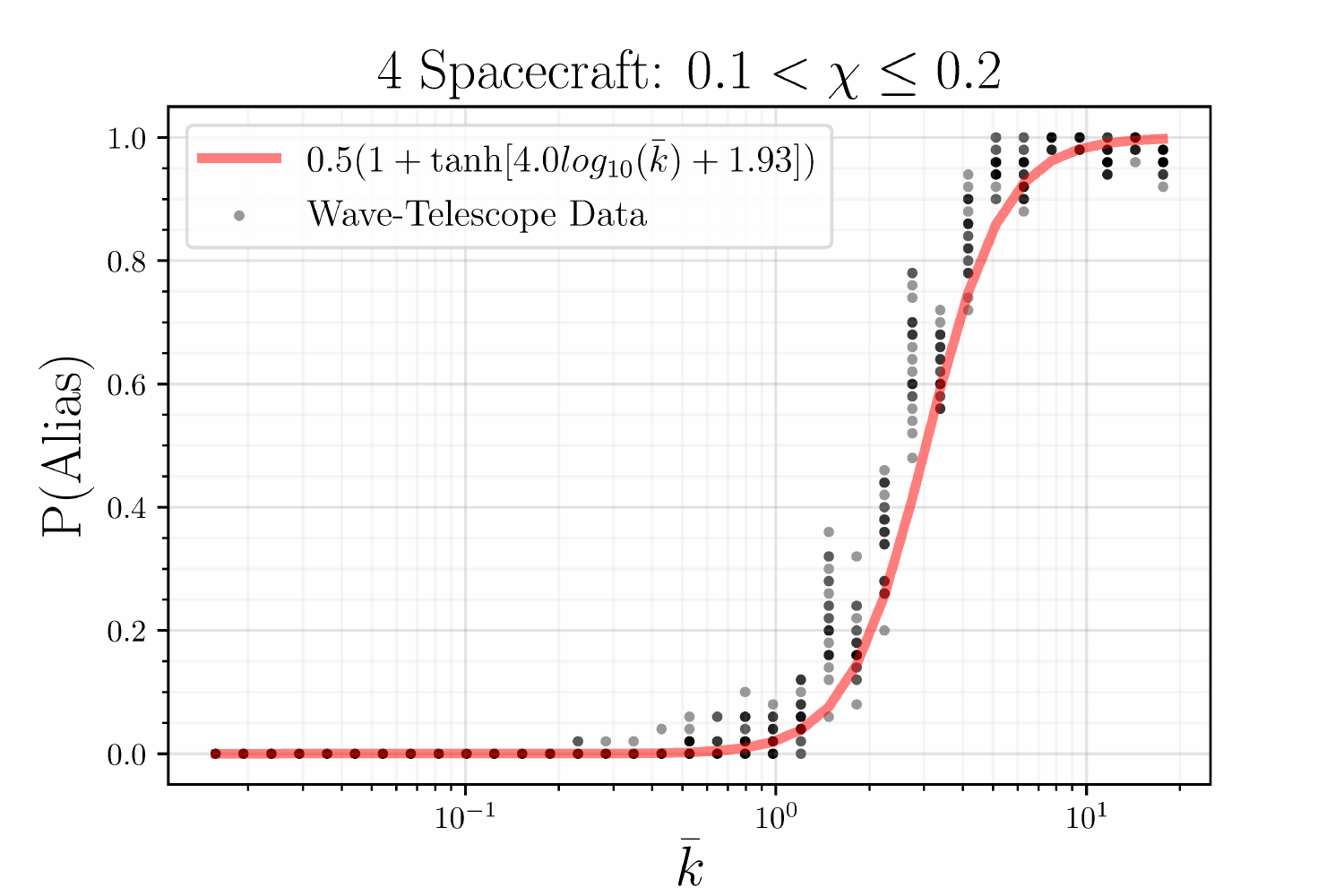}
\end{tabular}
\caption{{\small In the left panel we plot the errors found using four-spacecraft configurations with $0.1 \leq \chi < 0.2$ as black dots. We note that we can distinguish the aliased points with a simple threshold at $Error = 400\bar{k}^{-1}$. In the right panel we plot the probability that a wavevector reconstruction is aliased at each magnitude $\bar{k}$. }}
\label{fig:data_slice} 
\end{figure}

For both numbers of spacecraft found in Figure \ref{fig:mu_data} we see that high errors occur when a configuration is well-shaped but the the wavevector magnitude is too large. Recall that these cases should correspond to the incidences where the Nyquist sampling theorem is violated and $k_{calc}$ is incorrect. To gain a better understanding of how these aliased points differ from the rest of the smoothly varying dataset we plot a small subset of the data in the left panel of Figure \ref{fig:data_slice}, specifically for four-spacecraft configurations where $0.1 \leq \chi < 0.2$. 

We see that Figure \ref{fig:data_slice} has a different log-linear slope and $\bar{k}$ value where aliasing occurs compared to Figure \ref{fig:Sahrauri}. This is because Figure \ref{fig:data_slice} is using many configurations of spacecraft with a range of $\chi$ values that are not equal to zero. We see variations in error at a fixed $\bar{k}$ value in Figure \ref{fig:data_slice} because we are plotting errors computed over all wave directions in Figure \ref{fig:data_slice}, while in Figure \ref{fig:Sahrauri} we computed the median error value over all of the wave directions.

\subsection{Filtering Data}
\label{ssec:data_filter}
Comparing the left panel of Figure \ref{fig:data_slice} to Figure \ref{fig:Sahrauri} we see that there does not exist a threshold $k_{max}$ value that separates aliased and not aliased points. We instead see that the proportion of points falling into the aliased population changes as a function of $\bar{k}$. We manually tag the points which fall above the line $Error = 400\bar{k}^{-1}$ as points where aliasing has occurred in the wavevector reconstruction. These points will be omitted in the fitting of the estimated mean and standard deviation error equations. We will capture their influence by instead quantifying the probability that a wave-telescope computation aliases (and therefore falls into the population of points above $Error = 400\bar{k}^{-1}$). 

The small number of data points that fall between the two distinct log-linearly trending groups are due to numerical errors in the search algorithm used to find the peak power output of the wave-telescope technique. We did not perform a dense scan of $\mathbf{k}$ space for every wavevector and spacecraft combination. We instead used a 3-dimensional adaptive meshing scheme, described in Appendix \S \ref{sec:appendix.scan_alg}, to speed up the millions of computations. This small proportion of data points which have high errors due to the scanning algorithm do not seem to affect the overall statistics of the simulations (see table \ref{tab:PPC}).

Finally, we omit data that was generated using very poorly shaped spacecraft configurations ($\chi > 1$) from this point forward. When using configurations where $\chi > 1$ we found that the variations driven by changing the wave direction dominate the observed mean value trend. As modeling these large variations would require a different methodology then our approach, we are dropping these cases entirely. We are not interested in these poorly-shaped configuration cases as multi-spacecraft missions are typically designed to have very well-shaped tetrahedral configurations.

\subsection{Equation Forms}
\label{ssec:eqn_forms}
We view our dataset as having two independent variables (shape $\chi$ and relative wavevector magnitude $\bar{k}$) and one dependent variable (\% error). This means the propagation direction of the wave is responsible for the variation in errors at constant values of $\chi$ and $\bar{k}$. We also view our dataset as two distinct populations of points: non-aliased and aliased. We learn the equations that describe the non-aliased cases separately from those that describe the aliased ones. Once we have both equations, we combine them with a simple expected value formulation.

\subsubsection{Non-Aliased Case}
\label{sssec:non_alias_eqn}
In the left panel of Figure \ref{fig:data_slice} we plotted the error as a function of $\bar{k}$ for a small range of shapes and found that there is a strong log-linear trend for the non-aliased points. By using regression on an equation of the form
\begin{equation}
    Error = \alpha_0(\chi) \bar{k}^{\alpha_1(\chi)}, \label{eqn:alpha0_alpha1}
\end{equation}
we observed how the two coefficients ($\alpha_0$ and $\alpha_1$) vary for different values of $\chi$. This gave us a functional form that fits the average value in the dataset
\begin{equation}
    Error(\bar{k},\chi) = \left(a_0^{\chi-a_1} + a_2 \right)\bar{k}^{\left(a_3^{\chi - a_4} + a_5 \right)}. \label{eqn:mu}
\end{equation}

We then subtracted the mean value estimate given by eqn \ref{eqn:mu} from the actual error values found in the dataset. We found that $\log_{10}$ of this residual seemed to be normally distributed with zero mean, and had standard deviation which could be modelled by
\begin{equation}
    \sigma(\bar{k},\chi) = c_0 + c_1 \log_{10} \bar{k} + c_2 \chi^2. \label{eqn:std}
\end{equation}
Therefore, given the proper coefficients, the error in a given wave-telescope computation (assuming aliasing has not occurred) can be described by a normal distribution
\begin{equation}
    \log_{10}\left( Error(\bar{k},\chi) \right) \sim N(\mu, \sigma^2) \label{eqn:log_N}
\end{equation}
where
\begin{equation}
    \mu(\bar{k},\chi) = \log_{10}\left( a_0^{\chi-a_1} + a_2 \right) + \left(a_3^{\chi - a_4} + a_5 \right)\log_{10}\bar{k} \label{eqn:mu_log}
\end{equation}
and $\sigma$ is given by eqn \ref{eqn:std}.

\subsubsection{Aliased Case}
\label{sssec:alias_eqn}
Recall that the second population of points, which we are treating separately, can be identified using the threshold equation $Error = 400\bar{k}^{-1}$. For every four-spacecraft configuration with $0.1 < \chi \leq 0.2$, we plot the probability that a point is aliased (falls above this threshold) in the right panel of Figure \ref{fig:data_slice}. Because this probability is increasing for higher values of $\bar{k}$, we fit a sigmoid function of the form
\begin{equation}
    P(alias|\bar{k},\chi) = \frac{1}{2} \left(1 + \tanh\left[\beta_0(\chi) \log_{10} \bar{k} + \beta_1(\chi) \right] \right) \label{eqn:P_alias_beta}
\end{equation}
to the data. We then observe how the $\beta$ coefficients vary as a function of $\chi$ to find that the functional form 
\begin{equation}
    P(alias|\bar{k},\chi) = \frac{1}{2} \left(1 + \tanh\left[4 \log_{10} \bar{k} + b_0 \chi^2 + b_1 \chi + b_2 \right] \right) \label{eqn:P_alias}
\end{equation}
can represent the probability of aliasing for all values of $\chi$ and $\bar{k}$. 

Now that we have an equation for the mean value of error if no aliasing occurs, as well as the probability of aliasing, we can define an effective error equation that gives us the expected value given the two possible outcomes
\begin{equation}
    \mu_{eff}(\bar{k},\chi) = \left[1-P(alias|\bar{k},\chi) \right]\mu(\bar{k},\chi) + P(alias|\bar{k},\chi)\max\left(400\bar{k}^{-1}, 100\right). \label{eqn:mu_eff}
\end{equation}
The term $\max\left(400\bar{k}^{-1}, 100\right)$ assumes an error value of at least 100 for all aliased points, which may be a slight overestimate for large values of $\bar{k}$. We implement this minimum so that the effective error does not go to 0 as $\bar{k} \to \infty$.

\subsection{Bayesian Inference Verification}
\label{ssec:verify_bayes}
We wish to use Bayesian inference \citep{McElreath2016} to estimate the unknown coefficients in eqns \ref{eqn:mu_log}, \ref{eqn:std}, and \ref{eqn:P_alias} as well as their uncertainty. Before we attempt to do this we must verify that the dataset is rich enough to capture the dynamics of the functional forms that we are proposing to learn. We do this by generating a new verification dataset that follows the proposed functional distributions exactly and has known coefficient values. 

\begin{table}[ht]
\centering
\caption{Summary of results using Bayesian inference on our verification dataset. Because we were able to learn the coefficients of the proposed equations in this example, we have verified that it is possible to do so for our wave-telescope dataset.}
{
\begin{tabular}{|c|c|c|c|c|c|} \hline
Parameter & True & \multicolumn{2}{c|}{Prior Dist $\sim N(\mu, \sigma)$} & \multicolumn{2}{c|}{Posterior Dist $\sim N(\mu, \sigma)$} \\
\cline{3-6}
Name & Value & $\mu$ & $\sigma$ & $\mu$ & $\sigma$ \\
 \hline
$ a_0 $ & 576.409 & 576.410 & 200   & 530.111 & 53.881 \\
$ a_1 $ & 0.564   & 0.560   & 0.2   & 0.561   & 0.005 \\
$ a_2 $ & 2.823   & 2.820   & 0.5   & 2.786   & 0.020 \\
$ a_3 $ & 316.732 & 316.730 & 100   & 340.782 & 62.862 \\
$ a_4 $ & 1.139   & 1.140   & 0.1   & 1.139   & 0.007 \\
$ a_5 $ & -0.915  & -0.910  & 0.3   & -0.918  & 0.003 \\
$ c_0 $ & 0.119   & 0.12    & 0.02  & 0.117   & 0.002 \\
$ c_1 $ & 0.018   & 0.02    & 0.005 & 0.018   & 0.001 \\
$ c_2 $ & 0.143   & 0.14    & 0.03  & 0.141   & 0.004 \\
 \hline
\end{tabular}}
\label{tab:verification_model} 
\end{table}

This verification dataset uses the same spacecraft configurations and distribution of plane-waves that are described in \S \ref{ssec:data}. We randomly select a subset of 10,000 of these combinations as the training set. However, instead of simulating waves and estimating them using the wave-telescope technique, we instead directly relate the error in wavevector reconstruction to $\chi$ and $\bar{k}$ using eqns \ref{eqn:log_N}, \ref{eqn:mu_log}, and \ref{eqn:std}. The values that we have chosen for the nine coefficients of these equations are listed under the \textit{True Value} column in Table \ref{tab:verification_model}.

We then use Bayesian inference on this verification dataset to confirm that it can reconstruct the known values of coefficients with sufficient accuracy. To do this, we must first define a prior distribution for each coefficient value. We chose a normal distribution for each coefficient that is centered near the true value, but has a very large standard deviation. What we find when we perform the Bayesian inference is that after 5000 samples, the posterior distributions mean value is on average very close to the true value of the system. We also find that the standard deviation of the posterior distribution significantly narrows so that the true value of each parameter is well described using the posterior distribution's $\mu$ and $\sigma$. Table \ref{tab:verification_model} summarizes the true value, prior distribution, and posterior distribution of all nine coefficients in this verification dataset.

\subsection{Learning Equation Coefficients}
\label{ssec:coefs}
\begin{figure}[ht]
\centering
\includegraphics[width=0.7\textwidth]{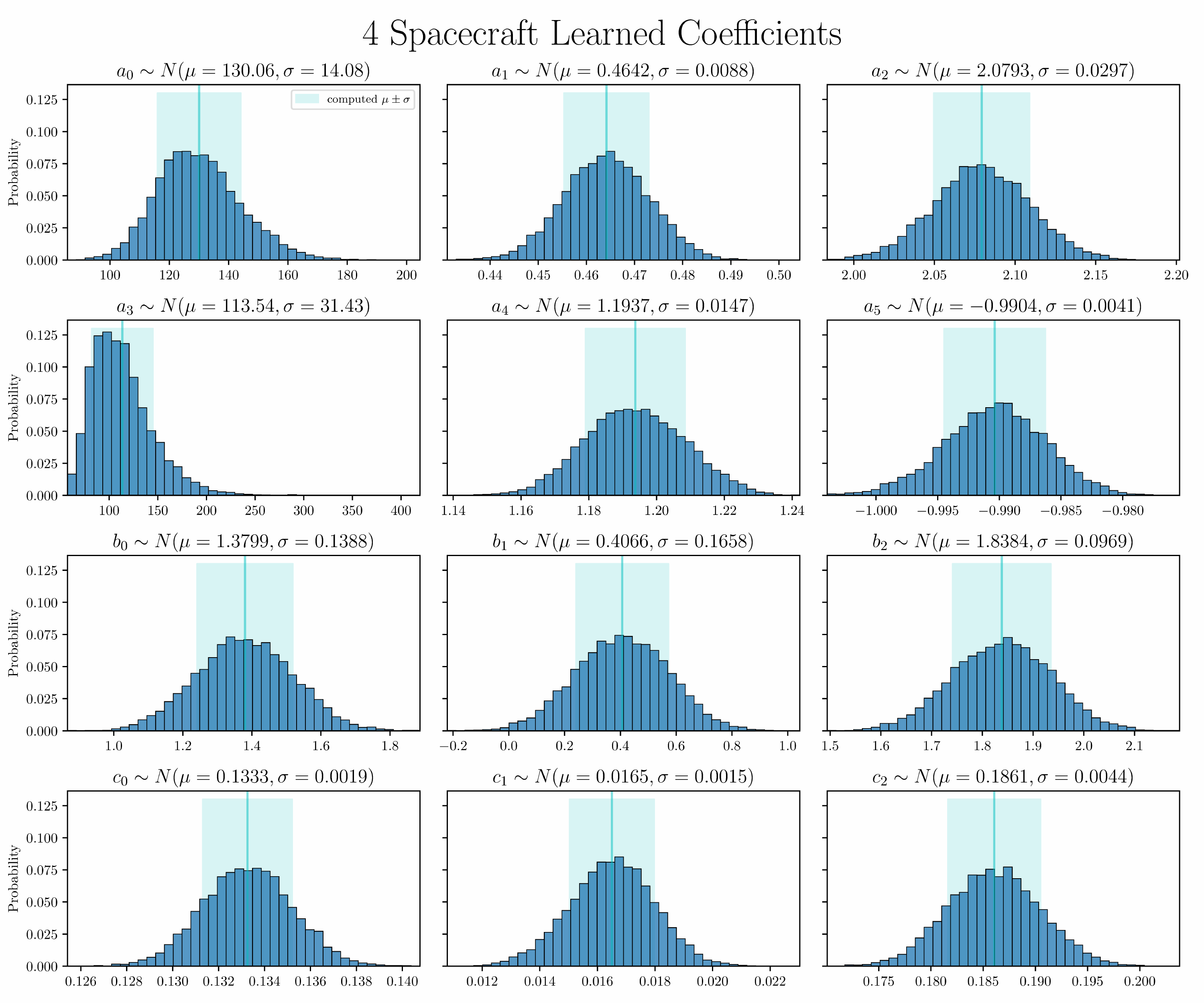}
\caption{{\small These panels represent the posterior distributions of the twelve coefficients in eqns \ref{eqn:mu_log}, \ref{eqn:std}, \ref{eqn:P_alias} for wave-telescope error from four-spacecraft configurations. The blue vertical line and shaded region represent the posterior distributions mean and standard deviation. See Table \ref{tab:coefs} for a summary of the distributions for all coefficients and numbers of spacecraft. A complete figure set (6 images) for four through nine spacecraft configurations is available in the online journal.}}
\label{fig:bayes_n4} 
\end{figure}

With the Bayesian inference approach verified, we compute the posterior distributions for the twelve coefficients $a$'s, $b$'s, $c$'s using the wave-telescope generated data. We again randomly select a subset of 10,000 datapoints to use as our training set. The posterior distributions for the twelve coefficients corresponding to the four-spacecraft configurations can be seen in Figure \ref{fig:bayes_n4}. The complete set of coefficient posterior distributions for $N\in\{4,5,6,7,8,9\}$ spacecraft can be found in the figure set corresponding to Figure \ref{fig:bayes_n4}. Table \ref{tab:coefs} summarizes the mean $\langle x_i \rangle$ and standard deviation $\sigma_{x_i}$ of the posterior distributions for all numbers of spacecraft $N$, for $x \in \{a,b,c\}$.


\figsetgrpstart
\figsetgrpnum{5.1}
\figsetgrptitle{Four-Spacecraft Posterior Distributions}
\figsetplot{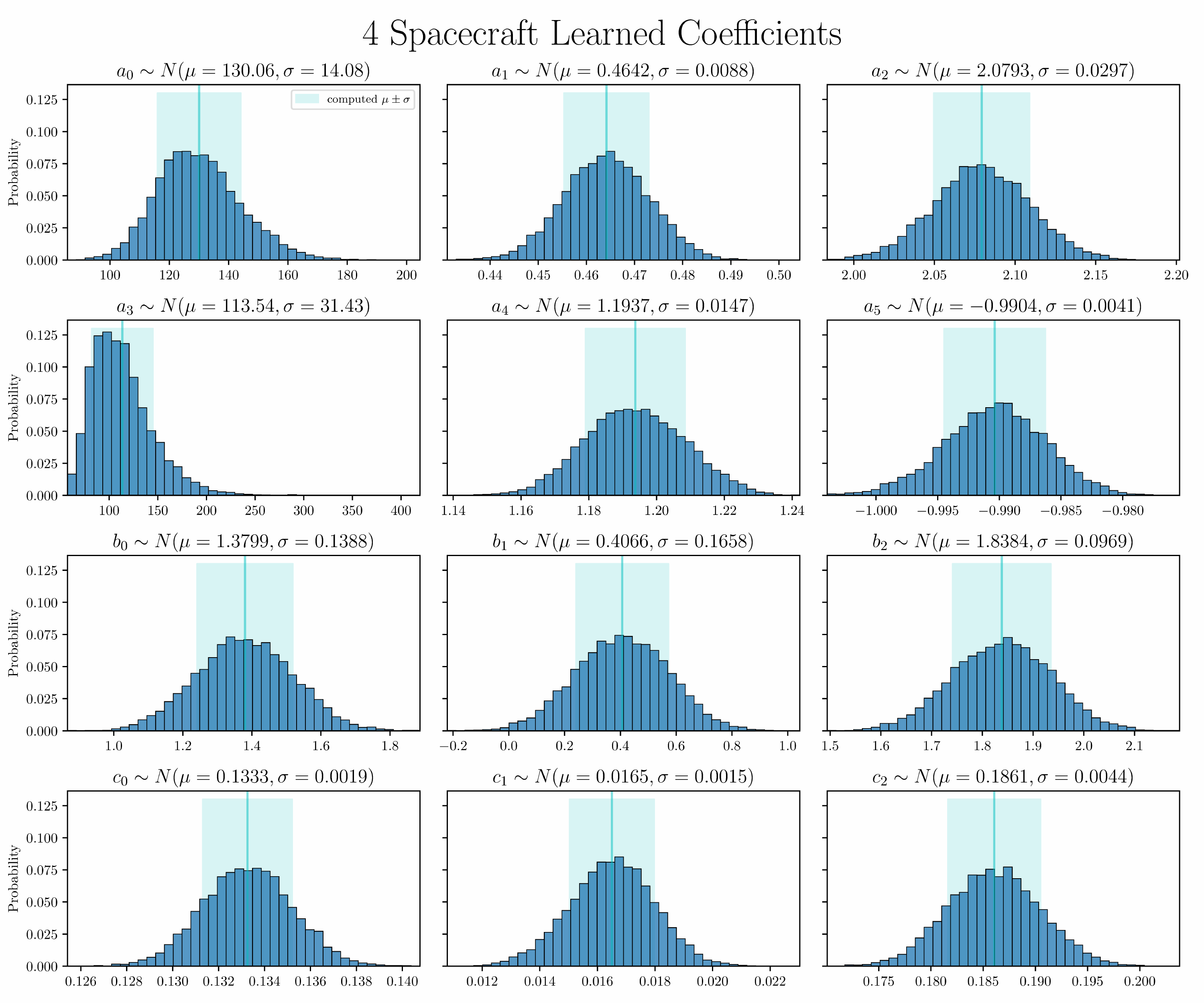}
\figsetgrpnote{These panels represent the posterior distributions of the twelve coefficients in eqns \ref{eqn:mu_log}, \ref{eqn:std}, \ref{eqn:P_alias} for wave-telescope error from four-spacecraft configurations. The blue vertical line and shaded region represent the posterior distributions mean and standard deviation.}
\figsetgrpend

\figsetgrpstart
\figsetgrpnum{5.2}
\figsetgrptitle{Five-Spacecraft Posterior Distributions}
\figsetplot{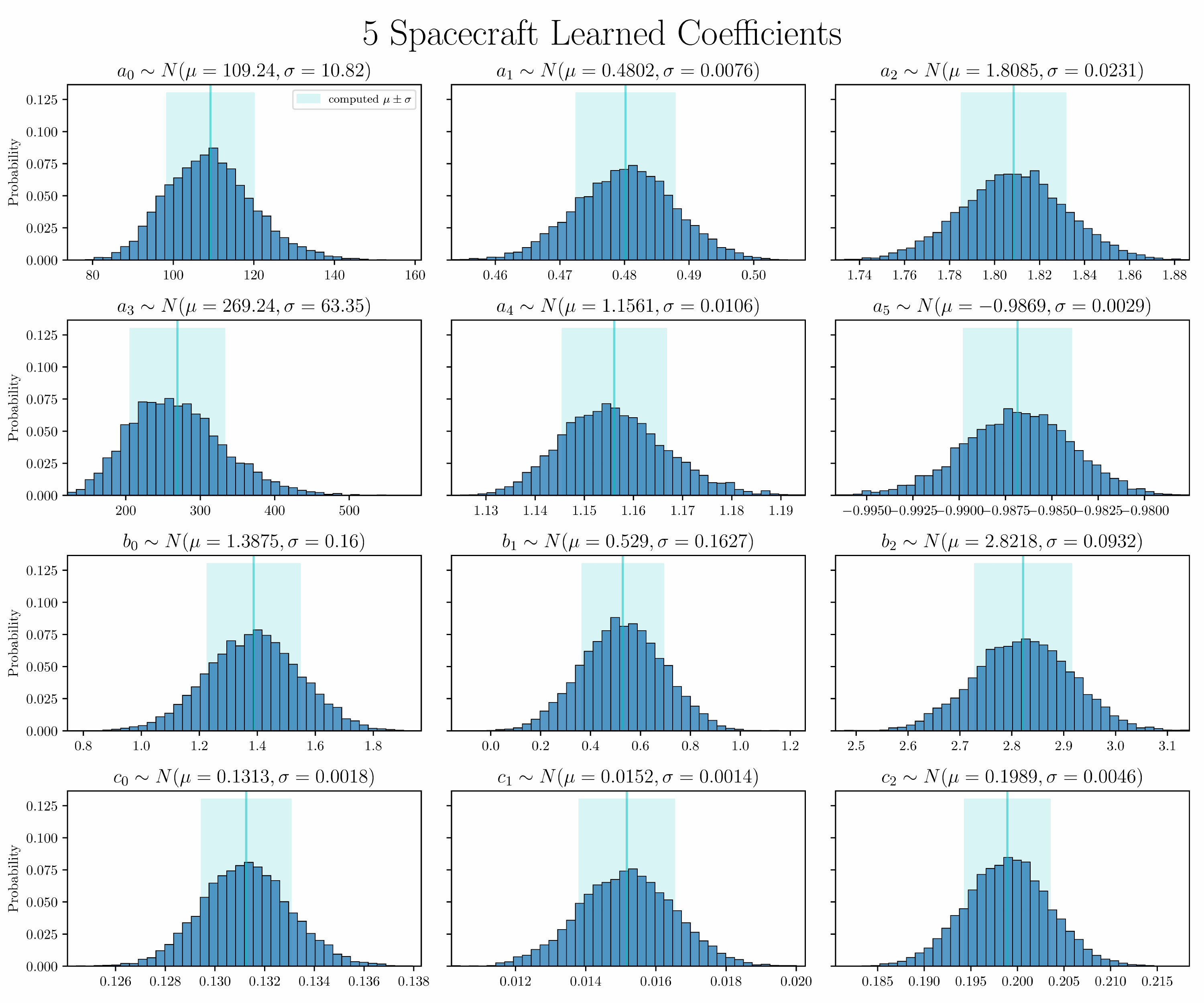}
\figsetgrpnote{These panels represent the posterior distributions of the twelve coefficients in eqns \ref{eqn:mu_log}, \ref{eqn:std}, \ref{eqn:P_alias} for wave-telescope error from five-spacecraft configurations. The blue vertical line and shaded region represent the posterior distributions mean and standard deviation.}
\figsetgrpend

\figsetgrpstart
\figsetgrpnum{5.3}
\figsetgrptitle{Six-Spacecraft Posterior Distributions}
\figsetplot{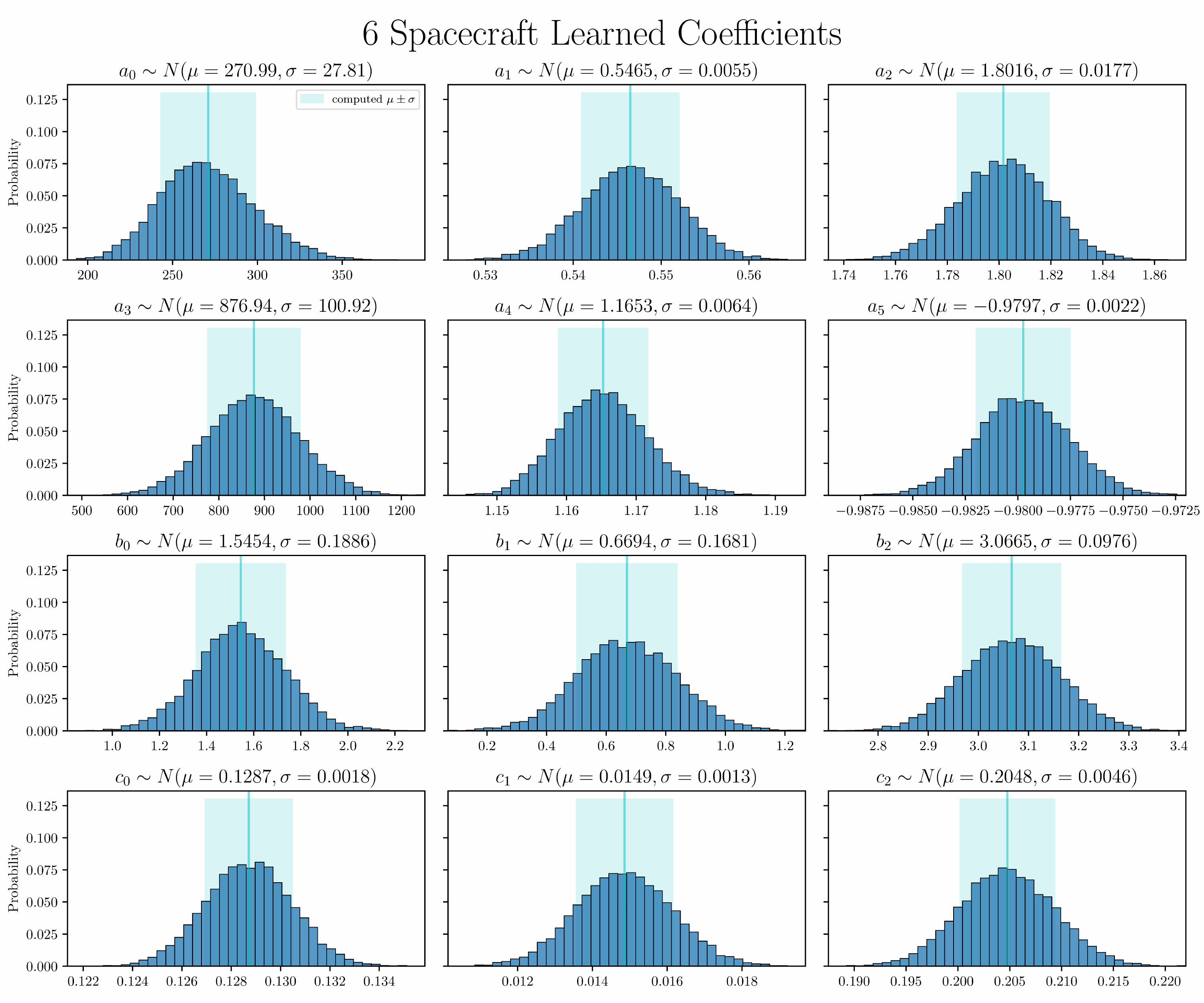}
\figsetgrpnote{These panels represent the posterior distributions of the twelve coefficients in eqns \ref{eqn:mu_log}, \ref{eqn:std}, \ref{eqn:P_alias} for wave-telescope error from six-spacecraft configurations. The blue vertical line and shaded region represent the posterior distributions mean and standard deviation.}
\figsetgrpend

\figsetgrpstart
\figsetgrpnum{5.4}
\figsetgrptitle{Seven-Spacecraft Posterior Distributions}
\figsetplot{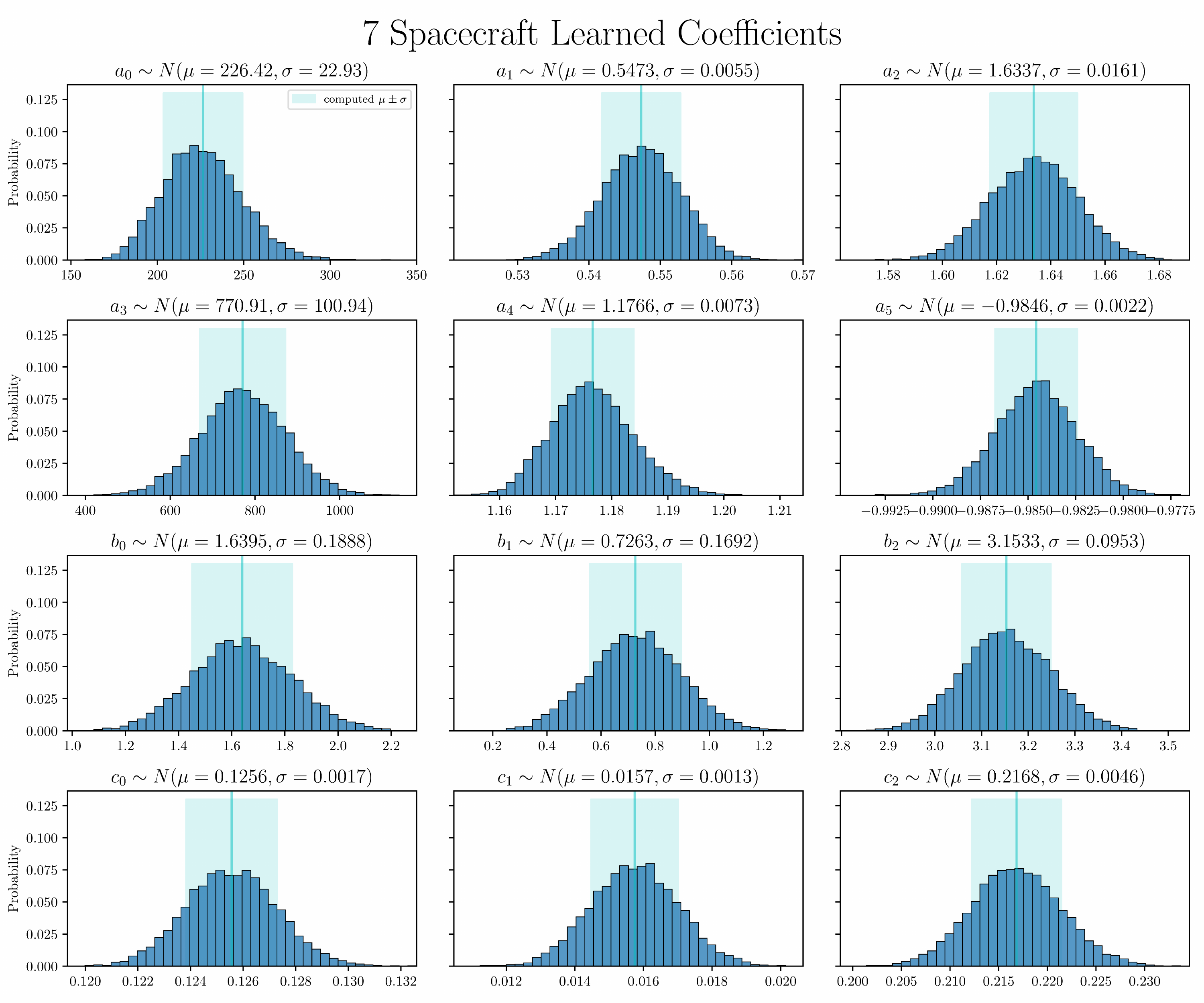}
\figsetgrpnote{These panels represent the posterior distributions of the twelve coefficients in eqns \ref{eqn:mu_log}, \ref{eqn:std}, \ref{eqn:P_alias} for wave-telescope error from seven-spacecraft configurations. The blue vertical line and shaded region represent the posterior distributions mean and standard deviation.}
\figsetgrpend

\figsetgrpstart
\figsetgrpnum{5.5}
\figsetgrptitle{Eight-Spacecraft Posterior Distributions}
\figsetplot{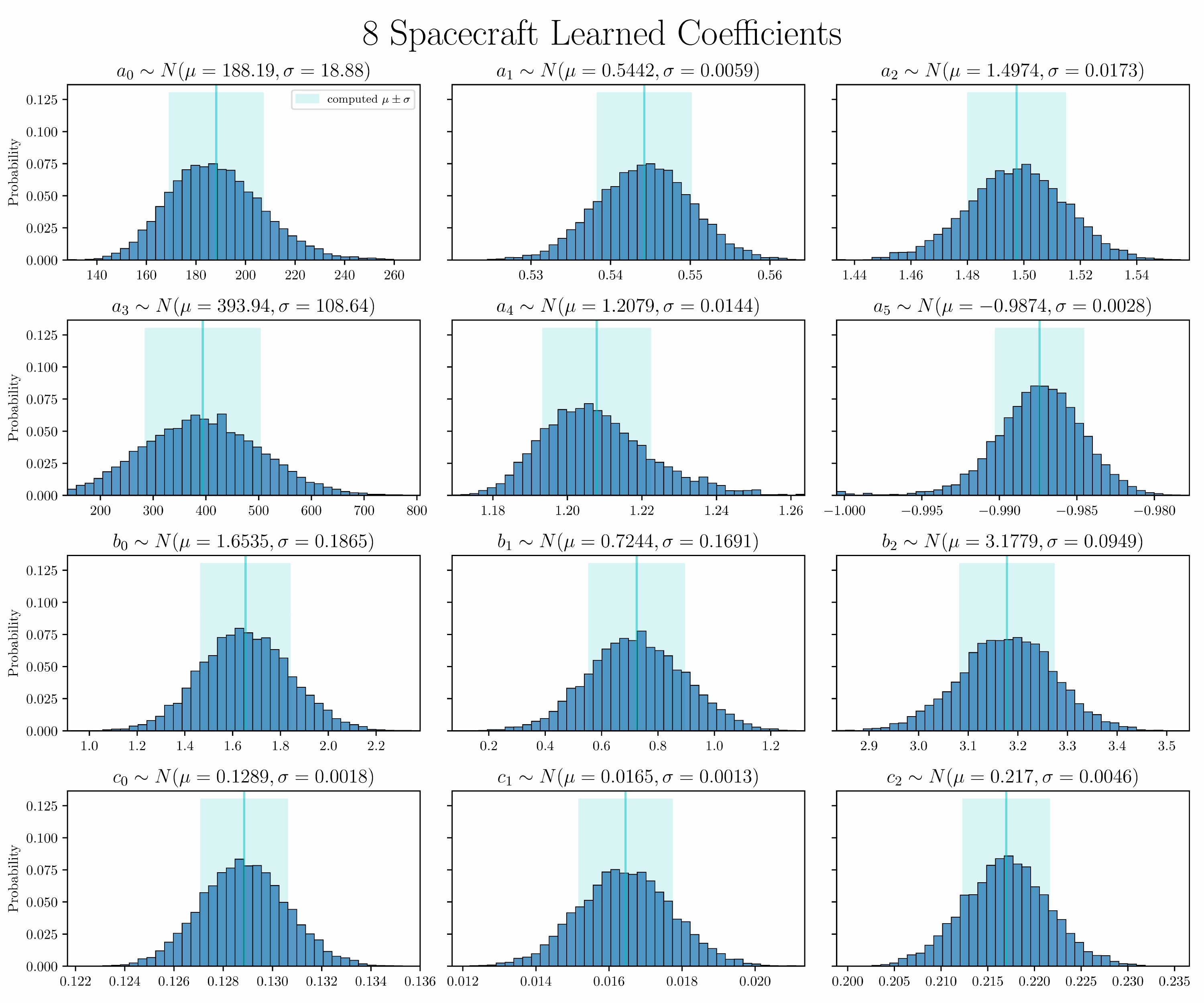}
\figsetgrpnote{These panels represent the posterior distributions of the twelve coefficients in eqns \ref{eqn:mu_log}, \ref{eqn:std}, \ref{eqn:P_alias} for wave-telescope error from eight-spacecraft configurations. The blue vertical line and shaded region represent the posterior distributions mean and standard deviation.}
\figsetgrpend

\figsetgrpstart
\figsetgrpnum{5.6}
\figsetgrptitle{Nine-Spacecraft Posterior Distributions}
\figsetplot{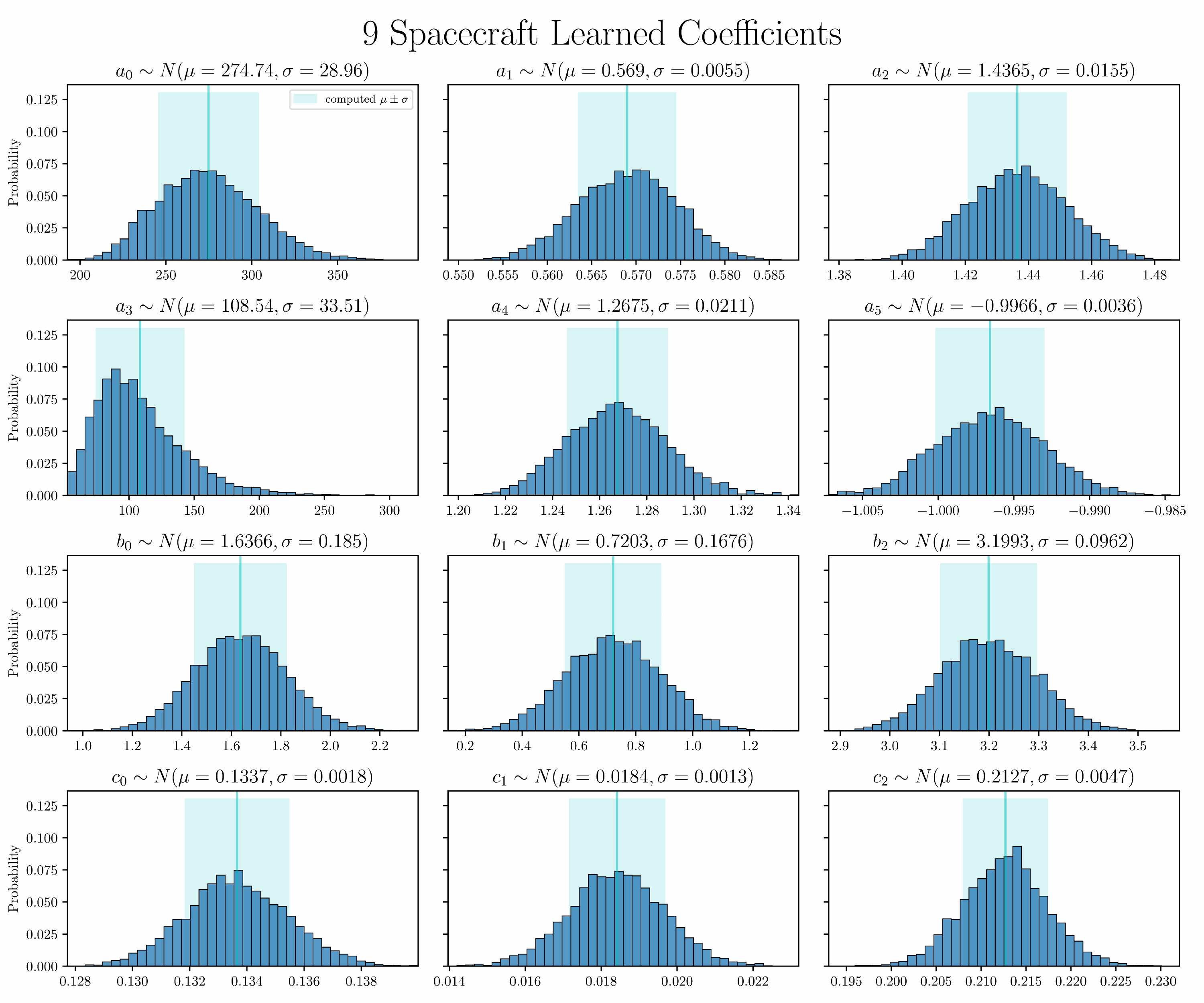}
\figsetgrpnote{These panels represent the posterior distributions of the twelve coefficients in eqns \ref{eqn:mu_log}, \ref{eqn:std}, \ref{eqn:P_alias} for wave-telescope error from nine-spacecraft configurations. The blue vertical line and shaded region represent the posterior distributions mean and standard deviation.}
\figsetgrpend

\figsetend

\begin{table}[ht]
\centering
\caption{Coefficient mean and standard deviation values, for each number of spacecraft $N$, found via Bayesian inference. The mean values of each coefficient (i.e. $\langle a_i \rangle$) are to be used in eqns \ref{eqn:mu_log}, \ref{eqn:std}, \ref{eqn:P_alias}.
 }{
\begin{tabular}{|c|c|c|c|c|c|c|c|}
\hline
Equation & Term & $N=4$ & $N=5$ & $N=6$ & $N=7$ & $N=8$ & $N=9$ \\
\hline 
$\mu(k,\chi)$ 
& $\langle a_0 \rangle$ & 130.06  & 109.24  & 270.99  & 226.42 & 188.19  & 274.74 \\
eqn \ref{eqn:mu_log}
& $\sigma_{a_0}$ & 14.1  & 10.8  & 27.8    & 22.9  & 18.9   & 29.0 \\
\cline{2-8}
& $\langle a_1 \rangle$ & 0.46    & 0.48    & 0.55    & 0.55   & 0.54    & 0.57 \\
& $\sigma_{a_1}$ & 0.009 & 0.008 & 0.006   & 0.006 & 0.006  & 0.006 \\
\cline{2-8}
& $\langle a_2 \rangle$ & 2.08    & 1.80    & 1.80    & 1.63   & 1.50    & 1.44 \\
& $\sigma_{a_2}$ & 0.030 & 0.023 & 0.018   & 0.016 & 0.017  & 0.016 \\
\cline{2-8}
& $\langle a_3 \rangle$ & 113.54  & 269.24  & 878.94  & 770.91 & 393.94  & 108.54 \\
& $\sigma_{a_3}$ & 31.4  & 63.4  & 100.9   & 100.9 & 108.6  & 33.5 \\
\cline{2-8}
& $\langle a_4 \rangle$ & 1.19    & 1.16    & 1.17    & 1.18   & 1.21    & 1.27 \\
& $\sigma_{a_4}$ & 0.015 & 0.011 &  0.006  & 0.007 & 0.014  & 0.021 \\
\cline{2-8}
& $\langle a_5 \rangle$ & -0.990  & -0.987  & -0.980  & -0.985 & -0.987  & -0.997 \\
& $\sigma_{a_5}$ & 0.004 & 0.003 & 0.002   & 0.002 & 0.038  & 0.004 \\
\hline
$P(alias|k,\chi)$ 
& $\langle b_0 \rangle$ & 1.38    & 1.39    & 1.55    & 1.64   & 1.65    & 1.64 \\
eqn \ref{eqn:P_alias}
& $\sigma_{b_0}$ & 0.139 & 0.160 & 0.189   & 0.189 & 0.187  & 0.185 \\
\cline{2-8}
& $\langle b_1 \rangle$ & 0.41    & 0.53    & 0.67    & 0.73   & 0.72    & 0.72 \\
& $\sigma_{b_1}$ & 0.166 & 0.163 & 0.168   & 0.169 & 0.169  & 0.168 \\
\cline{2-8}
& $\langle b_2 \rangle$ & 1.84    & 2.82    & 3.07    & 3.15   & 3.18    & 3.20 \\
& $\sigma_{b_2}$ & 0.097 & 0.093 & 0.098   & 0.095 & 0.095  & 0.096 \\
\hline
$\sigma(k,\chi)$ 
& $\langle c_0 \rangle$ & 0.133   & 0.131   & 0.129   & 0.126  & 0.129   & 0.134  \\
eqn \ref{eqn:std}
& $\sigma_{c_0}$ & 0.002 & 0.002 & 0.002   & 0.002 & 0.002  & 0.002 \\
\cline{2-8}
& $\langle c_1 \rangle$ & 0.017   & 0.015   & 0.015   & 0.016  & 0.017   & 0.018 \\
& $\sigma_{c_1}$ & 0.002 & 0.001 & 0.001   & 0.001 & 0.001  & 0.001 \\
\cline{2-8}
& $\langle c_2 \rangle$ & 0.186   & 0.199   & 0.205   & 0.217  & 0.217   & 0.213 \\ 
& $\sigma_{c_2}$ & 0.004 & 0.005 & 0.005   & 0.005 & 0.005  & 0.005 \\ 
\hline
\end{tabular}}
\label{tab:coefs} 
\end{table}

\subsection{Posterior Predictive Check}
\label{ssec:PPC}
\begin{figure}[ht]
\centering
\begin{tabular}{cc}
\includegraphics[width=0.42\textwidth]{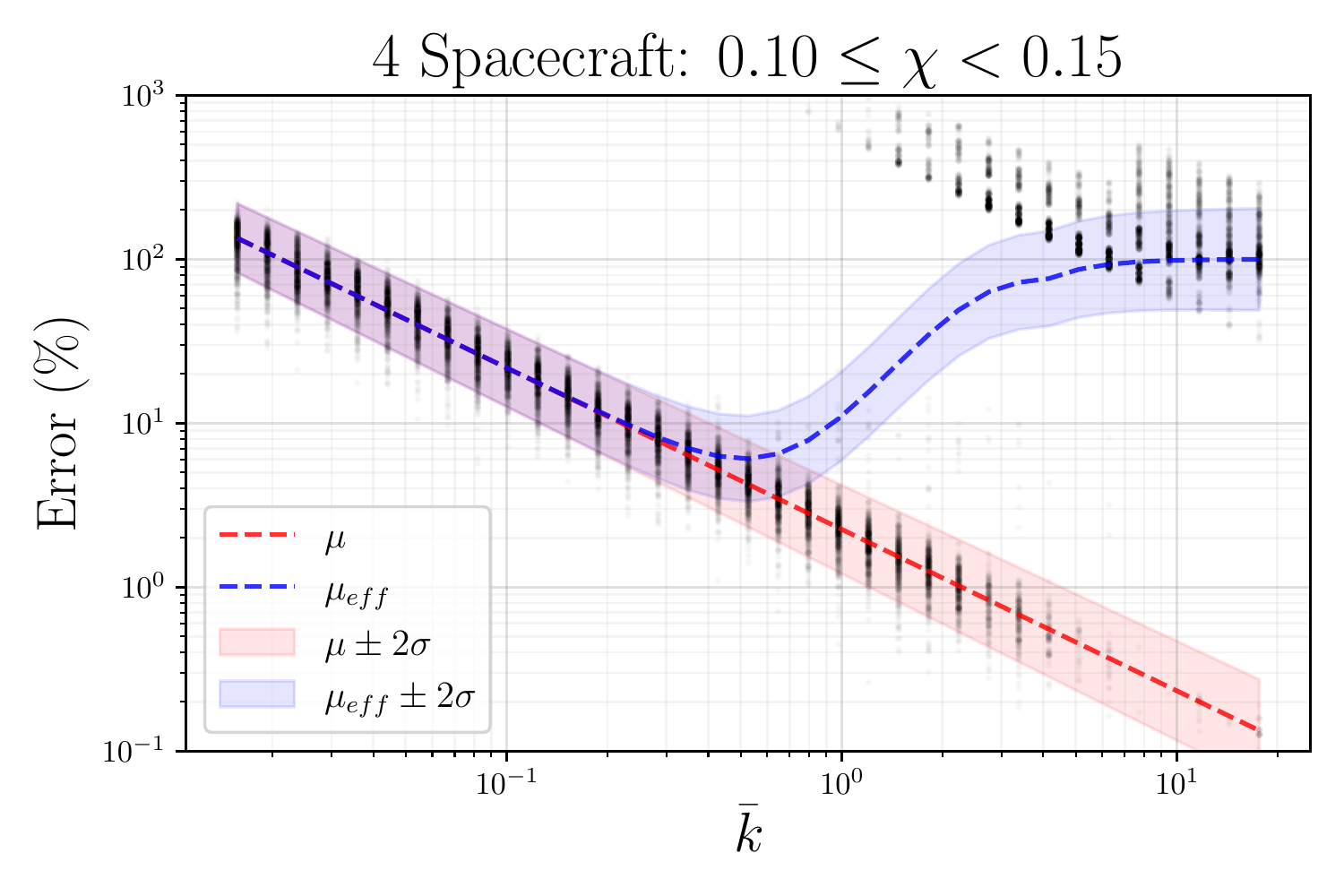} &
\includegraphics[width=0.42\textwidth]{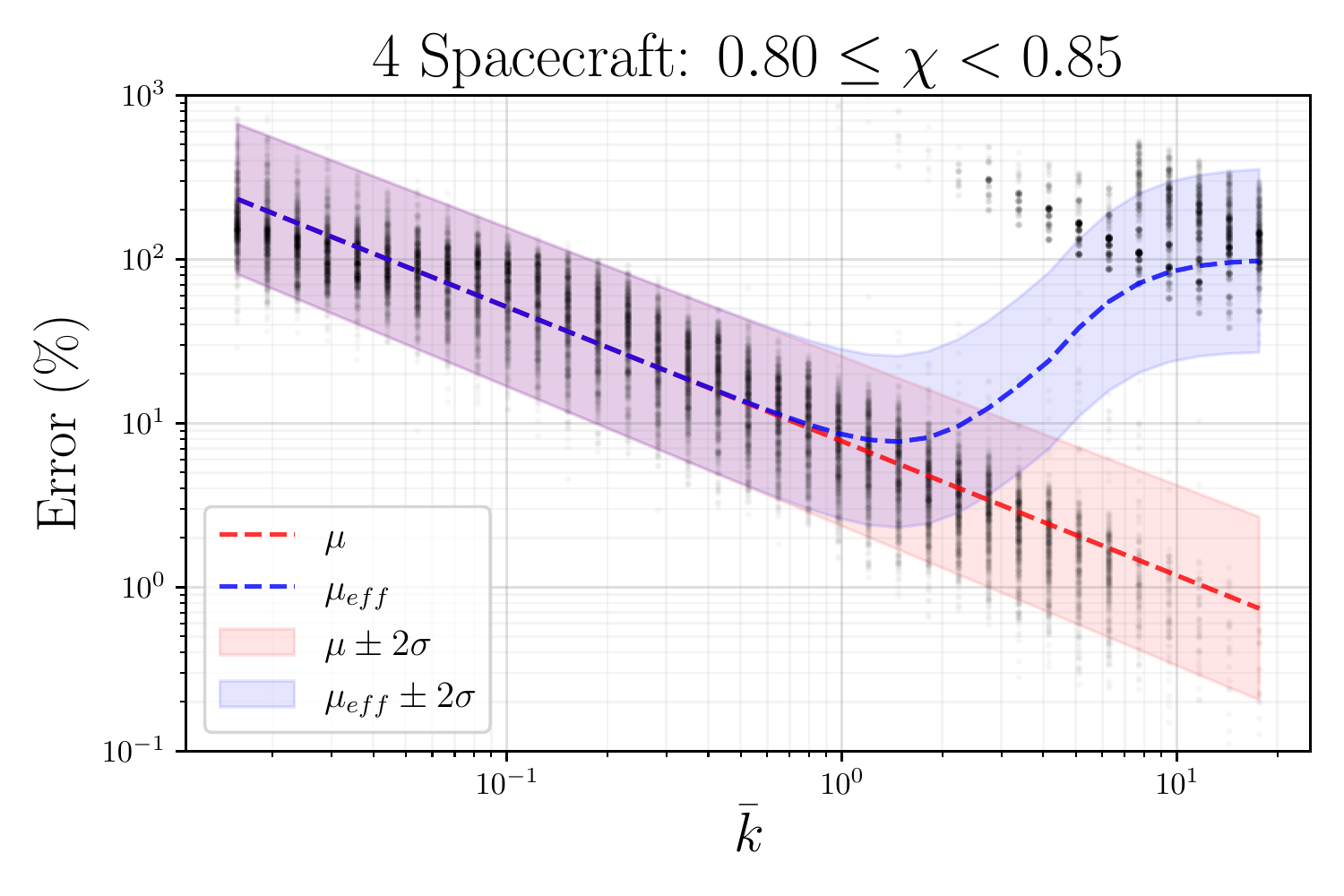} \\
\includegraphics[width=0.42\textwidth]{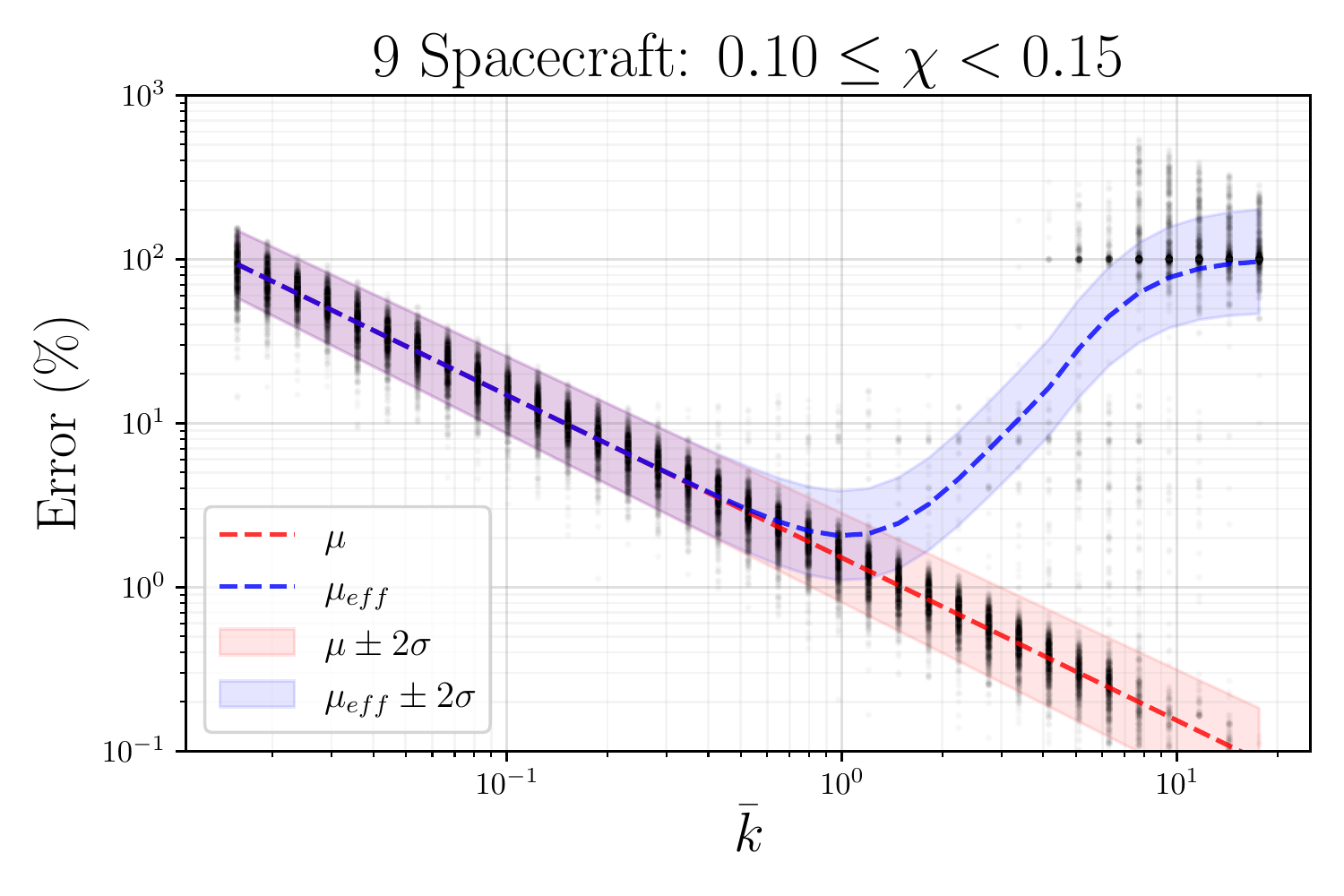} &
\includegraphics[width=0.42\textwidth]{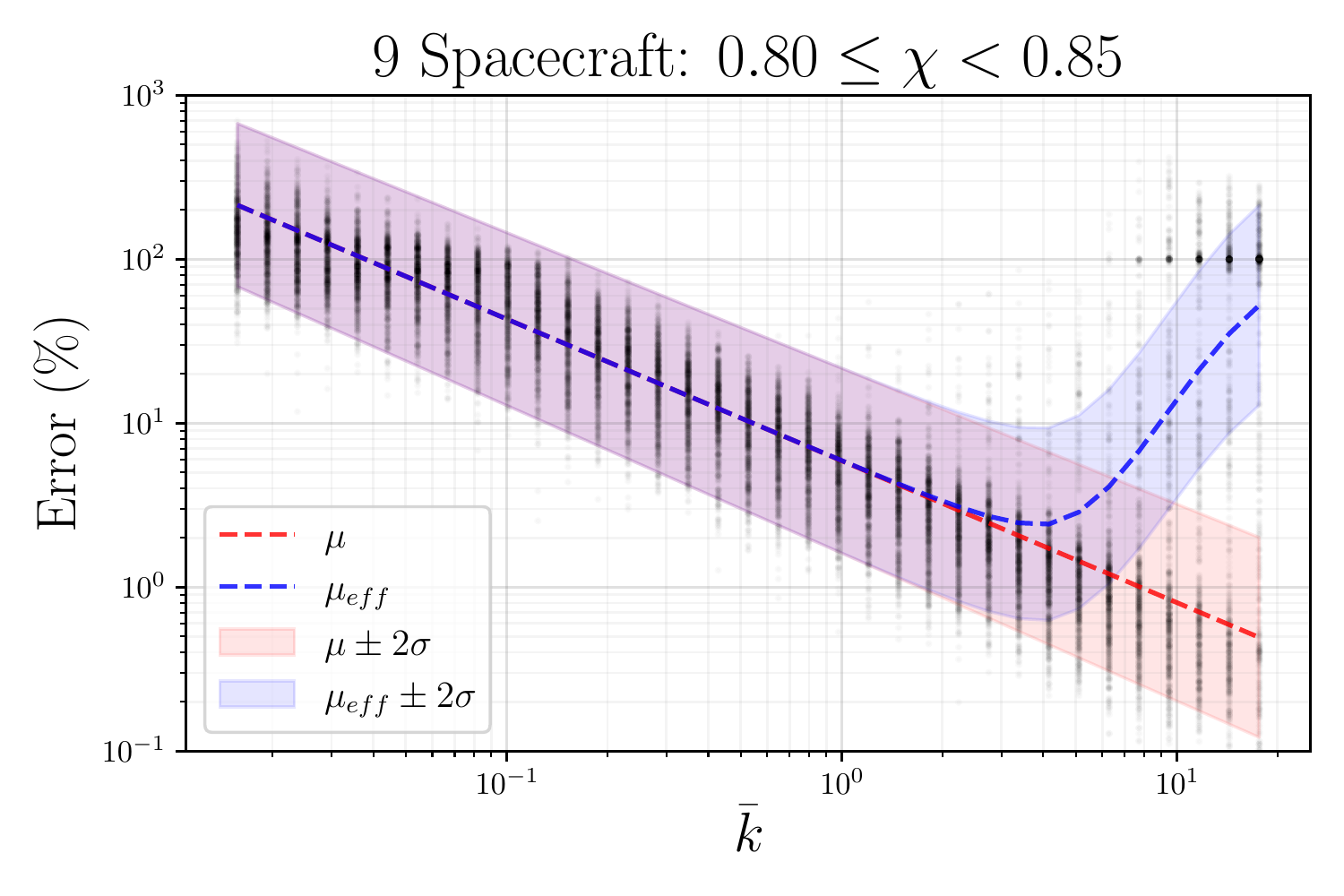} 
\end{tabular}
\caption{{\small We perform a posterior predictive check to verify that the coefficients found using Bayesian inference can correctly identify the mean value of error, $\mu$ (eqn \ref{eqn:mu_log}), along with the standard deviation $\sigma$ (eqn \ref{eqn:std}). We also show the effective error value $\mu_{eff}$ (eqn \ref{eqn:mu_eff}), which factors in the probability of aliasing. This is done for four and nine spacecraft configurations (top and bottom rows) with shape parameters between $0.1 \leq \chi < 0.15$ and $0.8 \leq \chi < 0.85$ (left and right columns). A figure set (18 images) for four through nine spacecraft configurations is available in the online journal. }}
\label{fig:PPC} 
\end{figure}

In the left panels of Figure \ref{fig:PPC} we plot the error found in wave-telescope simulations for four and nine-spacecraft configurations with shape $0.1 \leq \chi < 0.15$. These error values include the 10,000 points used in the training data, as well as those which were not used in training (i.e. the testing data). We overlay our computed equation that gives us the error distribution assuming no aliasing occurs (eqns \ref{eqn:mu_log} and \ref{eqn:std} in red) as well as our equation for effective error (eqn \ref{eqn:mu_eff} in blue), which factors in aliasing effects. In the right panels of Figure \ref{fig:PPC} we plot the same results but for four and nine-spacecraft configurations with shape $0.8 \leq \chi < 0.85$. A more complete set of figures displaying results from 4-9 spacecraft can be seen in the figure set corresponding to Figure \ref{fig:PPC}. 


\figsetgrpstart
\figsetgrpnum{6.1}
\figsetgrptitle{Four-Spacecraft: Well Shaped}
\figsetplot{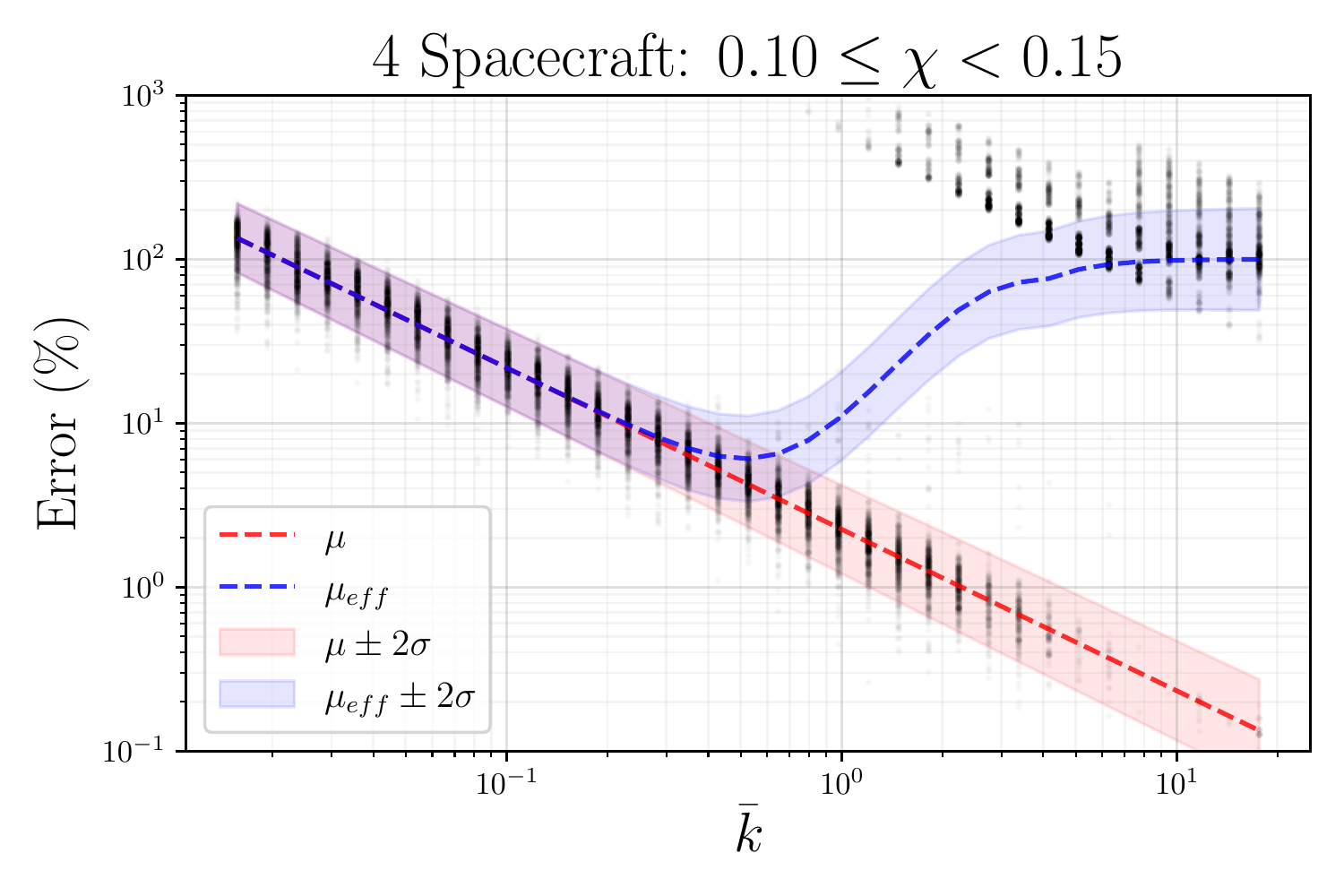}
\figsetgrpnote{We perform a posterior predictive check to verify that the coefficients found using Bayesian inference can correctly identify the mean value of error, $\mu$ (eqn \ref{eqn:mu_log}), along with the standard deviation $\sigma$ (eqn \ref{eqn:std}). We also show the effective error value $\mu_{eff}$ (eqn \ref{eqn:mu_eff}), which factors in the probability of aliasing. This is done for four-spacecraft configurations with shape parameters between $0.1 \leq \chi < 0.15$.}
\figsetgrpend

\figsetgrpstart
\figsetgrpnum{6.2}
\figsetgrptitle{Five-Spacecraft: Well Shaped}
\figsetplot{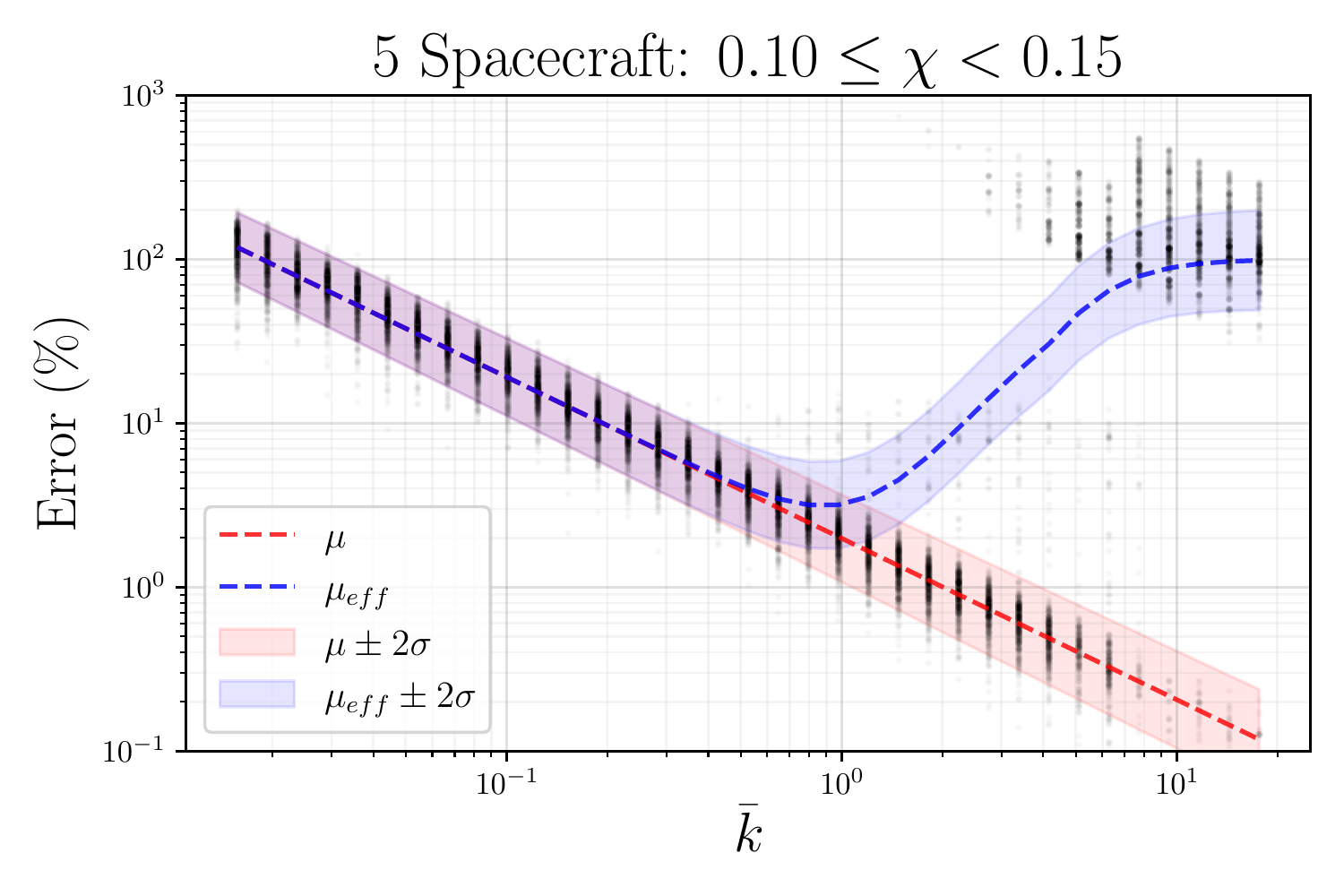}
\figsetgrpnote{We perform a posterior predictive check to verify that the coefficients found using Bayesian inference can correctly identify the mean value of error, $\mu$ (eqn \ref{eqn:mu_log}), along with the standard deviation $\sigma$ (eqn \ref{eqn:std}). We also show the effective error value $\mu_{eff}$ (eqn \ref{eqn:mu_eff}), which factors in the probability of aliasing. This is done for five-spacecraft configurations with shape parameters between $0.1 \leq \chi < 0.15$.}
\figsetgrpend

\figsetgrpstart
\figsetgrpnum{6.3}
\figsetgrptitle{Six-Spacecraft: Well Shaped}
\figsetplot{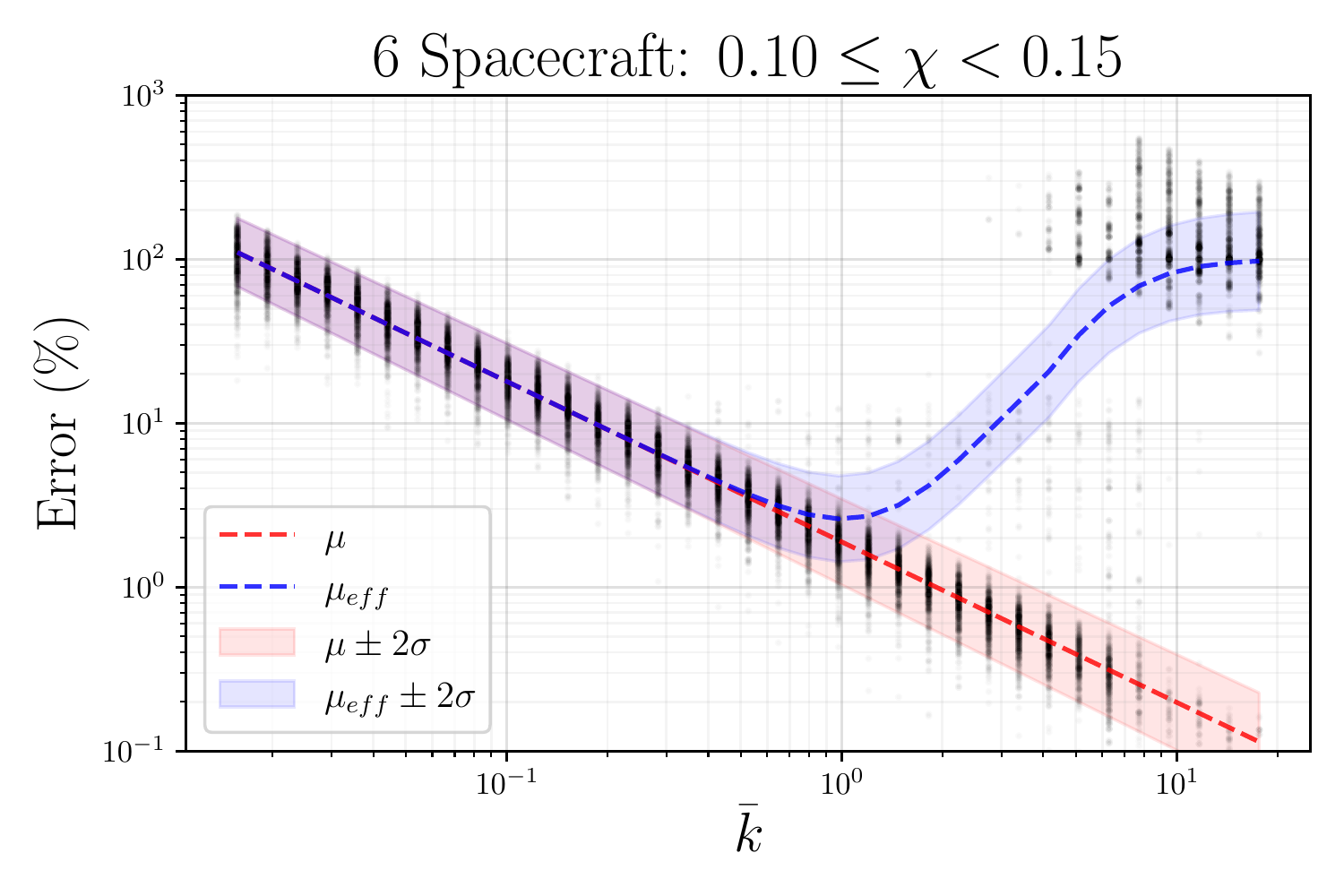}
\figsetgrpnote{We perform a posterior predictive check to verify that the coefficients found using Bayesian inference can correctly identify the mean value of error, $\mu$ (eqn \ref{eqn:mu_log}), along with the standard deviation $\sigma$ (eqn \ref{eqn:std}). We also show the effective error value $\mu_{eff}$ (eqn \ref{eqn:mu_eff}), which factors in the probability of aliasing. This is done for six-spacecraft configurations with shape parameters between $0.1 \leq \chi < 0.15$.}
\figsetgrpend

\figsetgrpstart
\figsetgrpnum{6.4}
\figsetgrptitle{Seven-Spacecraft: Well Shaped}
\figsetplot{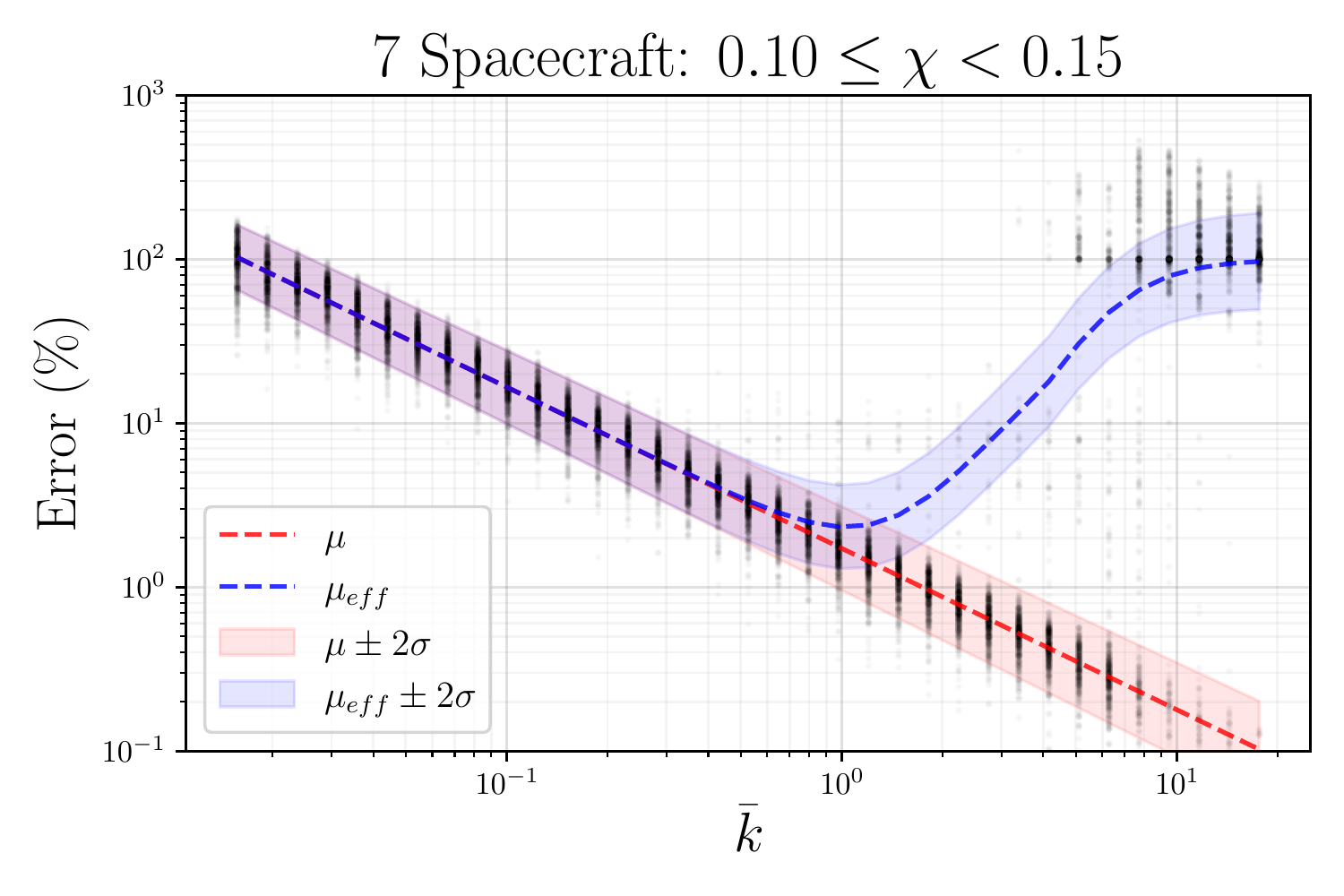}
\figsetgrpnote{We perform a posterior predictive check to verify that the coefficients found using Bayesian inference can correctly identify the mean value of error, $\mu$ (eqn \ref{eqn:mu_log}), along with the standard deviation $\sigma$ (eqn \ref{eqn:std}). We also show the effective error value $\mu_{eff}$ (eqn \ref{eqn:mu_eff}), which factors in the probability of aliasing. This is done for seven-spacecraft configurations with shape parameters between $0.1 \leq \chi < 0.15$.}
\figsetgrpend

\figsetgrpstart
\figsetgrpnum{6.5}
\figsetgrptitle{Eight-Spacecraft: Well Shaped}
\figsetplot{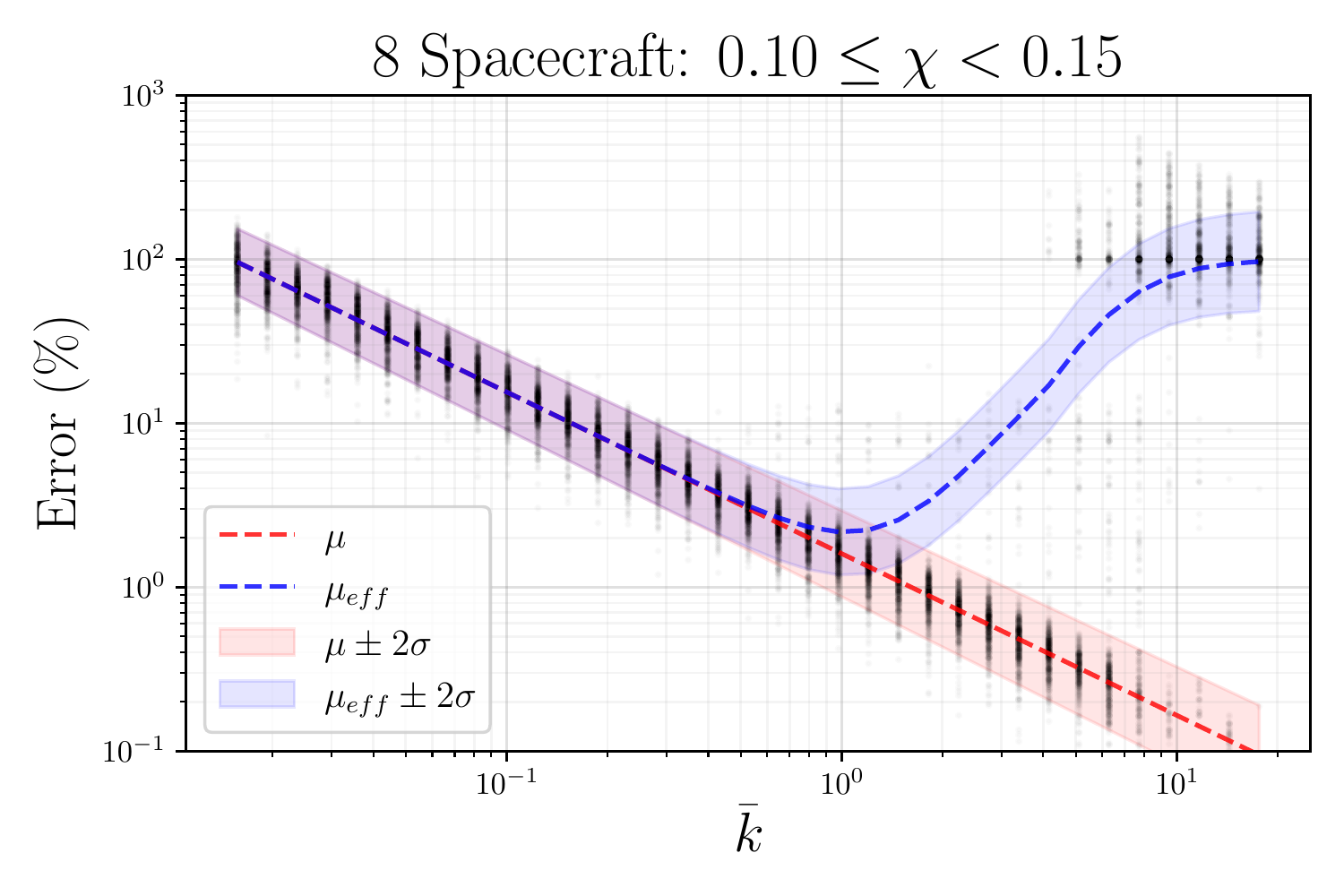}
\figsetgrpnote{We perform a posterior predictive check to verify that the coefficients found using Bayesian inference can correctly identify the mean value of error, $\mu$ (eqn \ref{eqn:mu_log}), along with the standard deviation $\sigma$ (eqn \ref{eqn:std}). We also show the effective error value $\mu_{eff}$ (eqn \ref{eqn:mu_eff}), which factors in the probability of aliasing. This is done for eight-spacecraft configurations with shape parameters between $0.1 \leq \chi < 0.15$.}
\figsetgrpend

\figsetgrpstart
\figsetgrpnum{6.6}
\figsetgrptitle{Nine-Spacecraft: Well Shaped}
\figsetplot{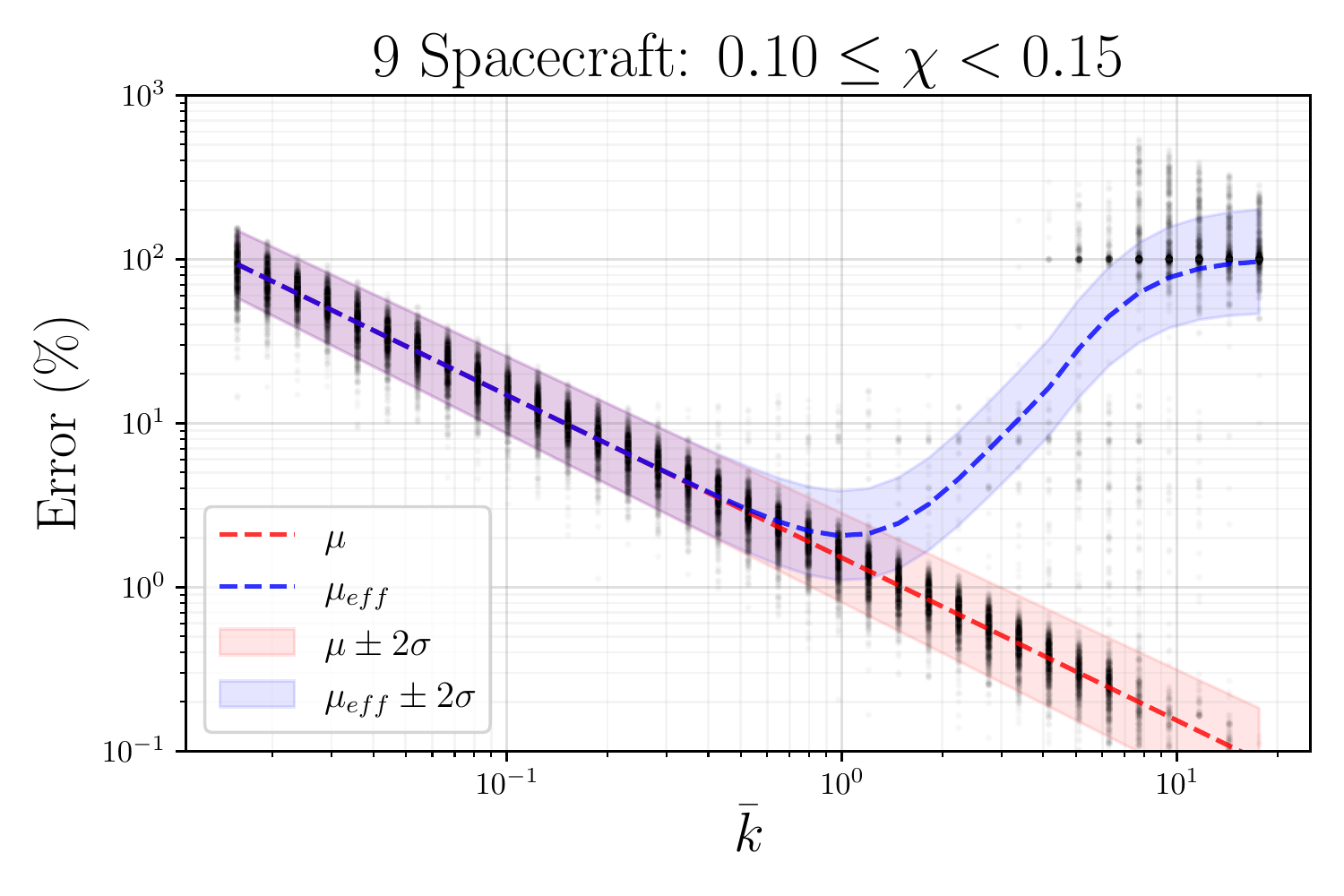}
\figsetgrpnote{We perform a posterior predictive check to verify that the coefficients found using Bayesian inference can correctly identify the mean value of error, $\mu$ (eqn \ref{eqn:mu_log}), along with the standard deviation $\sigma$ (eqn \ref{eqn:std}). We also show the effective error value $\mu_{eff}$ (eqn \ref{eqn:mu_eff}), which factors in the probability of aliasing. This is done for nine-spacecraft configurations with shape parameters between $0.1 \leq \chi < 0.15$.}
\figsetgrpend

\figsetgrpstart
\figsetgrpnum{6.7}
\figsetgrptitle{Four-Spacecraft: Average Shape}
\figsetplot{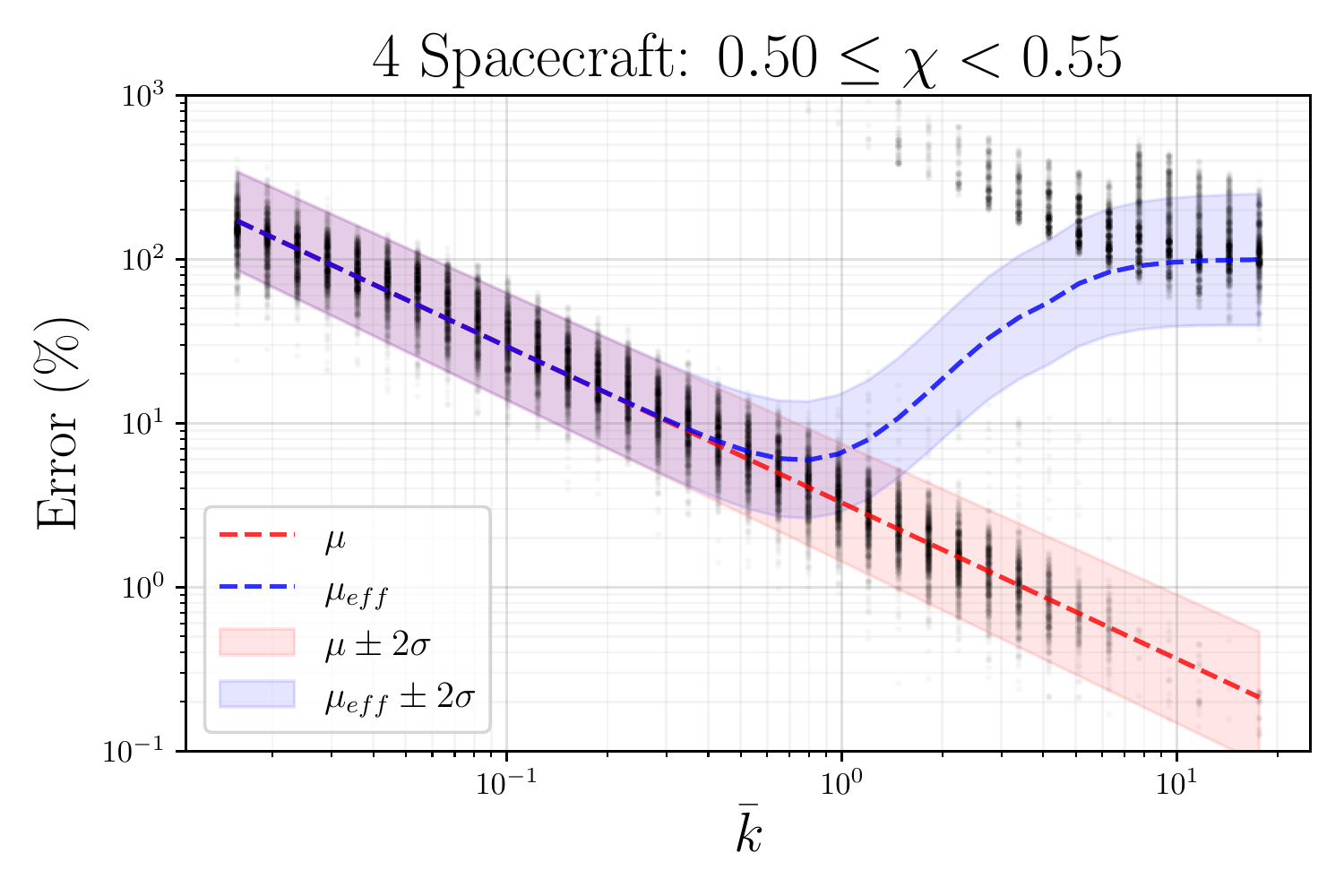}
\figsetgrpnote{We perform a posterior predictive check to verify that the coefficients found using Bayesian inference can correctly identify the mean value of error, $\mu$ (eqn \ref{eqn:mu_log}), along with the standard deviation $\sigma$ (eqn \ref{eqn:std}). We also show the effective error value $\mu_{eff}$ (eqn \ref{eqn:mu_eff}), which factors in the probability of aliasing. This is done for four-spacecraft configurations with shape parameters between $0.5 \leq \chi < 0.55$.}
\figsetgrpend

\figsetgrpstart
\figsetgrpnum{6.8}
\figsetgrptitle{Five-Spacecraft: Average Shape}
\figsetplot{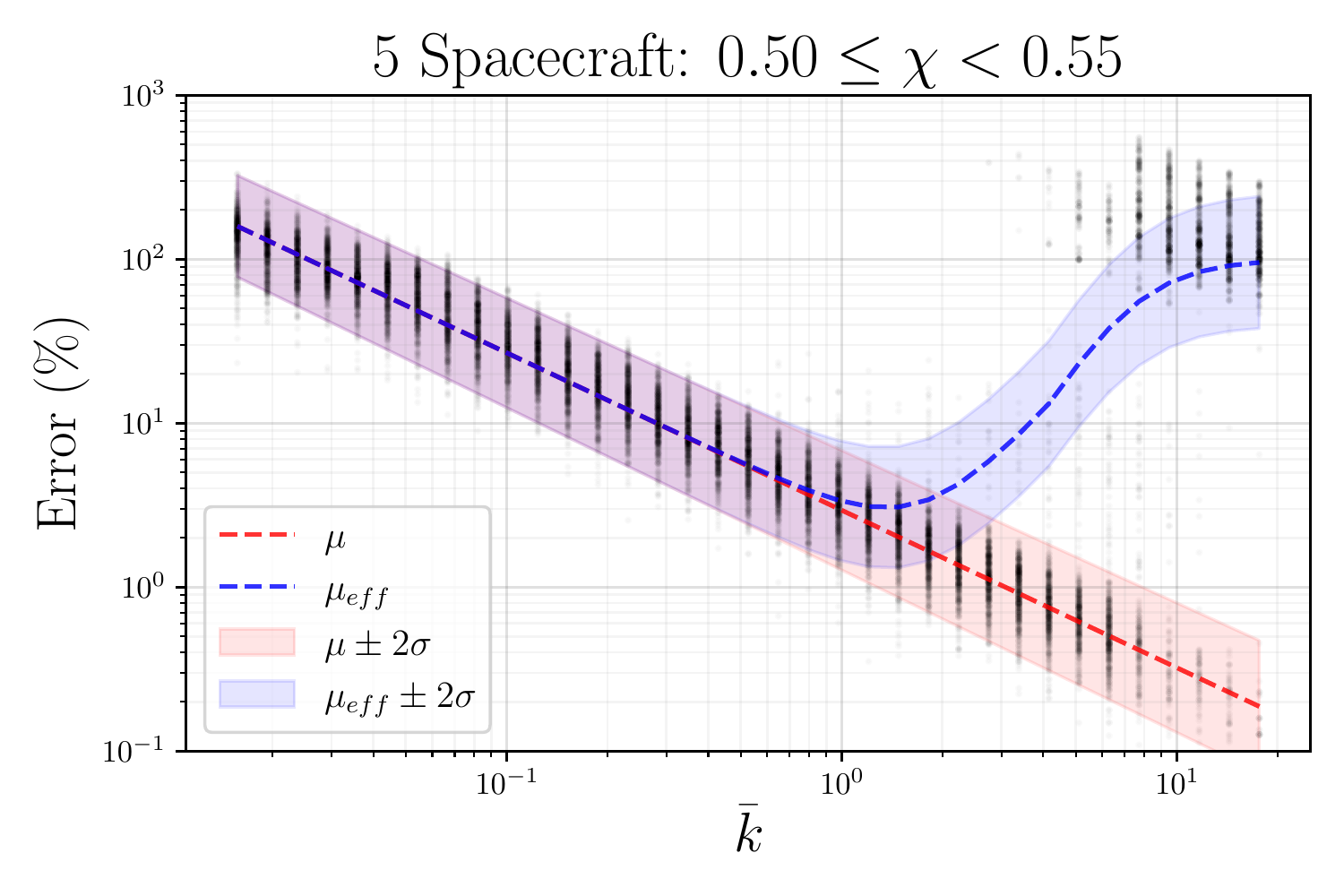}
\figsetgrpnote{We perform a posterior predictive check to verify that the coefficients found using Bayesian inference can correctly identify the mean value of error, $\mu$ (eqn \ref{eqn:mu_log}), along with the standard deviation $\sigma$ (eqn \ref{eqn:std}). We also show the effective error value $\mu_{eff}$ (eqn \ref{eqn:mu_eff}), which factors in the probability of aliasing. This is done for five-spacecraft configurations with shape parameters between $0.5 \leq \chi < 0.55$.}
\figsetgrpend

\figsetgrpstart
\figsetgrpnum{6.9}
\figsetgrptitle{Six-Spacecraft: Average Shape}
\figsetplot{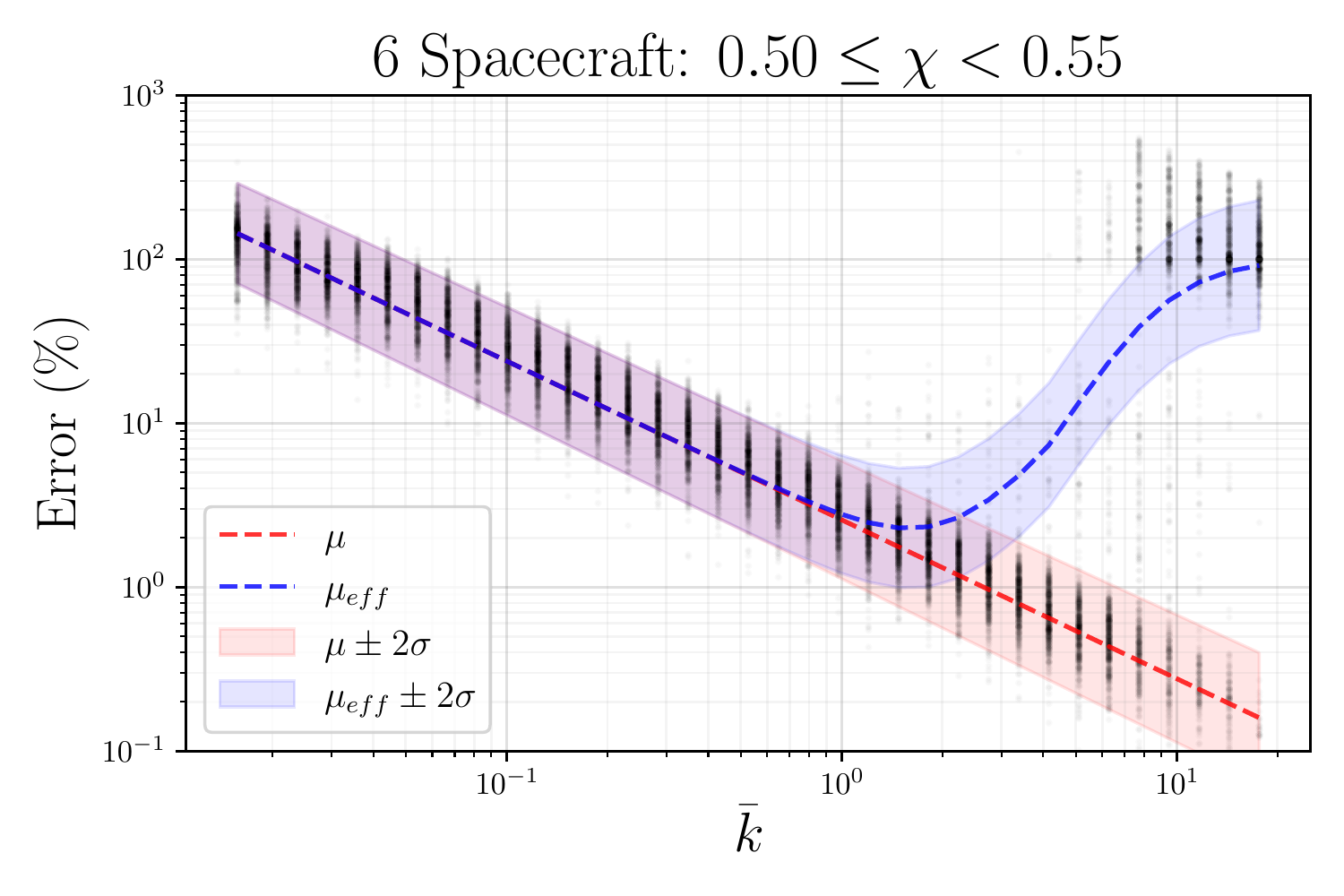}
\figsetgrpnote{We perform a posterior predictive check to verify that the coefficients found using Bayesian inference can correctly identify the mean value of error, $\mu$ (eqn \ref{eqn:mu_log}), along with the standard deviation $\sigma$ (eqn \ref{eqn:std}). We also show the effective error value $\mu_{eff}$ (eqn \ref{eqn:mu_eff}), which factors in the probability of aliasing. This is done for six-spacecraft configurations with shape parameters between $0.5 \leq \chi < 0.55$.}
\figsetgrpend

\figsetgrpstart
\figsetgrpnum{6.10}
\figsetgrptitle{Seven-Spacecraft: Average Shape}
\figsetplot{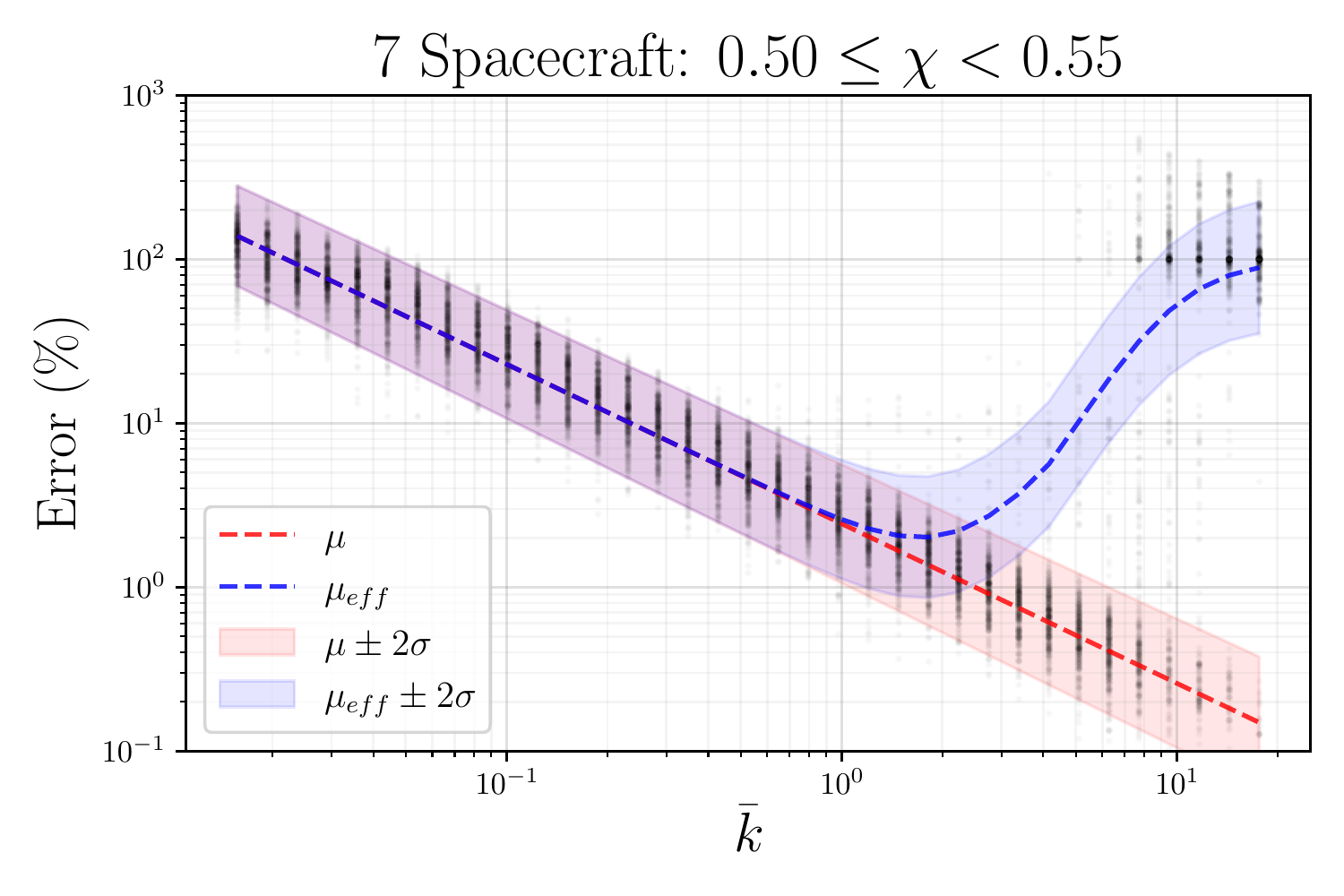}
\figsetgrpnote{We perform a posterior predictive check to verify that the coefficients found using Bayesian inference can correctly identify the mean value of error, $\mu$ (eqn \ref{eqn:mu_log}), along with the standard deviation $\sigma$ (eqn \ref{eqn:std}). We also show the effective error value $\mu_{eff}$ (eqn \ref{eqn:mu_eff}), which factors in the probability of aliasing. This is done for seven-spacecraft configurations with shape parameters between $0.5 \leq \chi < 0.55$.}
\figsetgrpend

\figsetgrpstart
\figsetgrpnum{6.11}
\figsetgrptitle{Eight-Spacecraft: Average Shape}
\figsetplot{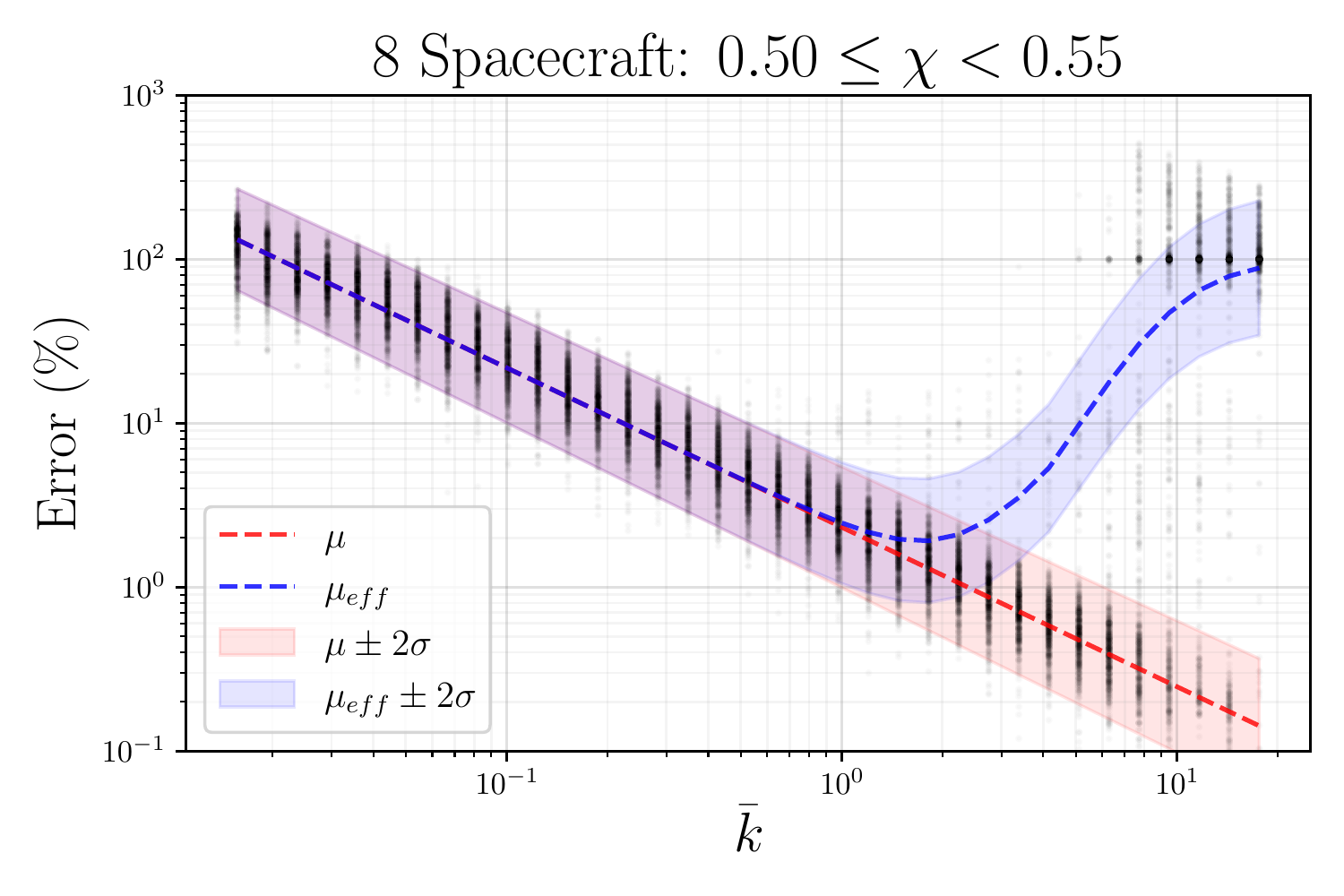}
\figsetgrpnote{We perform a posterior predictive check to verify that the coefficients found using Bayesian inference can correctly identify the mean value of error, $\mu$ (eqn \ref{eqn:mu_log}), along with the standard deviation $\sigma$ (eqn \ref{eqn:std}). We also show the effective error value $\mu_{eff}$ (eqn \ref{eqn:mu_eff}), which factors in the probability of aliasing. This is done for eight-spacecraft configurations with shape parameters between $0.5 \leq \chi < 0.55$.}
\figsetgrpend

\figsetgrpstart
\figsetgrpnum{6.12}
\figsetgrptitle{Nine-Spacecraft: Average Shape}
\figsetplot{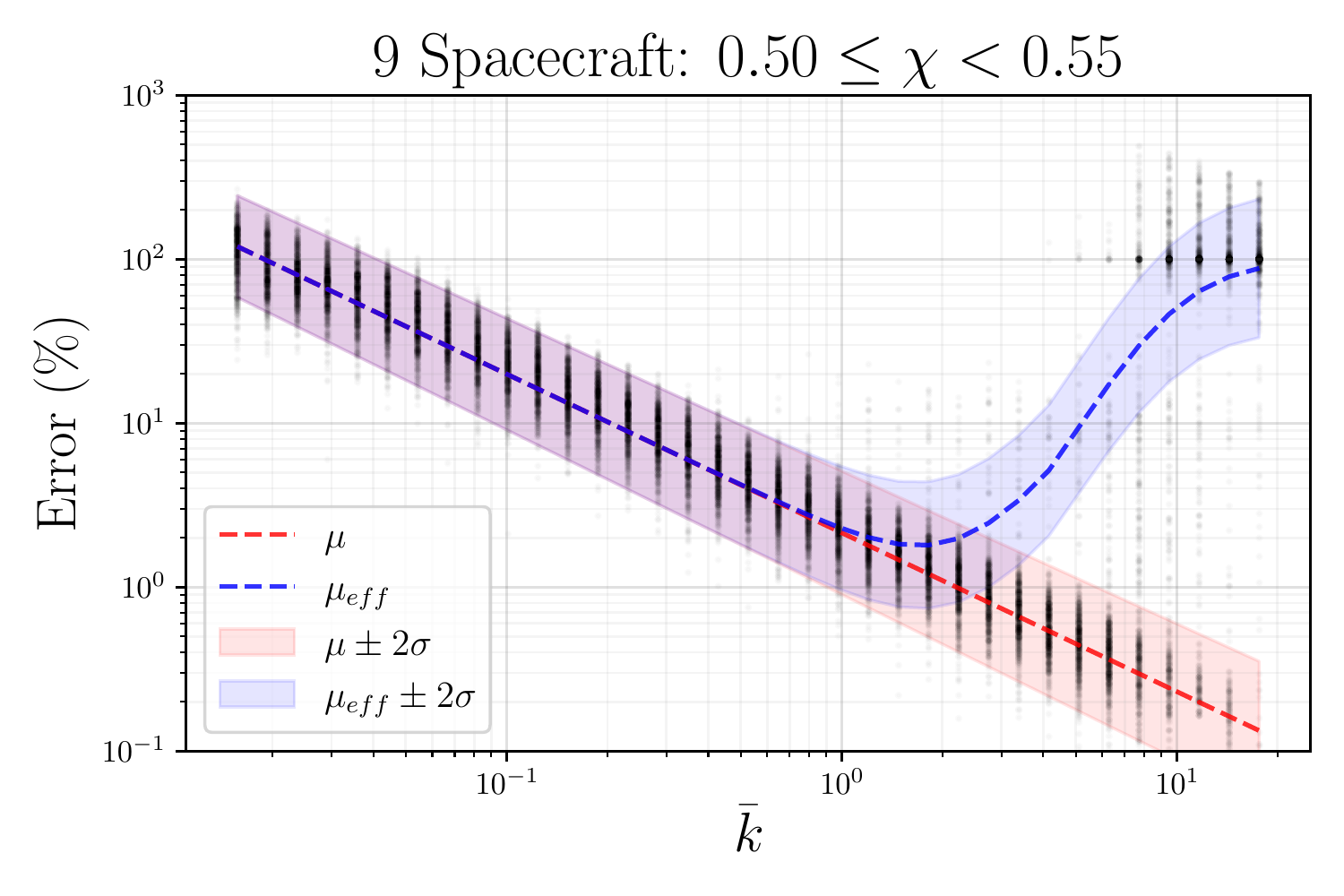}
\figsetgrpnote{We perform a posterior predictive check to verify that the coefficients found using Bayesian inference can correctly identify the mean value of error, $\mu$ (eqn \ref{eqn:mu_log}), along with the standard deviation $\sigma$ (eqn \ref{eqn:std}). We also show the effective error value $\mu_{eff}$ (eqn \ref{eqn:mu_eff}), which factors in the probability of aliasing. This is done for nine-spacecraft configurations with shape parameters between $0.5 \leq \chi < 0.55$.}
\figsetgrpend

\figsetgrpnum{6.13}
\figsetgrptitle{Four-Spacecraft: Poor Shape}
\figsetplot{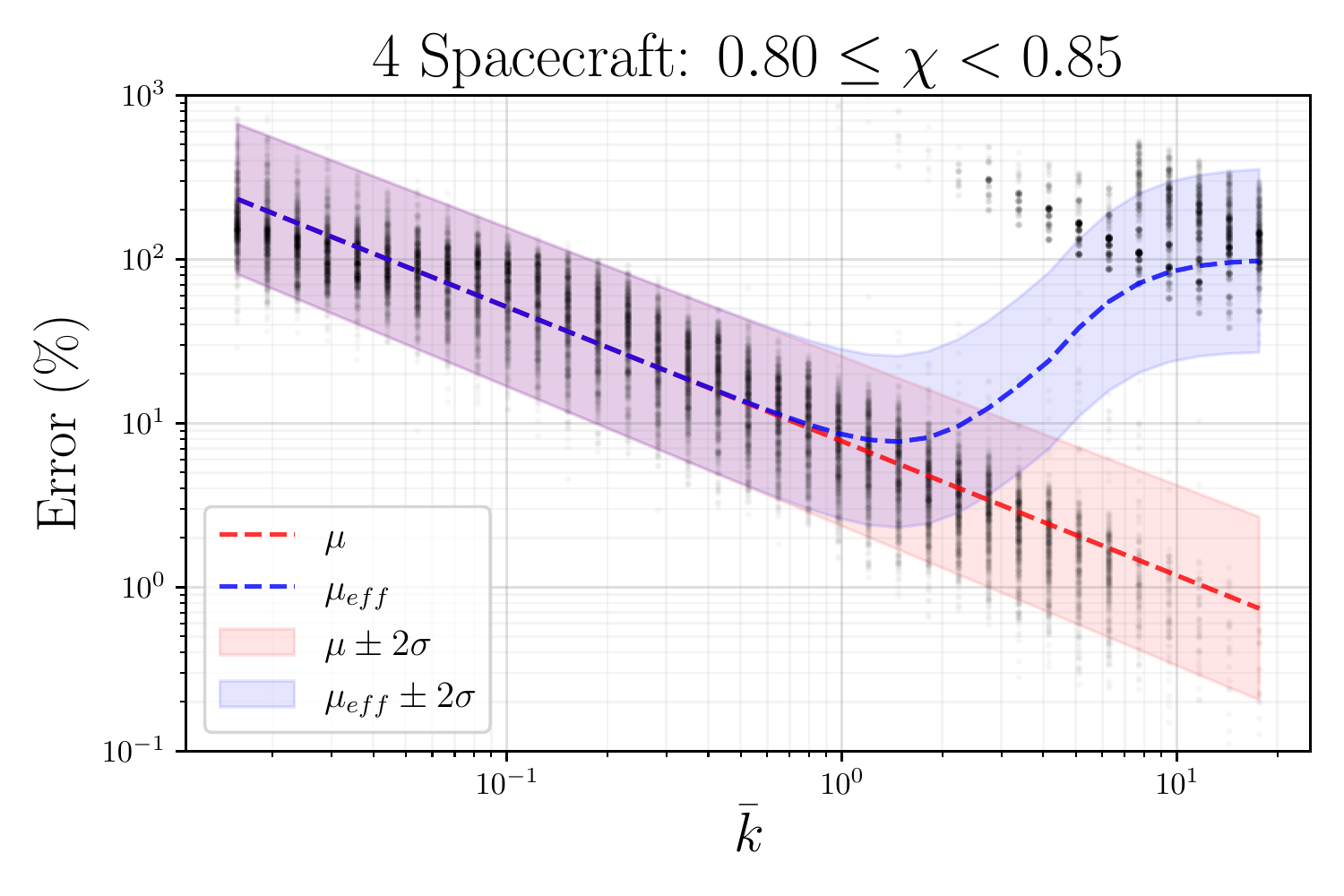}
\figsetgrpnote{We perform a posterior predictive check to verify that the coefficients found using Bayesian inference can correctly identify the mean value of error, $\mu$ (eqn \ref{eqn:mu_log}), along with the standard deviation $\sigma$ (eqn \ref{eqn:std}). We also show the effective error value $\mu_{eff}$ (eqn \ref{eqn:mu_eff}), which factors in the probability of aliasing. This is done for four-spacecraft configurations with shape parameters between $0.8 \leq \chi < 0.85$.}
\figsetgrpend

\figsetgrpstart
\figsetgrpnum{6.14}
\figsetgrptitle{Five-Spacecraft: Poor Shape}
\figsetplot{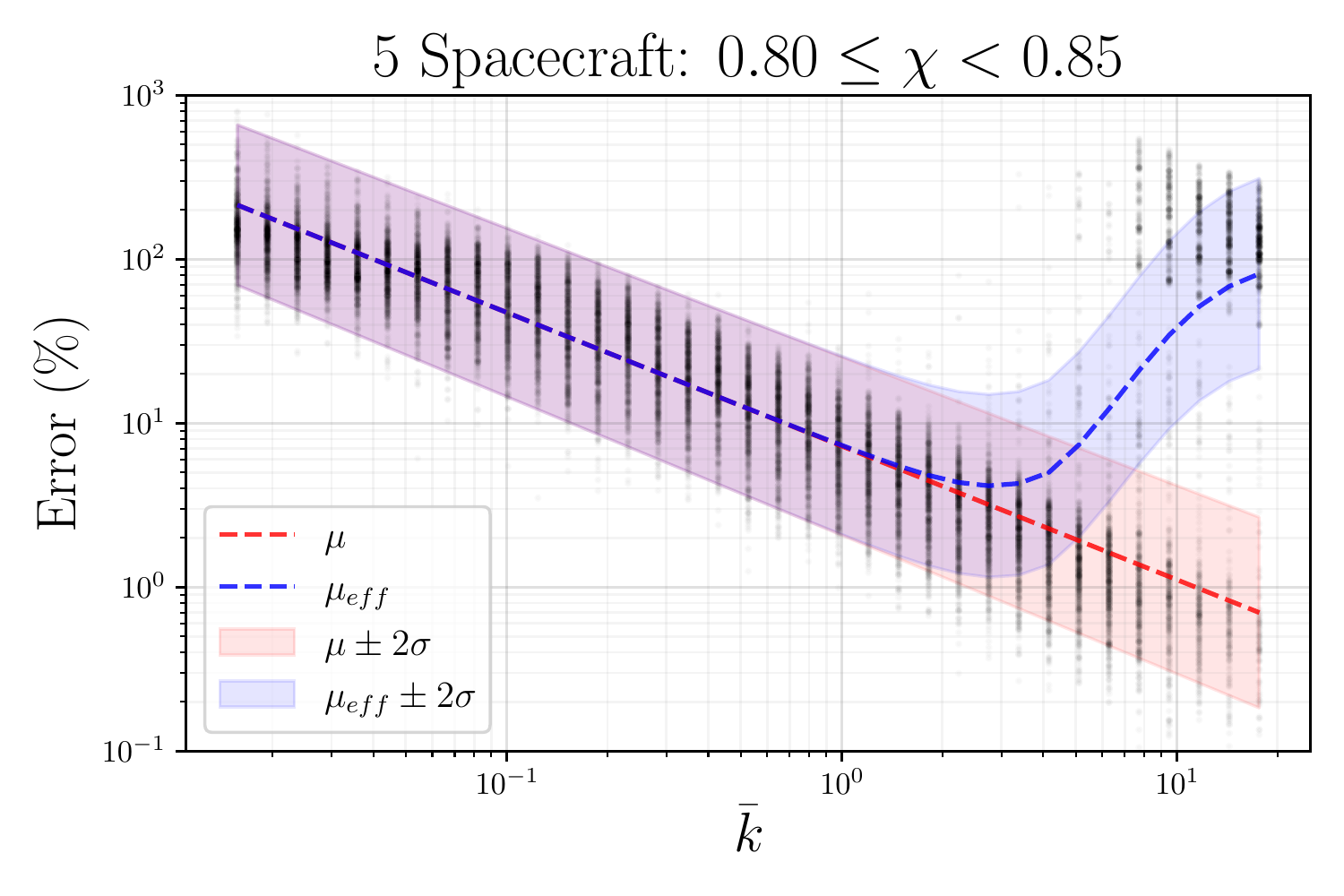}
\figsetgrpnote{We perform a posterior predictive check to verify that the coefficients found using Bayesian inference can correctly identify the mean value of error, $\mu$ (eqn \ref{eqn:mu_log}), along with the standard deviation $\sigma$ (eqn \ref{eqn:std}). We also show the effective error value $\mu_{eff}$ (eqn \ref{eqn:mu_eff}), which factors in the probability of aliasing. This is done for five-spacecraft configurations with shape parameters between $0.8 \leq \chi < 0.85$.}
\figsetgrpend

\figsetgrpstart
\figsetgrpnum{6.15}
\figsetgrptitle{Six-Spacecraft: Poor Shape}
\figsetplot{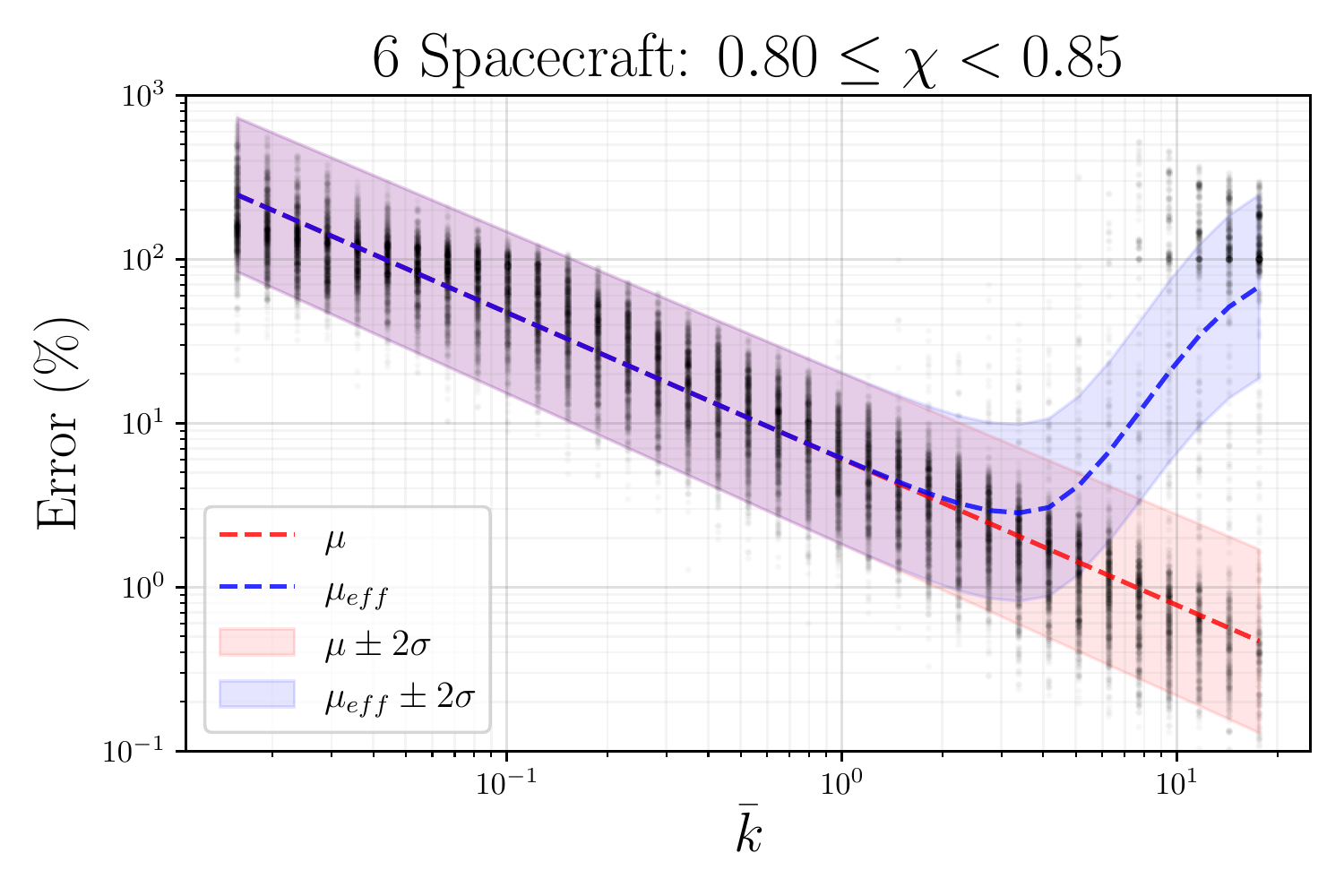}
\figsetgrpnote{We perform a posterior predictive check to verify that the coefficients found using Bayesian inference can correctly identify the mean value of error, $\mu$ (eqn \ref{eqn:mu_log}), along with the standard deviation $\sigma$ (eqn \ref{eqn:std}). We also show the effective error value $\mu_{eff}$ (eqn \ref{eqn:mu_eff}), which factors in the probability of aliasing. This is done for six-spacecraft configurations with shape parameters between $0.8 \leq \chi < 0.85$.}
\figsetgrpend

\figsetgrpstart
\figsetgrpnum{6.16}
\figsetgrptitle{Seven-Spacecraft: Poor Shape}
\figsetplot{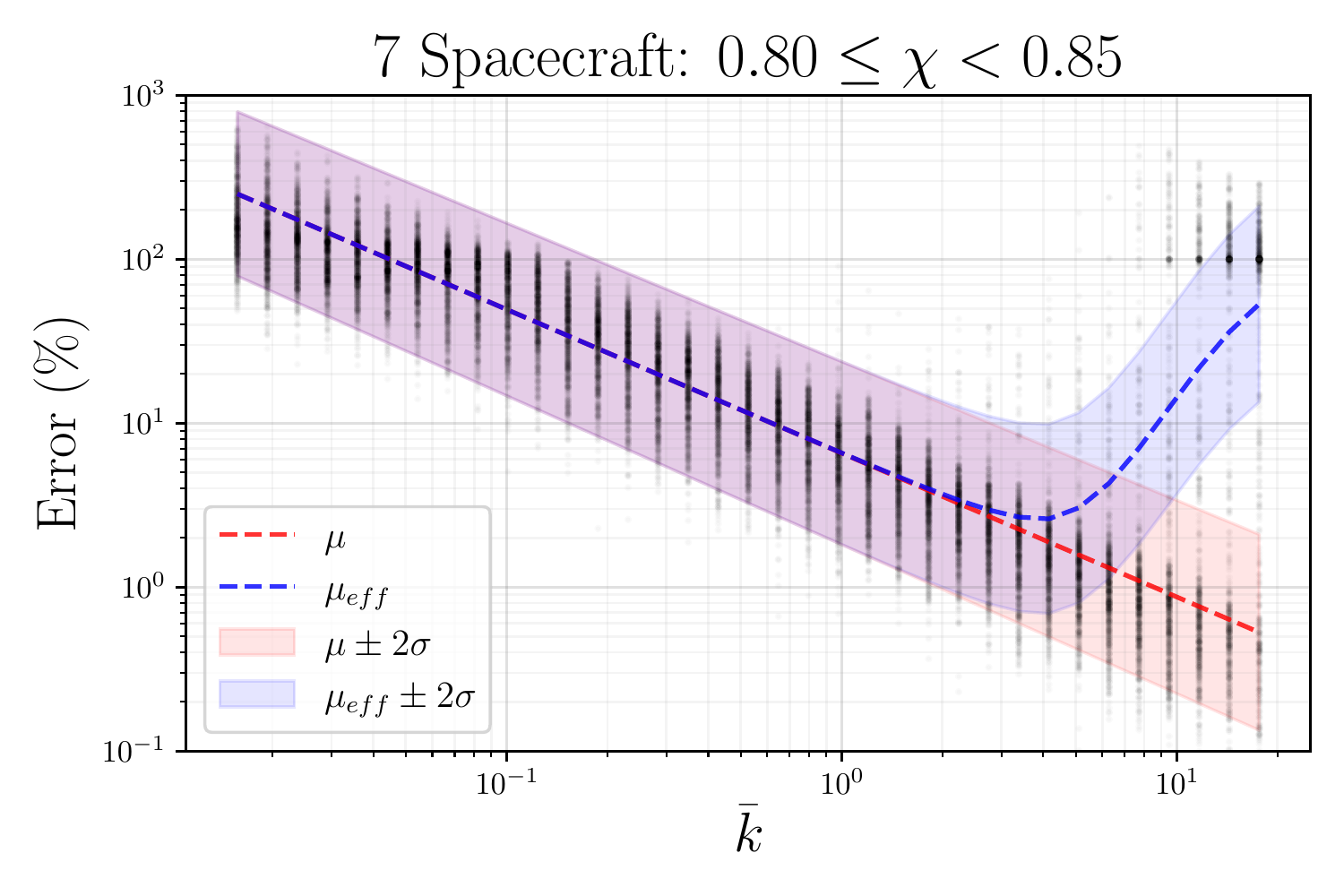}
\figsetgrpnote{We perform a posterior predictive check to verify that the coefficients found using Bayesian inference can correctly identify the mean value of error, $\mu$ (eqn \ref{eqn:mu_log}), along with the standard deviation $\sigma$ (eqn \ref{eqn:std}). We also show the effective error value $\mu_{eff}$ (eqn \ref{eqn:mu_eff}), which factors in the probability of aliasing. This is done for seven-spacecraft configurations with shape parameters between $0.8 \leq \chi < 0.85$.}
\figsetgrpend

\figsetgrpstart
\figsetgrpnum{6.17}
\figsetgrptitle{Eight-Spacecraft: Poor Shape}
\figsetplot{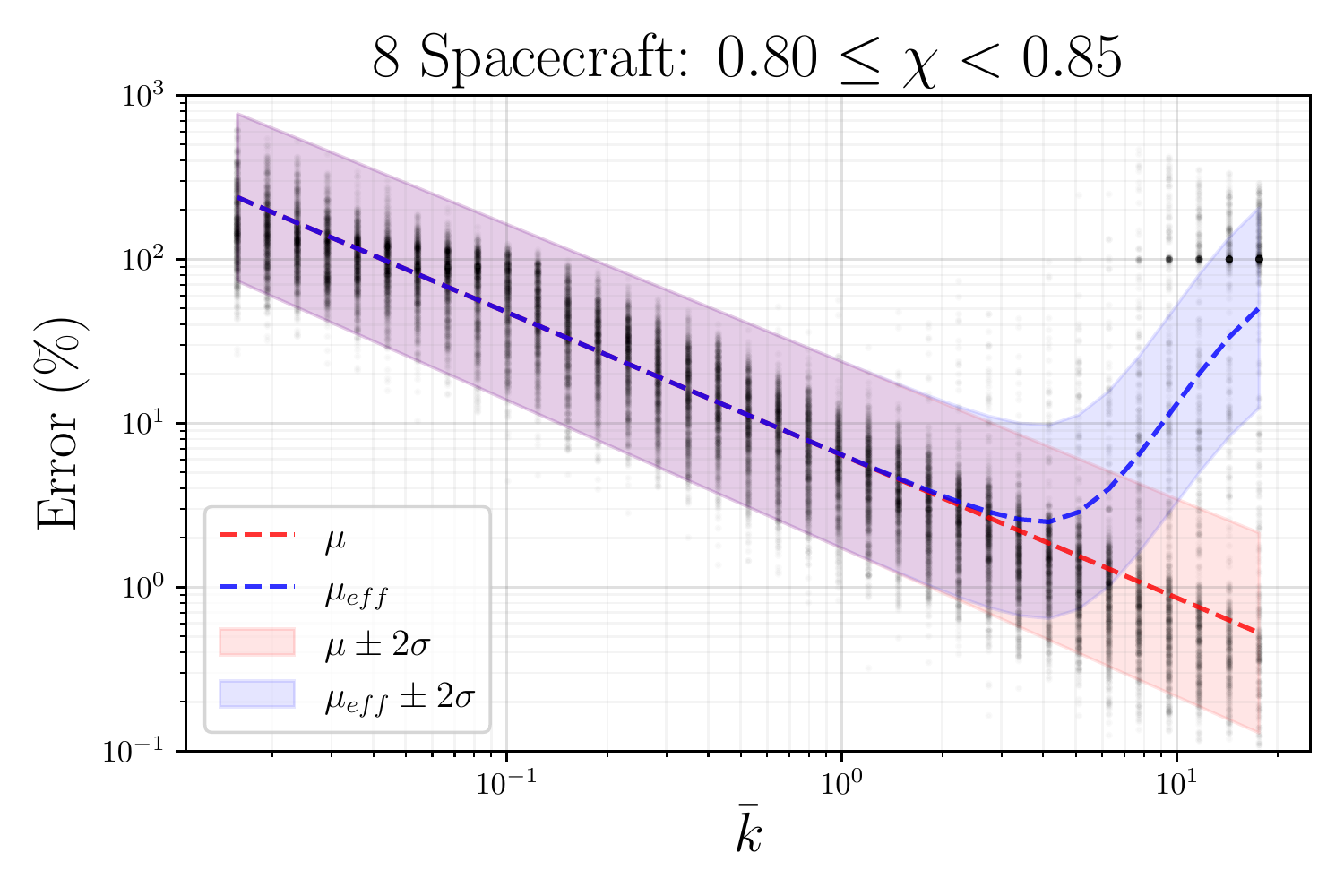}
\figsetgrpnote{We perform a posterior predictive check to verify that the coefficients found using Bayesian inference can correctly identify the mean value of error, $\mu$ (eqn \ref{eqn:mu_log}), along with the standard deviation $\sigma$ (eqn \ref{eqn:std}). We also show the effective error value $\mu_{eff}$ (eqn \ref{eqn:mu_eff}), which factors in the probability of aliasing. This is done for eight-spacecraft configurations with shape parameters between $0.8 \leq \chi < 0.85$.}
\figsetgrpend

\figsetgrpstart
\figsetgrpnum{6.18}
\figsetgrptitle{Nine-Spacecraft: Poor Shape}
\figsetplot{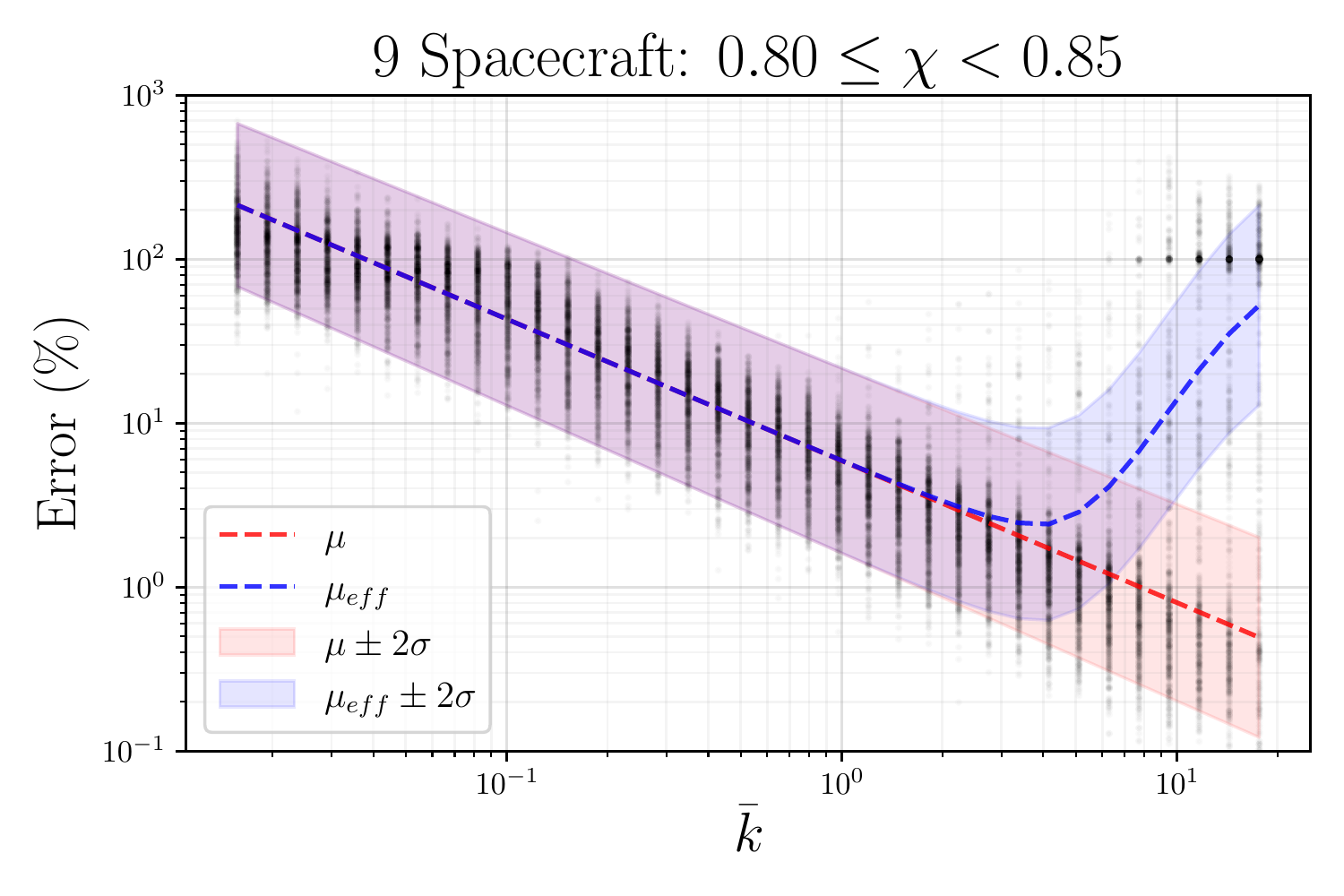}
\figsetgrpnote{We perform a posterior predictive check to verify that the coefficients found using Bayesian inference can correctly identify the mean value of error, $\mu$ (eqn \ref{eqn:mu_log}), along with the standard deviation $\sigma$ (eqn \ref{eqn:std}). We also show the effective error value $\mu_{eff}$ (eqn \ref{eqn:mu_eff}), which factors in the probability of aliasing. This is done for nine-spacecraft configurations with shape parameters between $0.8 \leq \chi < 0.85$.}
\figsetgrpend
\figsetend

By inspecting Figure \ref{fig:PPC} we find that the average error equation seems to model the data accurately, but we wish to numerically quantify how well the data follows the fitted equations. Recall that in \S \ref{ssec:eqn_forms} we supposed that (for a fixed $\bar{k}$ and $\chi$) the error was distributed log-normally, and we learned equations for the mean value $\mu$ and standard deviation $\sigma$ of error found computing wavevectors using the wave-telescope. If the hypothesis and learned equations for $\mu$ and $\sigma$ are correct, then we expect that the proportion of data points that fall within the intervals $\mu \pm \sigma$, $\mu \pm 2\sigma$, and $\mu \pm 3\sigma$ are $0.6827$, $0.9545$, and $0.9973$ respectively.

Using our simulation data and learned equations for $\mu$ and $\sigma$, we check how many points fall in each of the above intervals. Because aliasing causes large sudden deviations from the average error, we restrict ourselves to values of $\bar{k}$ and $\chi$ where $P(alias) < 0.01$ and $\chi < 1$. For spacecraft configurations and wavevector combinations that fit this restriction, Table \ref{tab:PPC} summarizes the proportions of wave-telescope computations that fall within the listed intervals. 

\begin{table}[ht]
\centering
\caption{Using our learned equations for $\mu$ and $\sigma$, we verify our hypothesis that the data is log-normally distributed and that our models confidence interval is correct for each number of spacecraft $N \in \{4,5,6,7,8,9\}$. We show the proportion of error values that fall into each range, as well as the size of the dataset $n$ that was used in the computation of these proportions (includes training and testing data). In the second row, we compare the computed proportions to those expected if the data was exactly log-normally distributed.}
{
\begin{tabular}{|c|ccc|} \hline
Interval & $\mu \pm \sigma$ & $\mu \pm 2\sigma$ & $\mu \pm 3\sigma$\\
\hline
 Expected Proportion & 0.6827 & 0.9545 & 0.9973 \\
 \hline
4 s/c ($n=240950$) & 0.6894 & 0.9577 & 0.9921  \\
5 s/c ($n=276100$) & 0.6993 & 0.9592 & 0.9926  \\
6 s/c ($n=275900$) & 0.6945 & 0.9553 & 0.9904  \\
7 s/c ($n=282750$) & 0.6942 & 0.9536 & 0.9896  \\
8 s/c ($n=287050$) & 0.7012 & 0.9547 & 0.9899  \\
9 s/c ($n=287850$) & 0.7016 & 0.9557 & 0.9897  \\
 \hline
\end{tabular}}
\label{tab:PPC} 
\end{table}

We see from Table \ref{tab:PPC} that the proportion of data points that fall within each specified interval is consistent with the expected proportion. This confirms that the log-normally distributed error hypothesis and the learned equations of $\mu$ and $\sigma$ accurately capture the errors in wave-telescope computations for all number of spacecraft $N \in \{4,5,6,7,8,9\}$. In the Appendix \S \ref{sec:appendix.verify} Figure \ref{fig:PPC_cdfs}, we expand this analysis by computing the proportion of points that satisfy $Error < \mu + \sigma_0 \sigma$ for $\sigma_0 \in [-3,3]$ and comparing it to the CDF of the standard normal distribution.

\section{Results}
\label{sec:Results}
We now have equations that estimate the effective error in the wave-telescope technique $\mu_{eff}(\bar{k},\chi)$ as well as the standard deviation from that mean value $\sigma(\bar{k},\chi)$. In \S \ref{ssec:k_orders} we study these equations to determine how shape $\chi$ and number of spacecraft $N$ impact the range of wavevectors which can be reconstructed. In \S \ref{ssec:app_HS} we demonstrate the utility of these equations by applying them to the nine-spacecraft configurations of the NASA mission HelioSwarm.

\subsection{Number and Shape Dependence}
\label{ssec:k_orders}
\begin{figure}[ht]
\centering
\begin{tabular}{cc}
\includegraphics[width=0.45\textwidth]{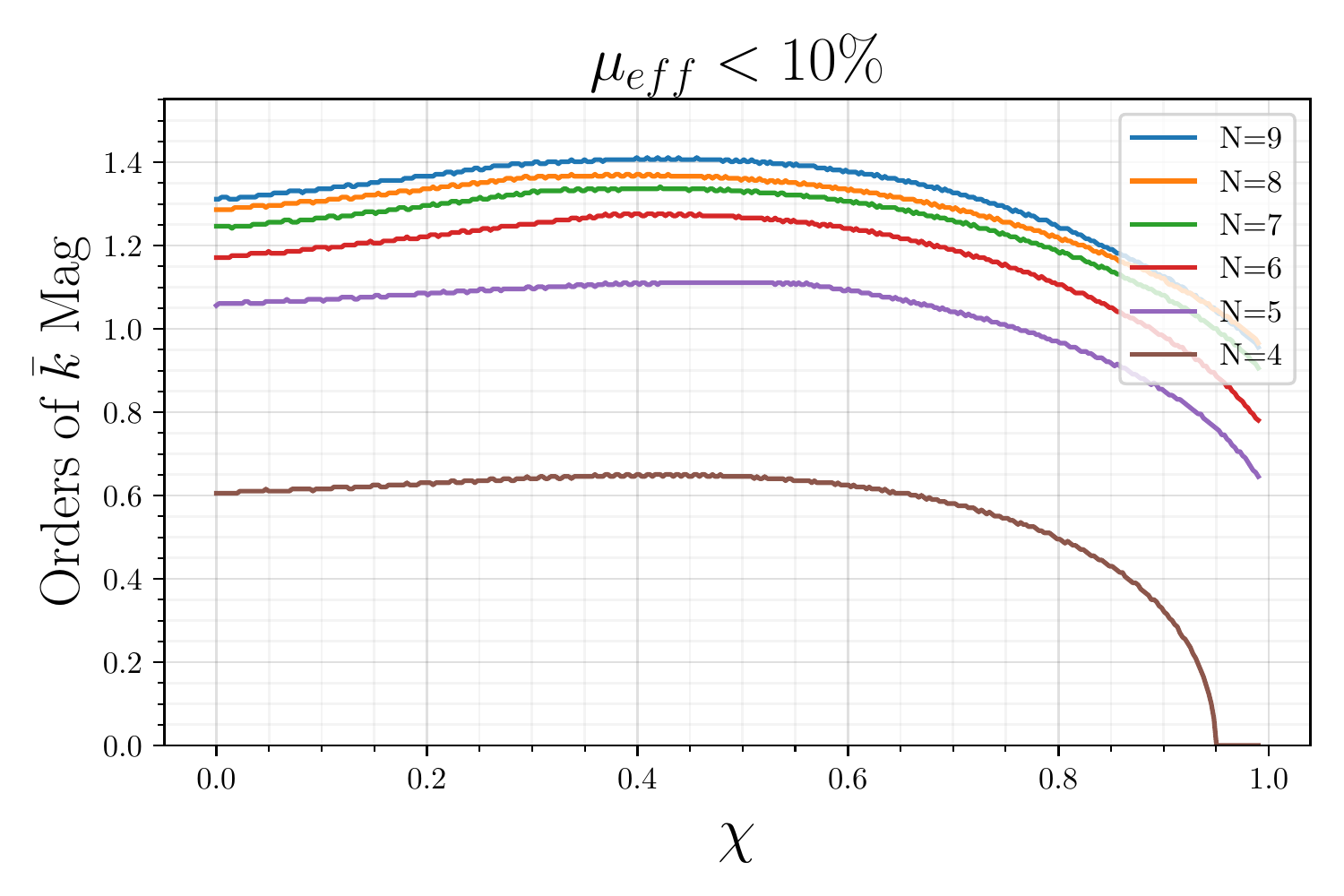} &
\includegraphics[width=0.45\textwidth]{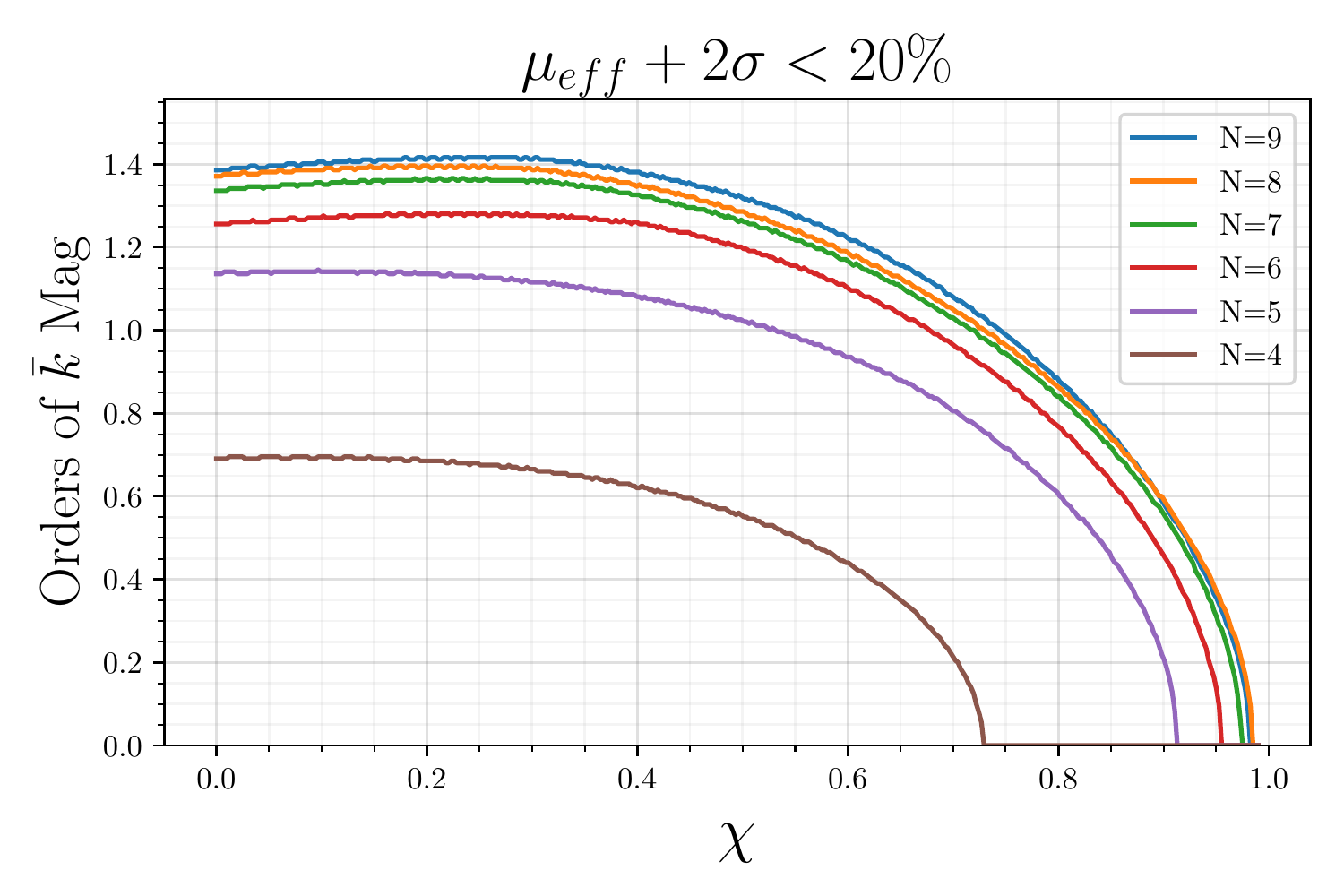} 
\end{tabular}
\caption{{\small Using our equations of effective error $\mu_{eff}$ and $\sigma$, we plot how many orders of magnitude of wavevectors $\bar{k}$ can be reconstructed with two different measures of accuracy and precision. The left panel shows how many orders of $\bar{k}$ can be reconstructed such that $\mu_{eff}$ is less than 10\%. The right panel shows how many order of $\bar{k}$ can be reconstructed such that $\mu_{eff} + 2\sigma$ is less than 20\%. }}
\label{fig:k_orders} 
\end{figure}

To quantify the magnitudes of wavevectors that can be reconstructed accurately as a function of number of spacecraft $N$ and shape of the configuration $\chi$, we use our effective error equation $\mu_{eff}(\bar{k},\chi)$ and standard deviation equation $\sigma(\bar{k},\chi)$. For every combination of $N \in \{4,5,6,7,8,9\}$ and $\chi \in [0,1]$ we examine the orders of $\bar{k}$-magnitude over which our learned equations predict the wave-telescope is accurate. Because the accuracy and precision of the technique are both important, we have quantified the accuracy using two separate metrics in Figure \ref{fig:k_orders}.

The left panel of Figure \ref{fig:k_orders} shows how many orders of magnitude of wavevectors an $N$-spacecraft configuration can reconstruct with an effective error less than 10\%. Due to the underlying distribution, this should be interpreted as the 50$^{\text{th}}$ percentile error is less than 10\%. The right panel of Figure \ref{fig:k_orders} shows how many orders of magnitude of wavevectors an $N$-spacecraft configuration can reconstruct with $\mu_{eff} + 2\sigma < 20\%$ (i.e. 97.7$^{\text{th}}$ percentile error is less than 20\%).

From Figure \ref{fig:k_orders} we conclude that increasing the number of spacecraft used in the wave-telescope monotonically increases the ranges of wavevectors that can be reconstructed. However, the largest increase in the orders of $\bar{k}$ that can be reconstructed accurately is gained by going from just four to five spacecraft. Exploring the data further, we conclude that this is due primarily to a sharp decline in aliasing occurring when increasing from four to five spacecraft. We suspect that the large decline in aliasing is likely caused by four-spacecraft configurations having just two spacecraft aligned in their reciprocal vector direction, while five-spacecraft configurations have at least three spacecraft aligned in all directions. The decrease in aliasing can be seen in the figure set corresponding to Figure \ref{fig:mu_data}. 

Figure \ref{fig:k_orders} also shows that spacecraft configurations with high shape parameter $\chi$ cannot be used to accurately reconstruct many orders of $\bar{k}$ magnitude. This is because the error variations, $\sigma$, imposed by waves traveling in different directions are quite high if a spacecraft configuration is poorly shaped.

\subsection{Application to HelioSwarm}
\label{ssec:app_HS}
The HelioSwarm mission will have nine spacecraft taking in situ measurements of plasmas in near Earth environments including the pristine solar wind, magnetosheath, and foreshock \citep{Klein:2019:WP}. We introduce NEWTSS (Numerically Estimated Wave-Telescope Subset Selector), which is our tool for a priori selection of the optimal subset of spacecraft to use at each wavevector magnitude $k$. NEWTSS uses the effective error equations presented in \S \ref{sec:learn_eqns} and performs an exhaustive search of the possible spacecraft subsets to determine which will minimizes the expected value of the $97.7^{\text{th}}$ percentile error ($\mu + 2 \sigma$).

\begin{figure}[ht]
\centering
\begin{tabular}{ccc}
\includegraphics[width=0.3\textwidth]{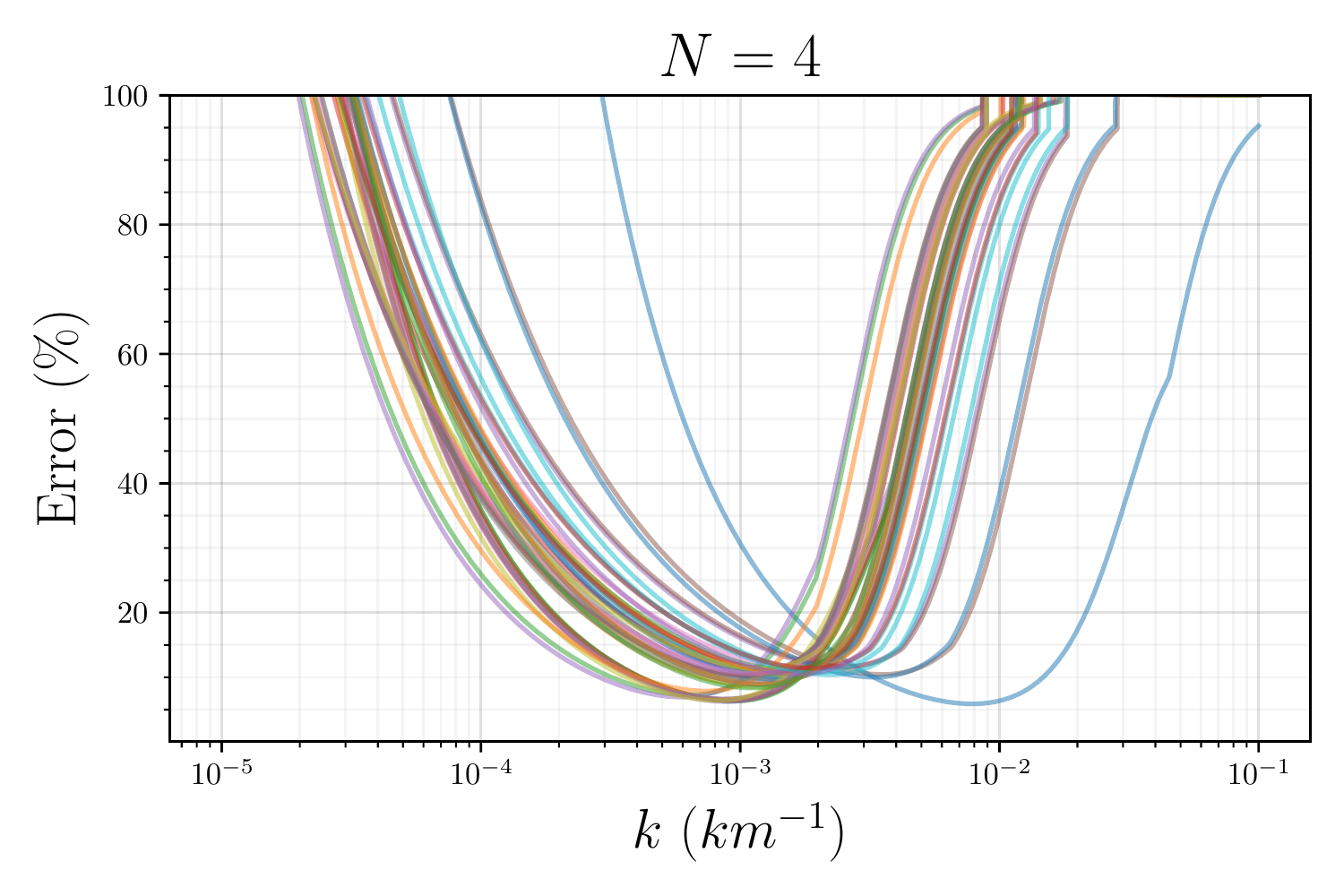} &
\includegraphics[width=0.3\textwidth]{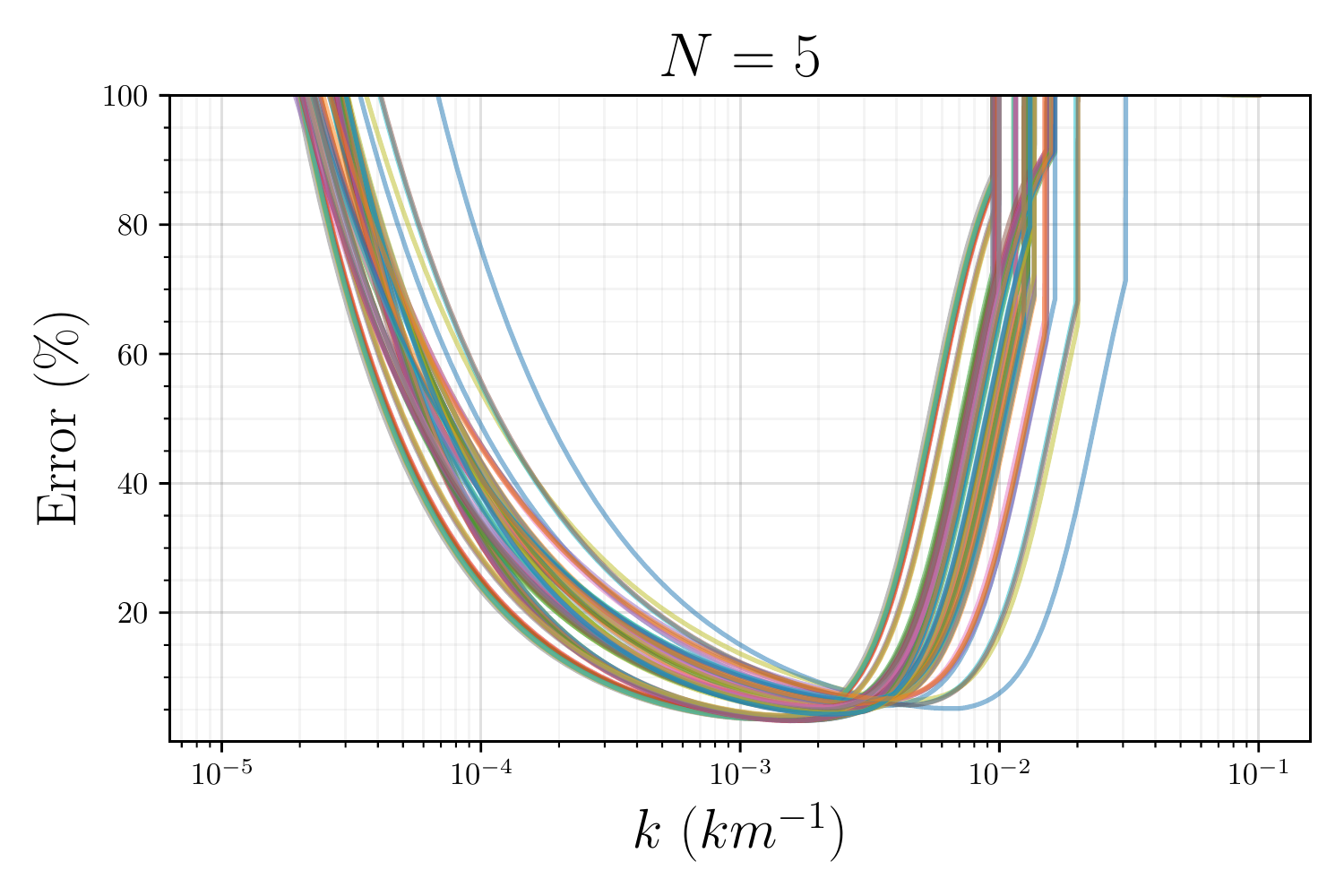} &
\includegraphics[width=0.3\textwidth]{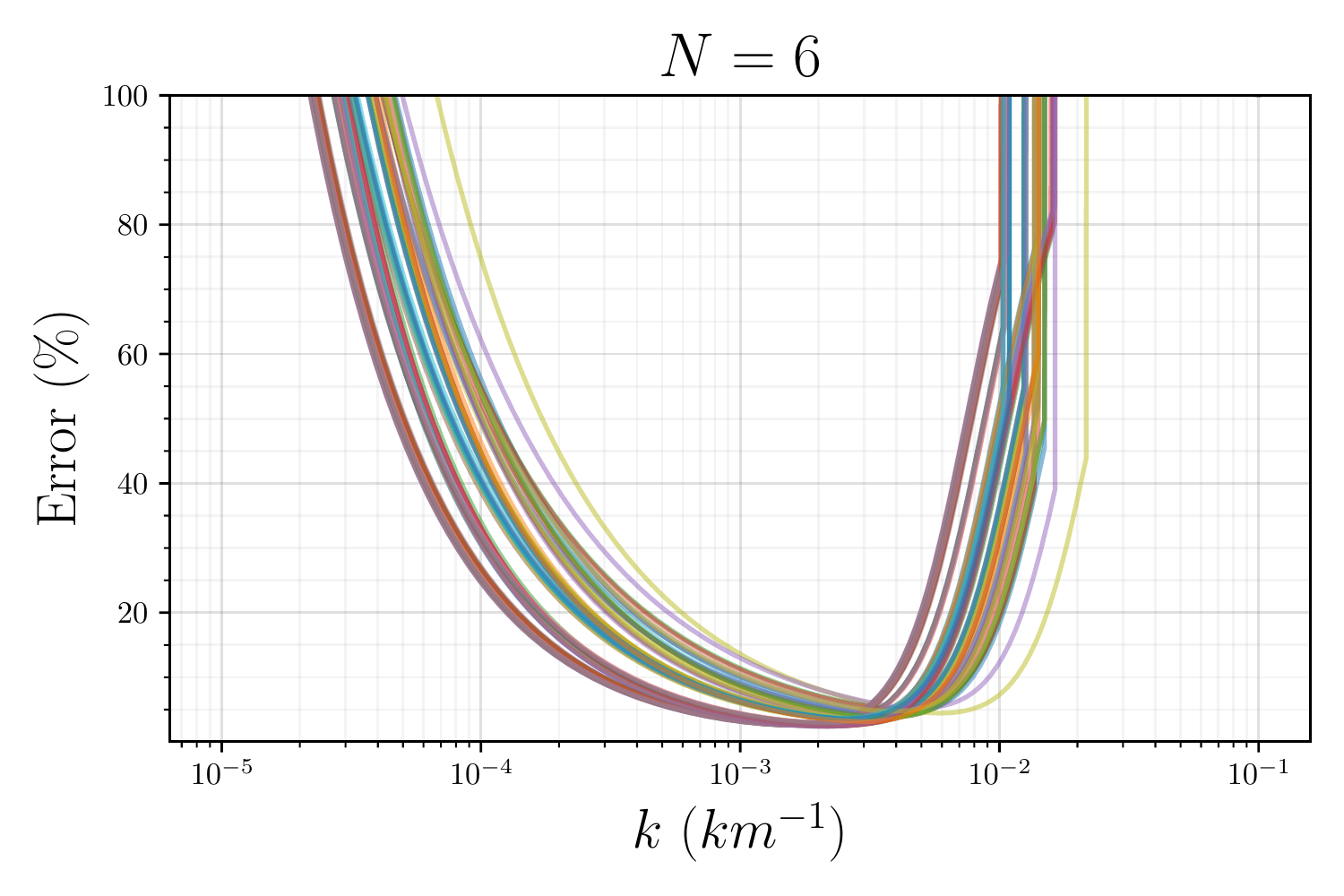} \\
\includegraphics[width=0.3\textwidth]{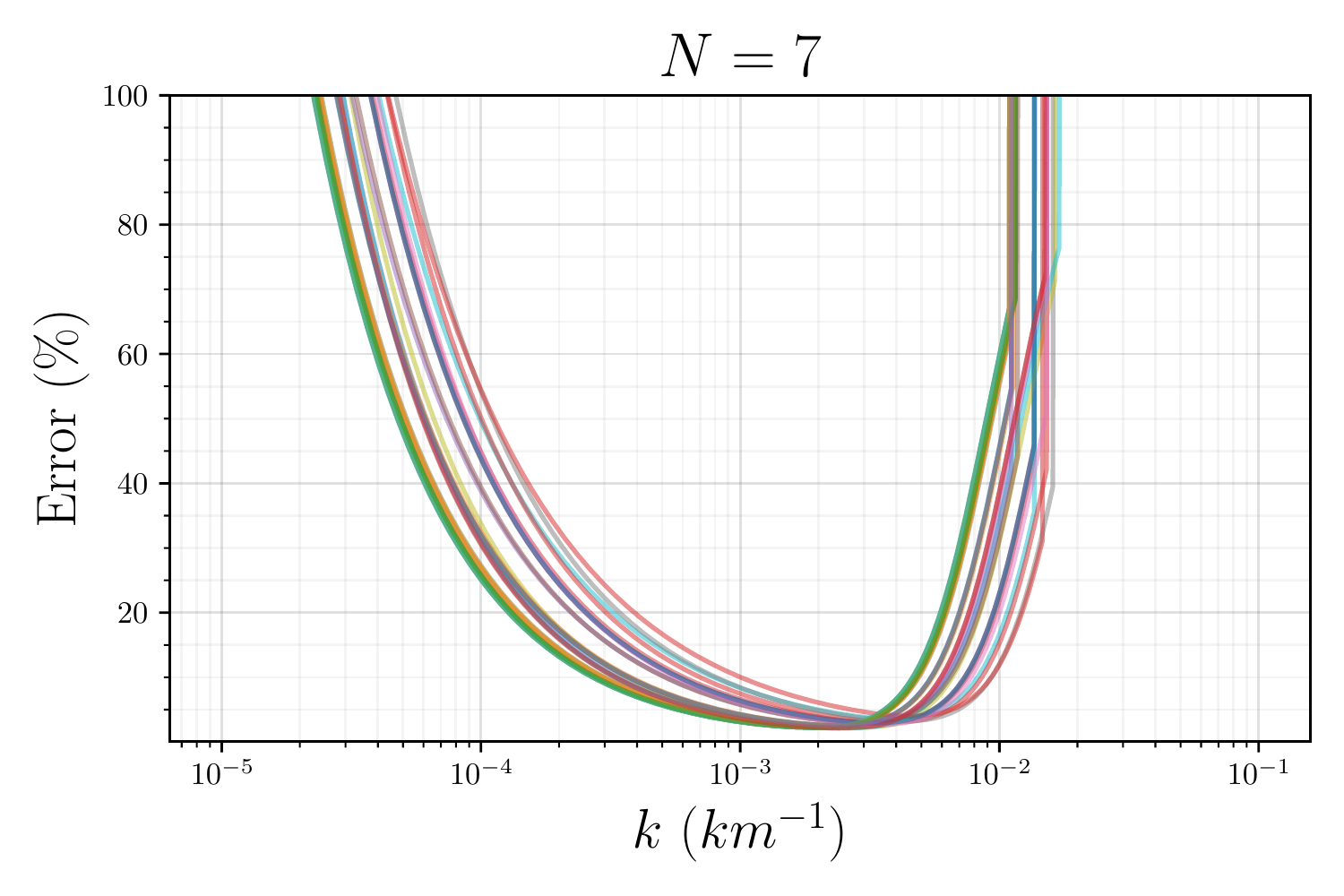} &
\includegraphics[width=0.3\textwidth]{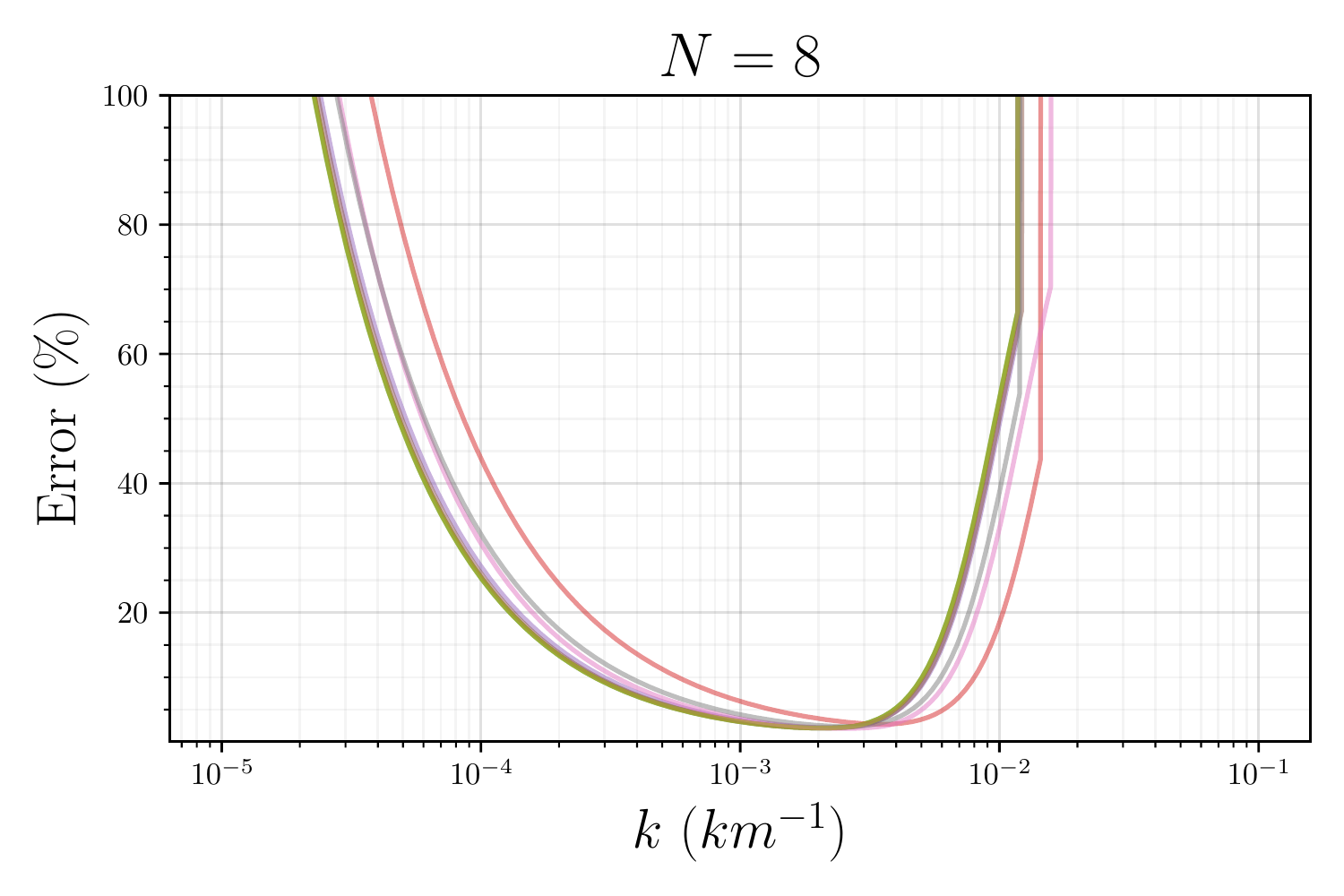} &
\includegraphics[width=0.3\textwidth]{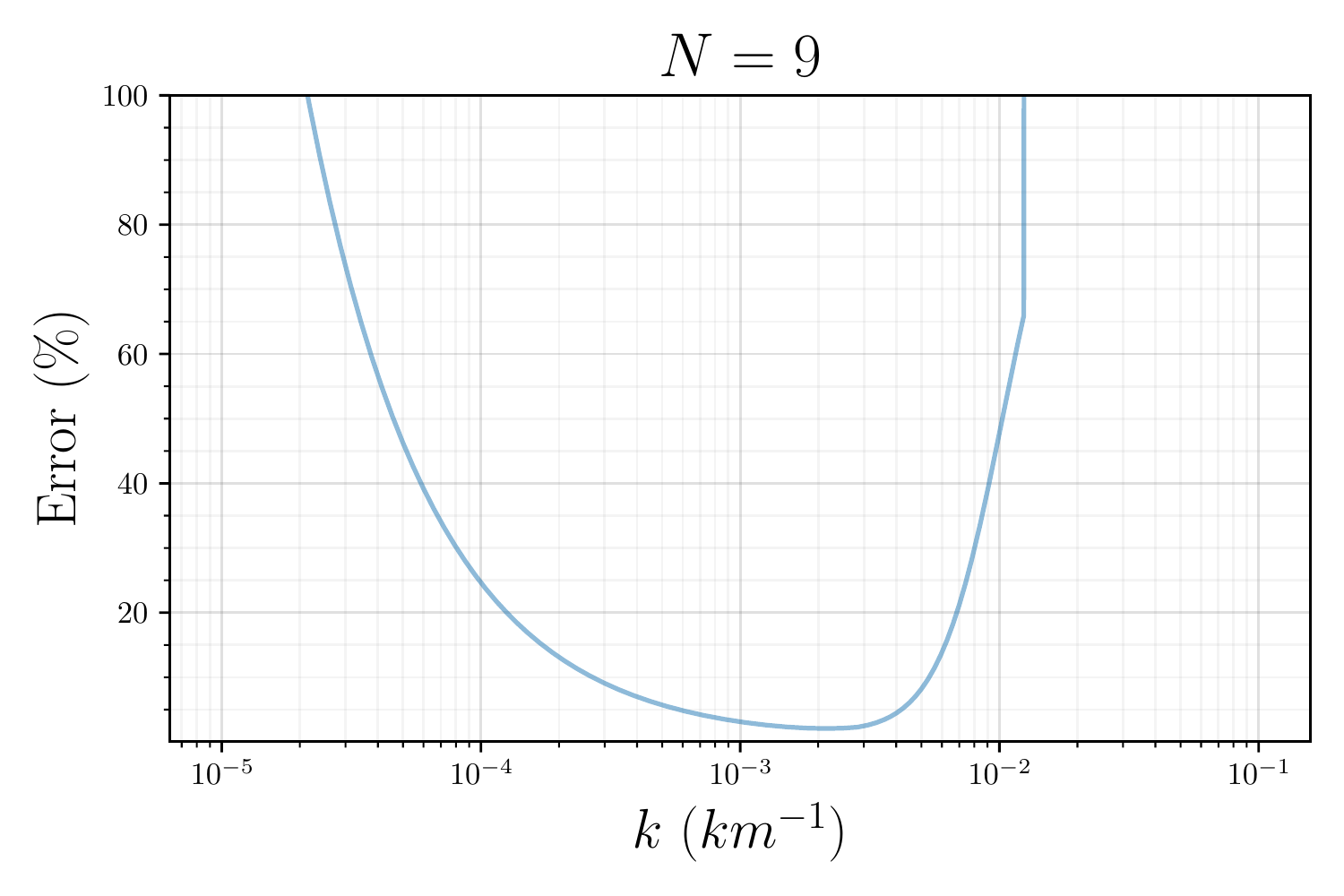}
\end{tabular}
\caption{{\small For a nine-spacecraft configuration there are $\sum_{n=4}^9 C_n^9=382$ subsets of at least four spacecraft. For each of these subsets, NEWTSS uses our learned equations to predict the effective error of the wave-telescope technique at each wavevector magnitude $k$. This is done for the example configuration of hour 205 of the HelioSwarm Phase B DRM. See Appendix \S \ref{sec:appendix.hour_205} for a breakdown of the shapes and sizes of all 382 of these $N \in \{4,5,6,7,8,9\}$ spacecraft subsets. }}
\label{fig:N_sc} 
\end{figure}

To demonstrate how this works, we note that the effective error and standard deviation for each $N \in \{4,5,6,7,8,9\}$ subset of spacecraft can be computed as a function of relative wavevector magnitude. Figure \ref{fig:N_sc} shows the effective error of each $N$-spacecraft subset for the configuration at hour 205 of the Design Reference Mission (DRM) science phase of the mission. By selecting the subset at each $k$-magnitude that minimizes the value of $\mu_{eff}(\bar{k},\chi) + 2 \sigma(\bar{k},\chi)$, we can increase the range of wavevectors over which we can accurately scan using the wave-telescope technique.

\begin{figure}[ht]
\centering
\includegraphics[width=0.9\textwidth]{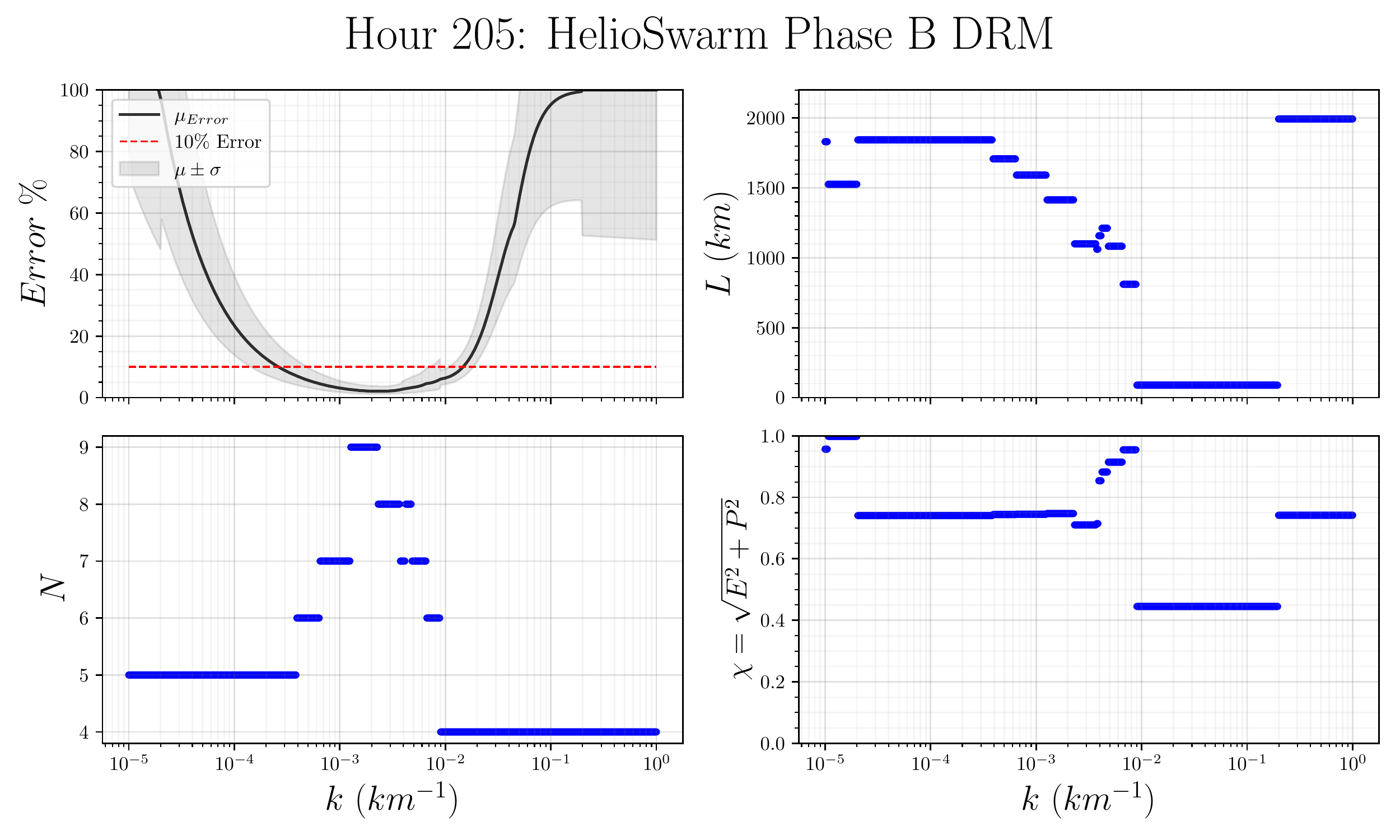}
\caption{{\small For each nine-spacecraft HelioSwarm configuration, NEWTSS selects the subset of spacecraft that minimizes the quantity $\mu_{eff} + 2 \sigma$ at each wavevector magnitude $k$. We plot the resulting minimized effective error and $2\sigma$ associated with the hour 205 HelioSwarm DRM configuration in the top left panel. In the other panels we show the number of spacecraft $N$ in the selected subset, the characteristic size $L$ of that subset, and the shape of that subset $\chi$. }}
\label{fig:HS_205} 
\end{figure}

Figure \ref{fig:HS_205} shows the expected error and noise once NEWTSS selects the subset of spacecraft that minimizes the value of $\mu_{eff} + 2 \sigma$ at each value of $k$. Displayed also in this figure is the number of spacecraft in this selected configuration ($N$), the characteristic size of the selected configuration ($L$), and the shape of the selected configuration ($\chi$). We see that for the magnitudes of wavevectors that can be identified accurately (where $\mu + 2\sigma < 20\%$), the optimal subset usually consists of more than 4 spacecraft. We also see that the characteristic size, $L$, of the optimal subset tends to be inversely proportional to wavevector magnitude.

\begin{figure}[ht]
\centering
\includegraphics[width=1.0\textwidth]{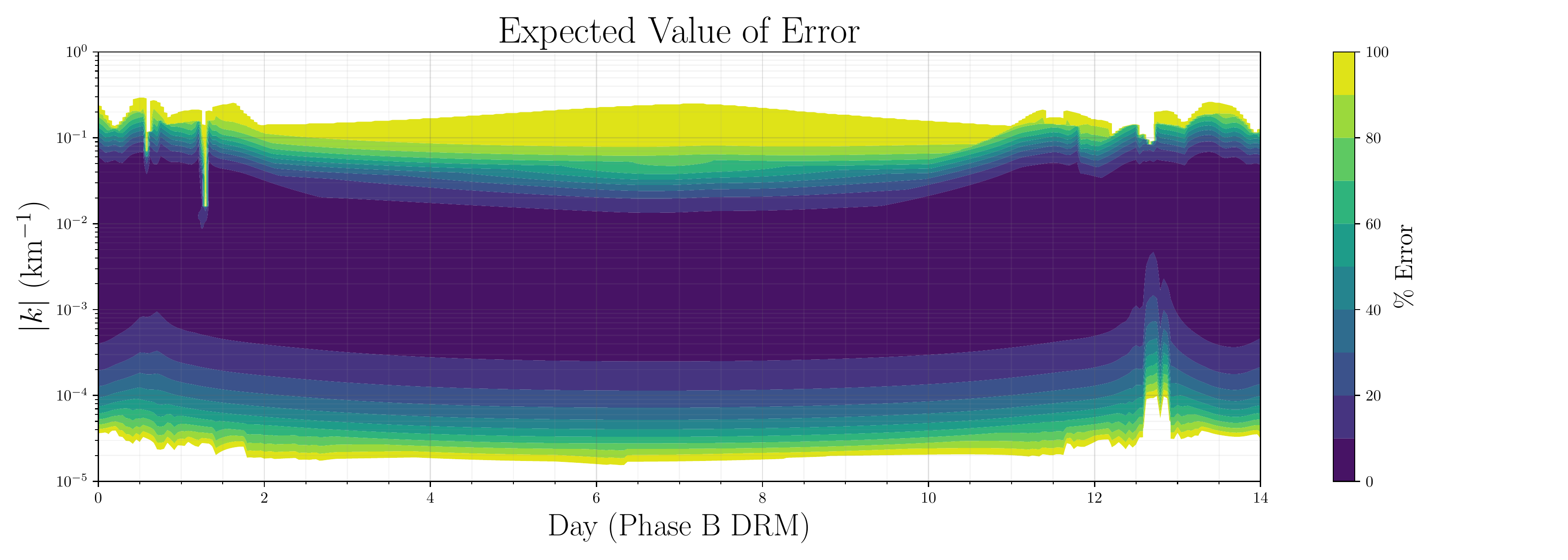}
\caption{{\small We use NEWTSS to select the subset of spacecraft that minimizes the value of $\mu_{eff} + 2 \sigma$ at each wavevector magnitude $k$ (as in Figure \ref{fig:HS_205}). We plot the resulting value of $\mu_{eff} + 2 \sigma$ over the first two weeks of HelioSwarm's Phase-B science mission for each wavevector magnitude $k$. The white dots outline the 20\% error value, the red dashes highlight the hour 205 configuration from Figure \ref{fig:HS_205}}}
\label{fig:HS_2weeks} 
\end{figure}

In Figure \ref{fig:HS_2weeks} we use NEWTSS to perform this optimization on the configurations at each hour over the first 2 weeks of the HelioSwarm mission science phase from the Phase-B DRM. By averaging over this time interval, we compute that we can reconstruct 1.86 orders of $k$-magnitude where the effective error is less than 10\%. We also see that we can reconstruct 1.52 orders of $k$-magnitude with $\mu_{eff} + 2\sigma < 20\%$. By comparing these values to those found in Figure \ref{fig:k_orders}, we conclude that selecting optimal subsets using HelioSwarm configurations can reconstruct a larger range of wavevectors than using any individual $N$-spacecraft configuration with $N \in \{4,5,6,7,8,9\}$.

\section{Summary and Discussion}
\label{sec:summary}
We have used data from many wave-telescope simulations with four-to-nine spacecraft to learn a functional form for the expected error of the technique, $\mu_{eff}(\bar{k},\chi)$ (eqn \ref{eqn:mu_eff}). For a given collection of spacecraft, expected error is a function of the shape of the spacecraft configuration shape $\chi$ (eqn \ref{eqn:chi}) and the relative magnitude of the wavevector selected $\bar{k}$ (eqn \ref{eqn:k_unitless}). Along with the expected error equation, we have learned an equation that estimates the standard deviation of the error, $\sigma(\bar{k},\chi)$ (eqn \ref{eqn:std}). These results factor in the chance of aliasing (eqns \ref{eqn:P_alias}) and not aliasing (eqn \ref{eqn:mu_log}). The 12 coefficients of equations \ref{eqn:mu_log}, \ref{eqn:std}, \ref{eqn:P_alias} ($a_i$'s, $b_i$'s, and $c_i$'s), along with their uncertainties, can be found in Table \ref{tab:coefs}.

Using these equations we have shown that using more spacecraft in the wave-telescope does minimize the error in the wavevector reconstruction, as well as expand the ranges of wavevectors that can be reconstructed. However, we have observed that the greatest increase in accuracy was achieved by the increase from just four to five spacecraft. Because these equations are quick to evaluate, we have shown that NEWTSS can leverage that information to break down any configuration of $N \geq 4$ spacecraft into its optimal subsets for each $\bar{k}$. We have demonstrated that this expands the ranges of detectable wavevectors by applying it to the first two weeks of positional data for HelioSwarm's nine-spacecraft Design Reference Mission configuration.

Future work should include extrapolating this analysis to configurations of many more spacecraft. Importantly, more spacecraft in a configuration means the more resilient the configuration is to changing its shape parameter value $\chi$. Further extrapolation will likely have to identify additional independent variables to use in their analysis. Theoretically, the direction of the wavevector should also be included as an independent variable, but we were unable to incorporate it into the method in such a way that improved the quality of the fitted equations. Comparison of these derived equations for the probability of aliasing to recently developed analytic descriptions \citep{Schulz:preprint} is also a potential focus of future work.

\section*{Acknowledgments}
Construction and analysis of the HelioSwarm Observatory Design Reference Mission trajectories was supported in part by the HelioSwarm Project funded under NASA’s Prime contract no. 80ARC021C0001. K.G.K was also supported by NASA Early Career Grant 80NSSC19K0912. This research was supported by the International Space Science Institute in Bern, through ISSI International Team project \#556 (\href{https://teams.issibern.ch/energtransferspaceplasmas/}{Cross-Scale Energy Transfer in Space Plasmas}).

\newpage
\appendix
\section{Lagrange Multiplier Minimization}
\label{sec:appendix.lagrange}
Dropping the independent variables from our notation, the minimization problem to be solved is
\begin{equation}
    P = \min_{W} \text{Tr}\left[\mathbf{W}^\dagger \M \mathbf{W} \right] 
   \text{ \hspace{.2cm} s.t. \hspace{.2cm} }    \mathbf{W}^\dagger \HH = \mathbf{I} \text{ \hspace{.1cm} and \hspace{.1cm} }     \HH^\dagger \mathbf{W} = \mathbf{I}. \label{eqn:min_append}
\end{equation}
We reformulate \ref{eqn:min_append} using the $3\times 3$ Lagrange multiplier matrices $\mathbf{\Lambda}$ and $\mathbf{\Gamma}$
\begin{equation}
    P = \min_{W} \text{Tr}\left[\mathbf{W}^\dagger \M \mathbf{W}  + \left(  \mathbf{W}^\dagger \HH - \mathbf{I} \right)\mathbf{\Lambda} + \mathbf{\Gamma} \left( \HH^\dagger \mathbf{W} - \mathbf{I} \right) \right].
\end{equation}
Because $\mathbf{P}$ is a $3\times 3$ matrix, the equation for the trace $P$ has three terms. Each of these terms on the diagonal has the form
\begin{equation}
    P_{ii} = W^\dagger_{ij} M_{jk} W_{ki}  + W^\dagger_{ij} H_{jk} \Lambda_{ki} - \Lambda_{ii} + \Gamma_{ij} H^\dagger_{jk} W_{ki} - \Gamma_{ii}.
\end{equation}
We now find the local minimums be setting $\partial P_{ii}/\partial \mathbf{W}$ to zero. This implies that both
\begin{equation}
    \frac{\partial P_{ii}}{\partial W_{ki}} = W^\dagger_{ij} M_{jk} + \Gamma_{ij} H^\dagger_{jk} = 0 \label{eqn:L1}
\end{equation}
and
\begin{equation}
    \frac{\partial P_{ii}}{\partial W^\dagger_{ij}} = M_{jk} W_{ki} + H_{jk} \Lambda_{ki} = 0. \label{eqn:L2}
\end{equation}
Dropping the subscripts, we see from multiplying eqn \ref{eqn:L1} on the right by $\mathbf{W}$ that $\mathbf{W}^\dagger \M \mathbf{W} = - \mathbf{\Gamma} \HH^\dagger \mathbf{W}$. We then apply the constraint that $\HH^\dagger \mathbf{W} = \mathbf{I}$ to find that 
\begin{equation}
    -\mathbf{\Gamma} = \mathbf{W}^\dagger \M \mathbf{W}.
\end{equation}
We also note that if we multiply eqn \ref{eqn:L2} by $\mathbf{W}^\dagger$ on the left we have that $\mathbf{W}^\dagger \M \mathbf{W} = - \mathbf{W}^\dagger \HH \mathbf{\Lambda}$. Using the constraint that $\mathbf{W}^\dagger \HH = \mathbf{I}$ gives us
\begin{equation}
    -\mathbf{\Lambda} = \mathbf{W}^\dagger \M \mathbf{W}. \label{eqn:lambda1}
\end{equation}
Simultaneously, we see that eqn \ref{eqn:L2} implies
\begin{align}
    \M \mathbf{W} &= -\HH\mathbf{\Lambda}\\
     \left( \HH^\dagger \M^{-1}\right) \M \mathbf{W} &= -\left( \HH^\dagger \M^{-1}\right) \HH\mathbf{\Lambda} \\
     \HH^\dagger \mathbf{W} &= - \left(\HH^\dagger \M^{-1} \HH \right) \mathbf{\Lambda}.
\end{align}
Using the constraint of $\HH^\dagger W = \mathbf{I}$ gives us
\begin{equation}
    -\mathbf{\Lambda} = \left(\HH^\dagger \M^{-1} \HH \right)^{-1}. \label{eqn:lambda2}
\end{equation}
Equating our two formulas for $\mathbf{\Lambda}$ (\ref{eqn:lambda1} and \ref{eqn:lambda2}) we find that at the critical point we have
\begin{equation}
    \mathbf{W}^\dagger \M \mathbf{W} =  \left(\HH^\dagger \M^{-1} \HH \right)^{-1}.
\end{equation}
Plugging this critical point back into the argument of the original minimization problem \ref{eqn:min_append}, we find that the solution to this problem which satisfies the constraints exactly is
\begin{equation}
    P = \text{Tr}\left[\left(\HH^\dagger \M^{-1} \HH \right)^{-1} \right] .
\end{equation}

\section{Wave-Telescope Scanning Algorithm}
\label{sec:appendix.scan_alg}
The wave-telescope technique's mechanism of identifying the waves present given magnetic field timeseries is to evaluate $P(\omega,\mathbf{k})$ over a dense range of $\mathbf{k}$ values. The value of $\mathbf{k}$ that yields the maximal value of $P(\omega,\mathbf{k})$ is then identified as the calculated wavevector. Because we wish to build a large dataset consisting of millions of wavevector identifications, we wish to minimize the number of $P(\omega,\mathbf{k})$ function evaluations required. We note that for a single plane-wave, there is a sharp peak in $P(\omega,\mathbf{k})$ at the correct $\mathbf{k}$ value using the wave-telescope, which exponentially decays with distance away from the peak (see Figure 2 of \cite{Narita:2022}). If we can identify the region of $\mathbf{k}$ space that houses this peak using a course scan, we can then adaptively refine our $\mathbf{k}$ scan around the region containing the peak value of $P(\omega,\mathbf{k})$.

\RestyleAlgo{ruled}
\begin{algorithm}[ht]
\caption{Adaptive $\mathbf{k}$ Search}
\label{alg:k_search}
\While{$P(\omega,\mathbf{k}^*) < P_{\text{thres}}$}{
Define grid resolution $(I,J,L)$\;
Define scan range: $k \in [k_{min}, k_{max}]$, $k^\theta \in [0,2\pi]$, $k^\phi \in [0,\pi]$\;
Define scan grid: $k_{\text{grid}} = [k_{0},...,k_{I}]$, $k^\theta_{\text{grid}} = [k^\theta_{0},...,k^\theta_{J}]$, $k^\phi_{\text{grid}} = [k^\phi_{0},...,k^\phi_{L}]$\;
\While{$|k_{max} - k_{min}| < \Delta k$}{
Evaluate $P(\omega,\mathbf{k})$ on grid\;
Identify peak: $(k_{i^*},k^\theta_{j^*},k^\phi_{l^*}) = \argmax_{\mathbf{k}\in \mathbf{k}_{\text{grid}}} P(\omega,\mathbf{k})$\;
Update scan range: $k \in [k^*_{i^*-1}, k^*_{i^*+1}]$, $k^\theta \in [k^\theta_{j^*-1}, k^\theta_{j^*+1}]$, $k^\phi \in [k^\phi_{l^*-1}, k^\phi_{l^*+1}]$\;
Update scan grid\;
}
Increase grid resolution $(I,J,L)$\;
}
\Return $(k_{i^*},k^\theta_{j^*},k^\phi_{l^*})$
\end{algorithm}

We implement Algorithm \ref{alg:k_search} to perform this adaptive mesh refinement. First, we divide our wavevector $\mathbf{k} \in [k,k_{\theta},k_{\phi}]$ (in spherical coordinates) domain into a $16 \times 13 \times 13$ grid. Then, we evaluate $P(\omega,\mathbf{k})$ on all grid points. We identify the $[k,k_{\theta},k_{\phi}]$ where the maximal value of $P(\omega,\mathbf{k})$ was achieved and center a new, more refined scan at that location. This scan and refinement process repeats until the magnitudes of the $\mathbf{k}$ vectors that are being scanned reach a $k$-magnitude resolution threshold ($\Delta k = 0.01k$). 

\begin{figure}[ht]
\centering
\includegraphics[width=0.75\textwidth]{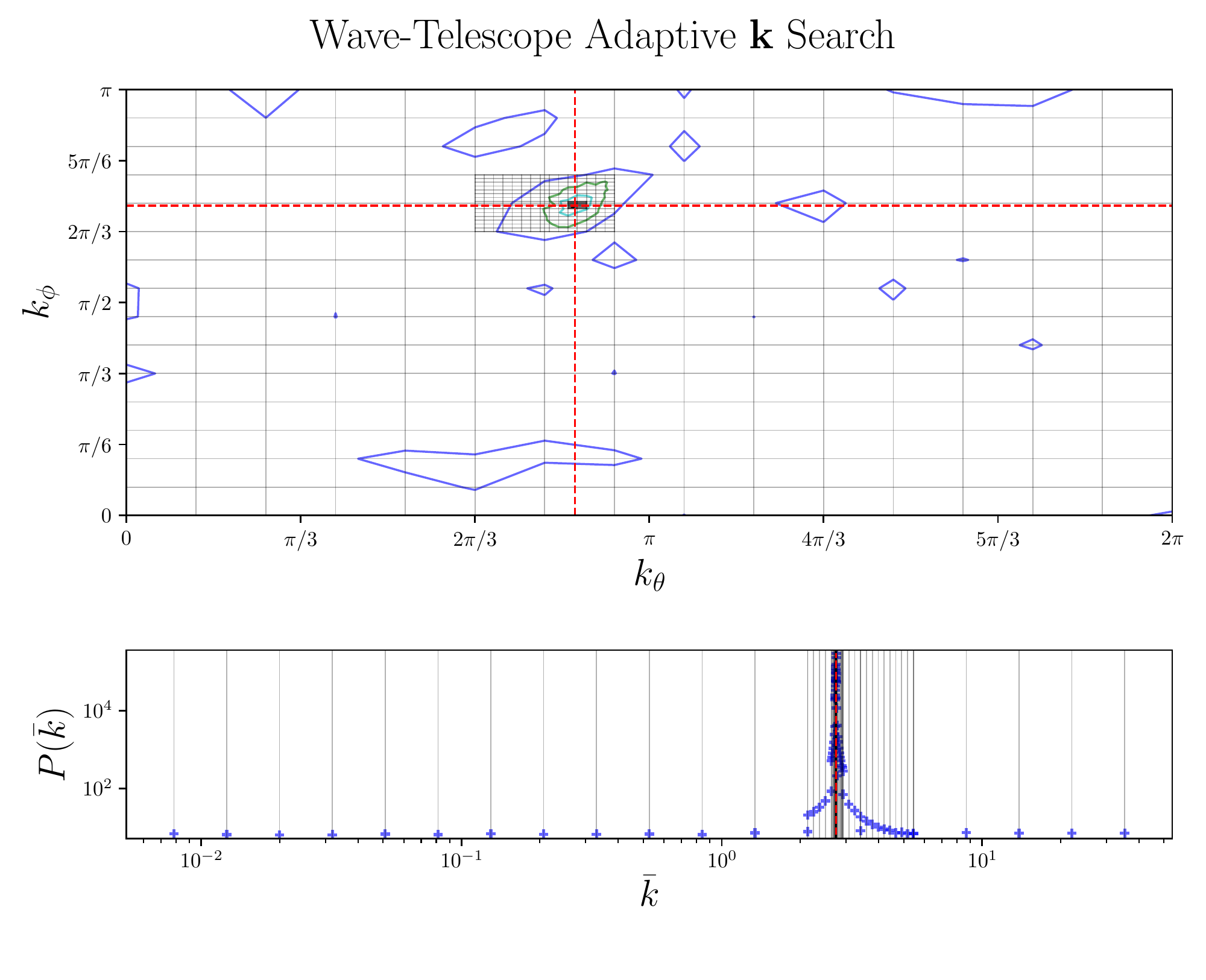}
\caption{{\small An example plane-wave is simulated using six spacecraft. We use the adaptive $\mathbf{k}$ search Algorithm \ref{alg:k_search} to define an adaptive mesh that allows us to identify the peak value of $P(\omega,\mathbf{k})$ while minimizing function evaluations. We display the values of $\mathbf{k} \in [\bar{k},k_{\theta},k_{\phi}]$ that are scanned using this procedure as the black lines. We see that the identified peak corresponds to the true wavevector of the plane-wave (shown as red dashes). We overlay contours at three values $P$ (blue, green, cyan lines) on the upper panel. We also display the computed values of $P(\bar{k})$ (blue crosses) on the lower panel. }}
\label{fig:k_search} 
\end{figure}

As a final check, we evaluate whether the peak value of $P(\omega,\mathbf{k})$ is greater than a power threshold ($P_{thres}=1000$). If the peak $P(\omega,\mathbf{k})$ found is not above this threshold, we restart the adaptive scanning process with an increased resolution. We repeat the adaptive scanning and increasing resolution process until either we find a peak value above $P_{thres}$ or until we reach a predefined maximal resolution ($22 \times 19 \times 19$). Figure \ref{fig:k_search} shows the $\mathbf{k}$ values which are scanned using this adaptive method for one example plane-wave. We see that after successive iterations, the grid focuses where the peak value of $P(\omega,\mathbf{k})$ was identified.

\section{HelioSwarm Hour 205}
\label{sec:appendix.hour_205}
\begin{figure}[ht]
\centering
\includegraphics[width=0.9\textwidth]{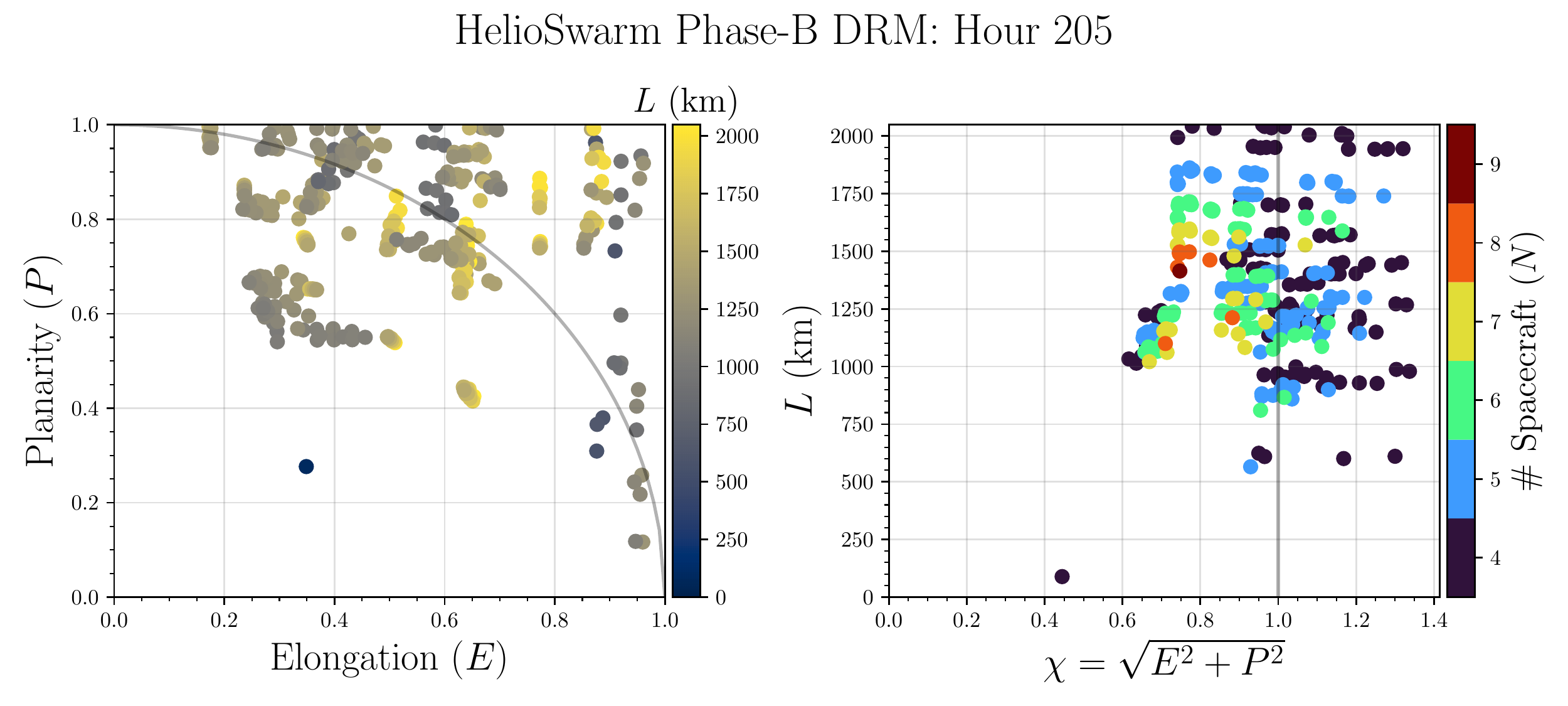}
\caption{{\small From hour 205 of the science phase of the Phase-B Design Reference Mission of HelioSwarm we compute the elongation, planarity, and characteristic sizes of all subsets of the nine-spacecraft configuration which contain $N \in \{4,5,6,7,8,9\}$ spacecraft. The black lines represent the shape threshold $\chi=1$. }}
\label{fig:hour_205_sup} 
\end{figure}

In Figures \ref{fig:N_sc} and \ref{fig:HS_205} of \S \ref{ssec:app_HS} we highlight the application of NEWTSS to one example nine-spacecraft configuration from the HelioSwarm Phase-B Design Reference Mission. In Figure \ref{fig:hour_205_sup} we display the elongation, planarity, and characteristic sizes of all subsets of this configuration (which contain $N \in \{4,5,6,7,8,9\}$ spacecraft) that is planned for hour 205 of the science portion of the mission.

\section{Verification of Fitted Equations}
\label{sec:appendix.verify}
In Table \ref{tab:PPC} of \S \ref{ssec:PPC} we counted how many data points fell within the intervals $\mu \pm \sigma$, $\mu \pm 2\sigma$, and $\mu \pm 3\sigma$ where $\mu$ and $\sigma$ were defined using our learned equations. We now now wish to quantify the proportion of data points that fall within an arbitrary interval near $\mu$.  We do this by checking what proportion of data points fall within the interval $\mu \pm \sigma_0 \sigma$ for values of $\sigma_0$ in $[-3,3]$. If our learned equations are valid, this quantification should match with a log-normally distributed Cumulative Distribution Function (CDF), because that was our initial hypothesis. We display the results of this comparison for each of the $N \in \{4,5,6,7,8,9\}$ spacecraft learned equations below. Because aliasing causes large sudden deviations from the average error, we restrict ourselves to points in the dataset where $P(alias) < 0.01$ and $\chi < 1$.

\begin{figure}[ht]
\centering
\begin{tabular}{ccc}
\includegraphics[width=0.30\textwidth]{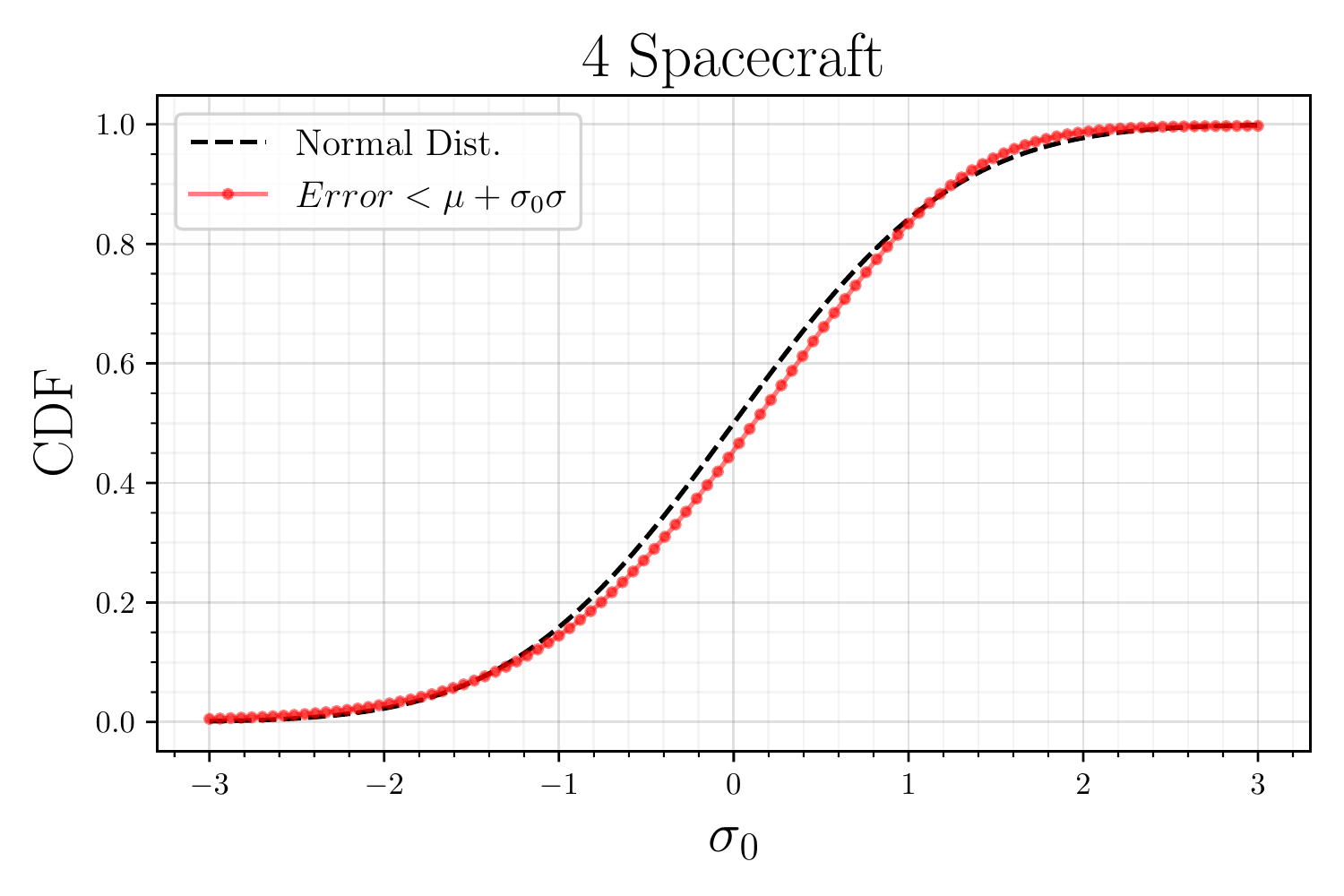} &
\includegraphics[width=0.30\textwidth]{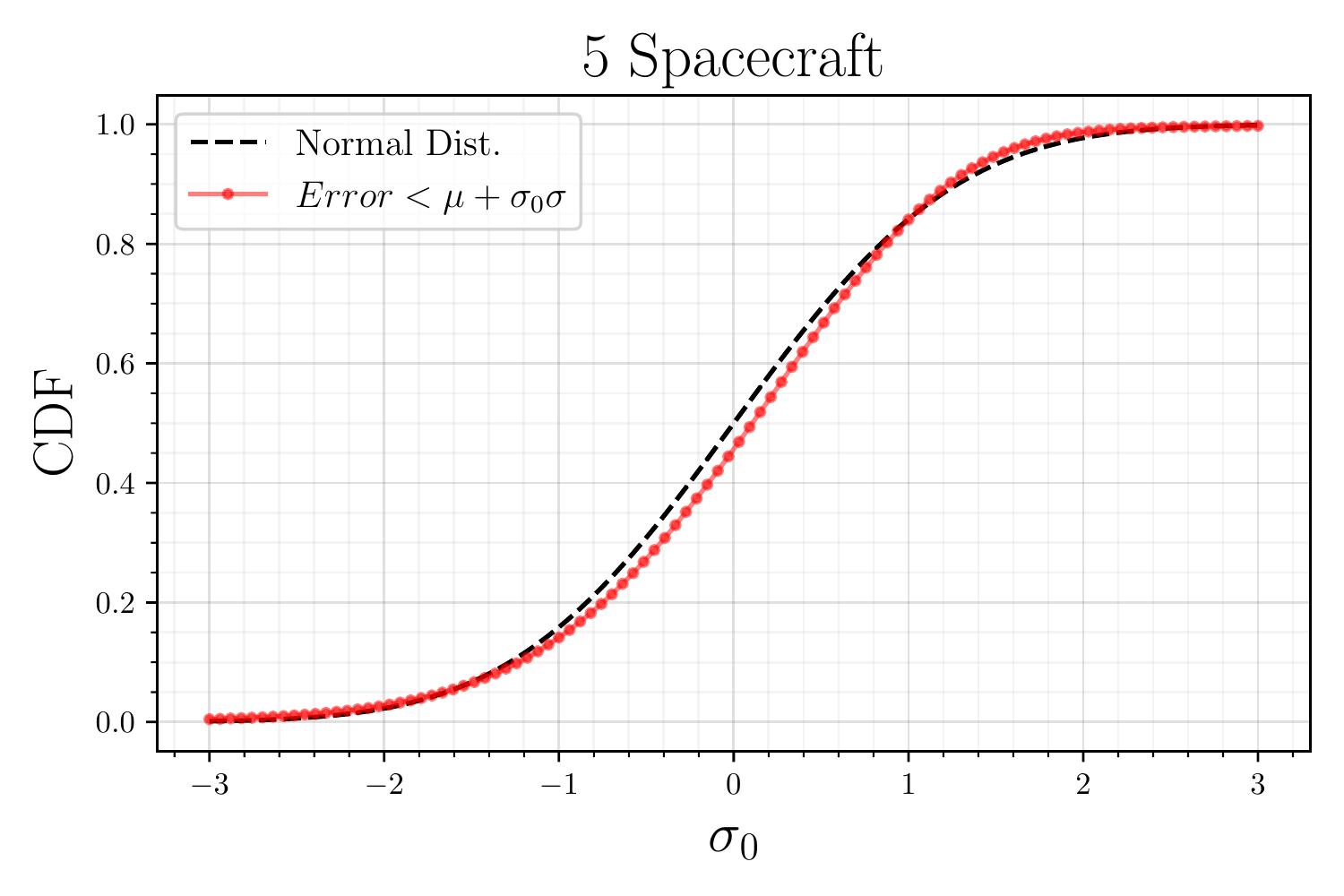} &
\includegraphics[width=0.30\textwidth]{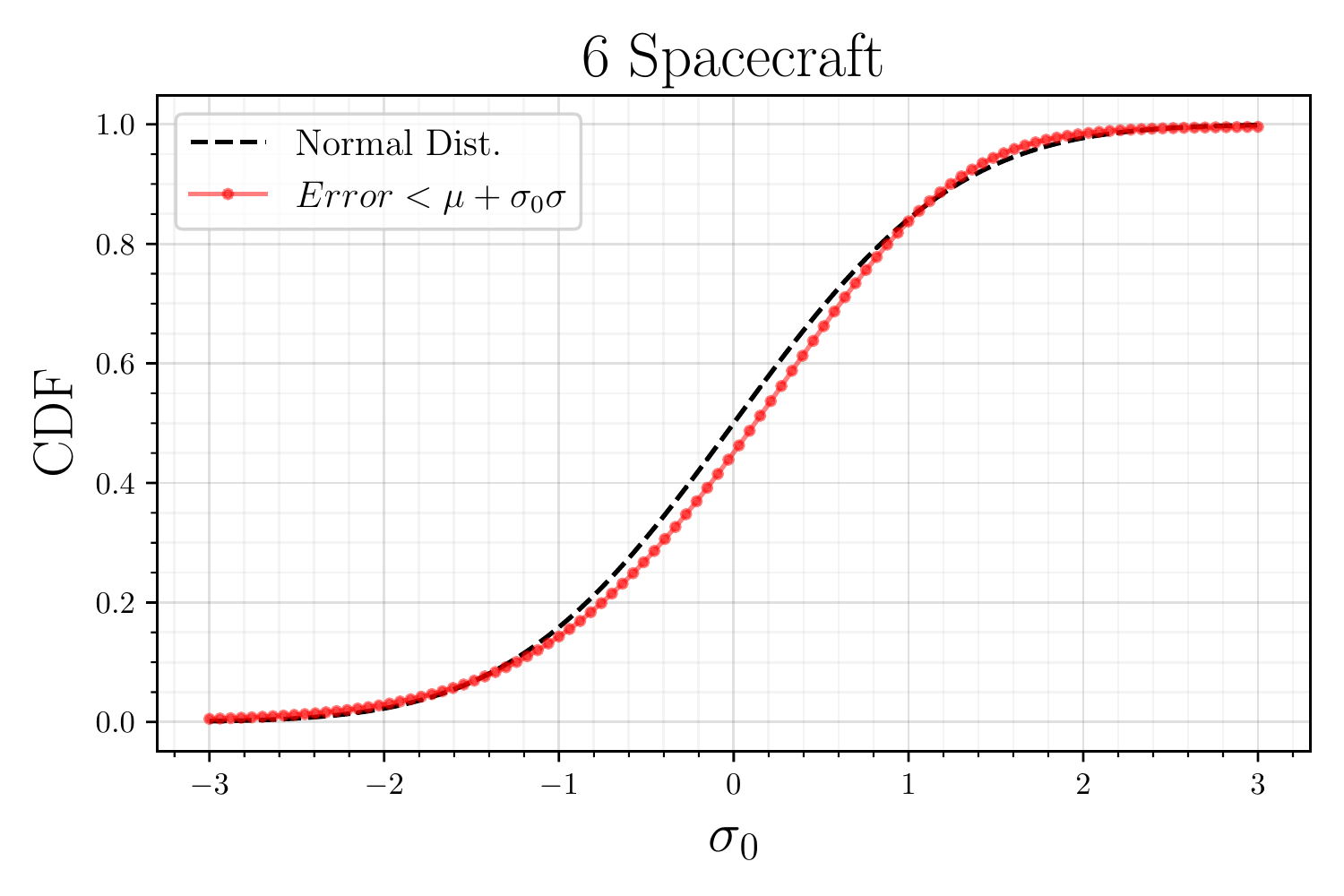} \\
\includegraphics[width=0.30\textwidth]{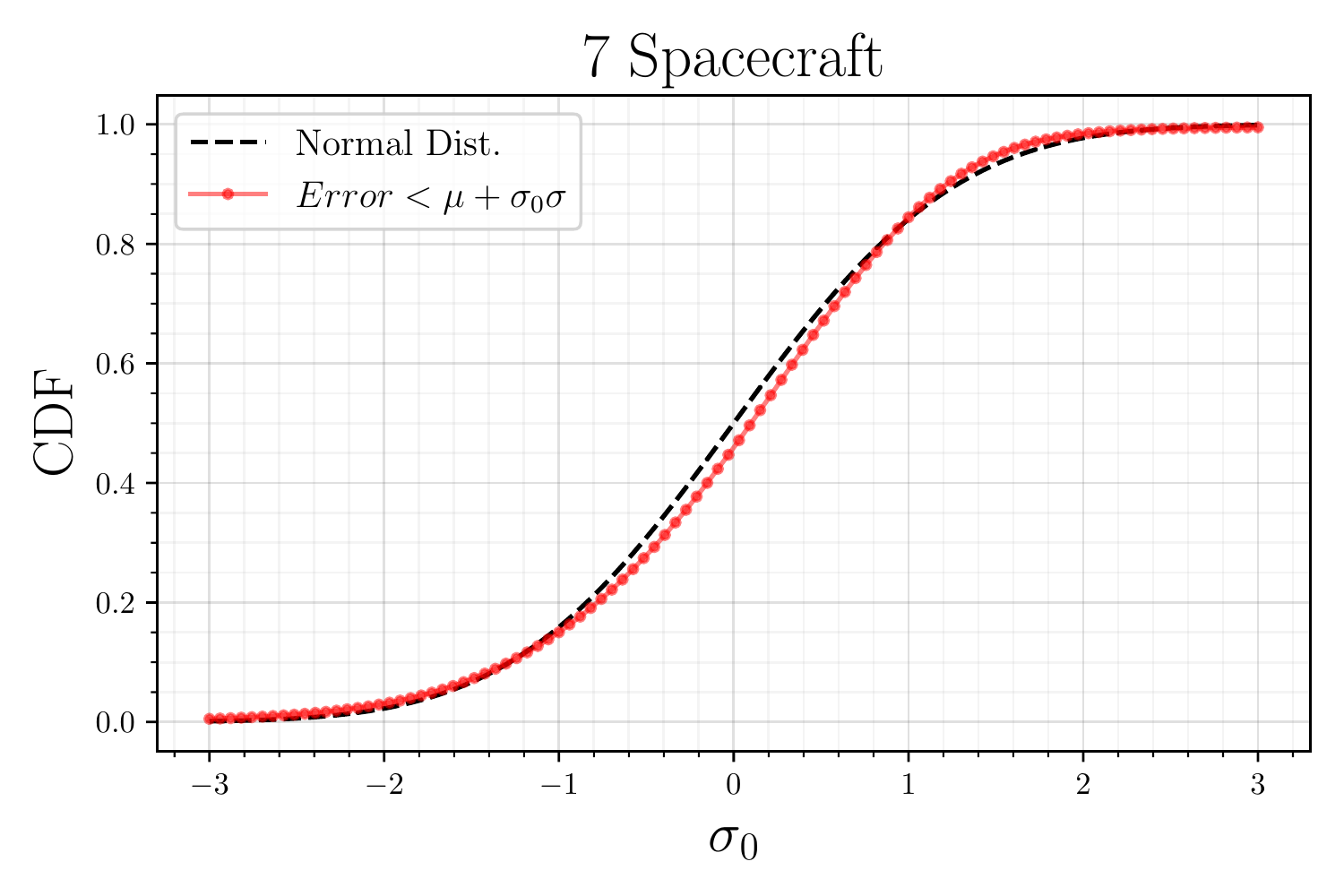} &
\includegraphics[width=0.30\textwidth]{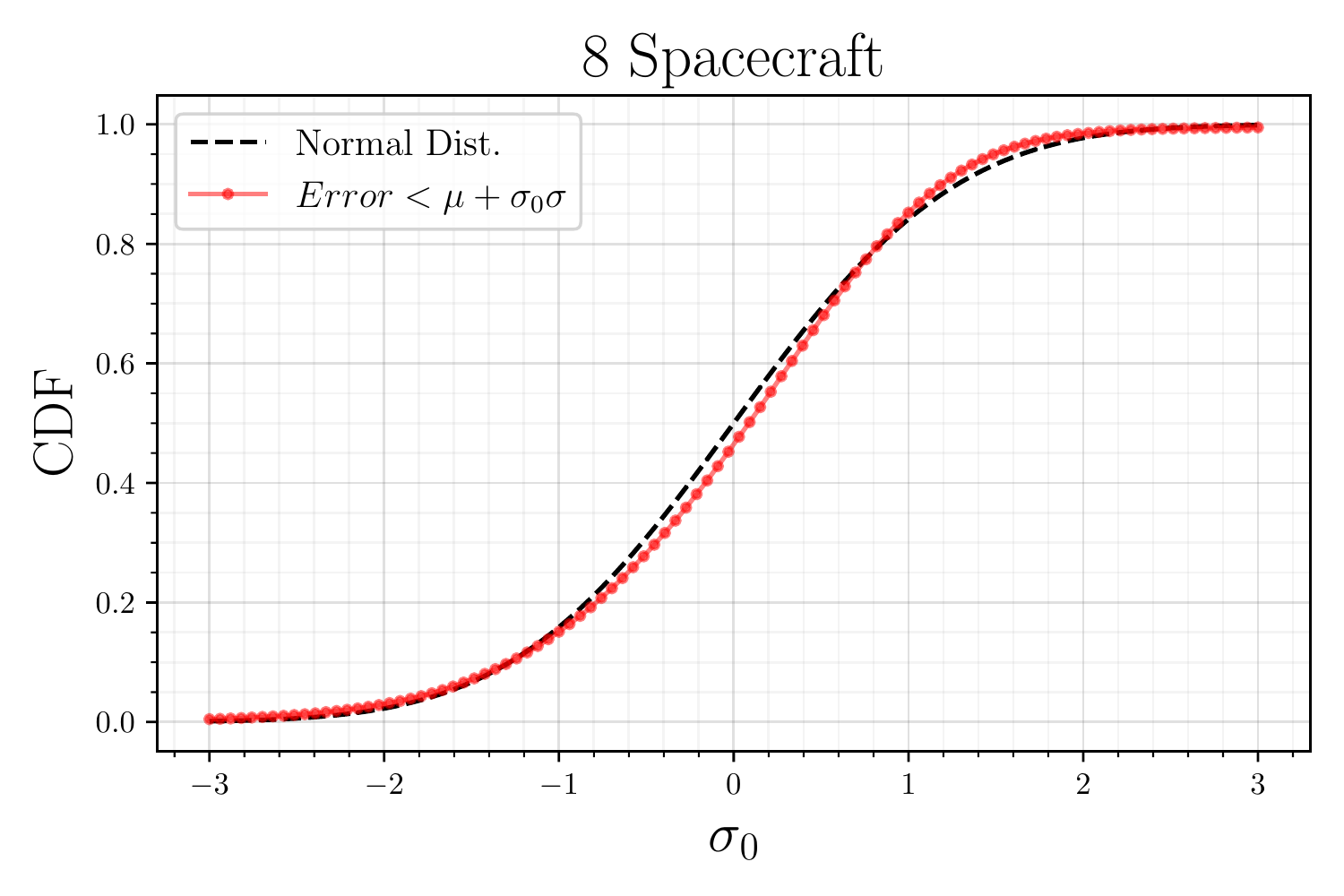} &
\includegraphics[width=0.30\textwidth]{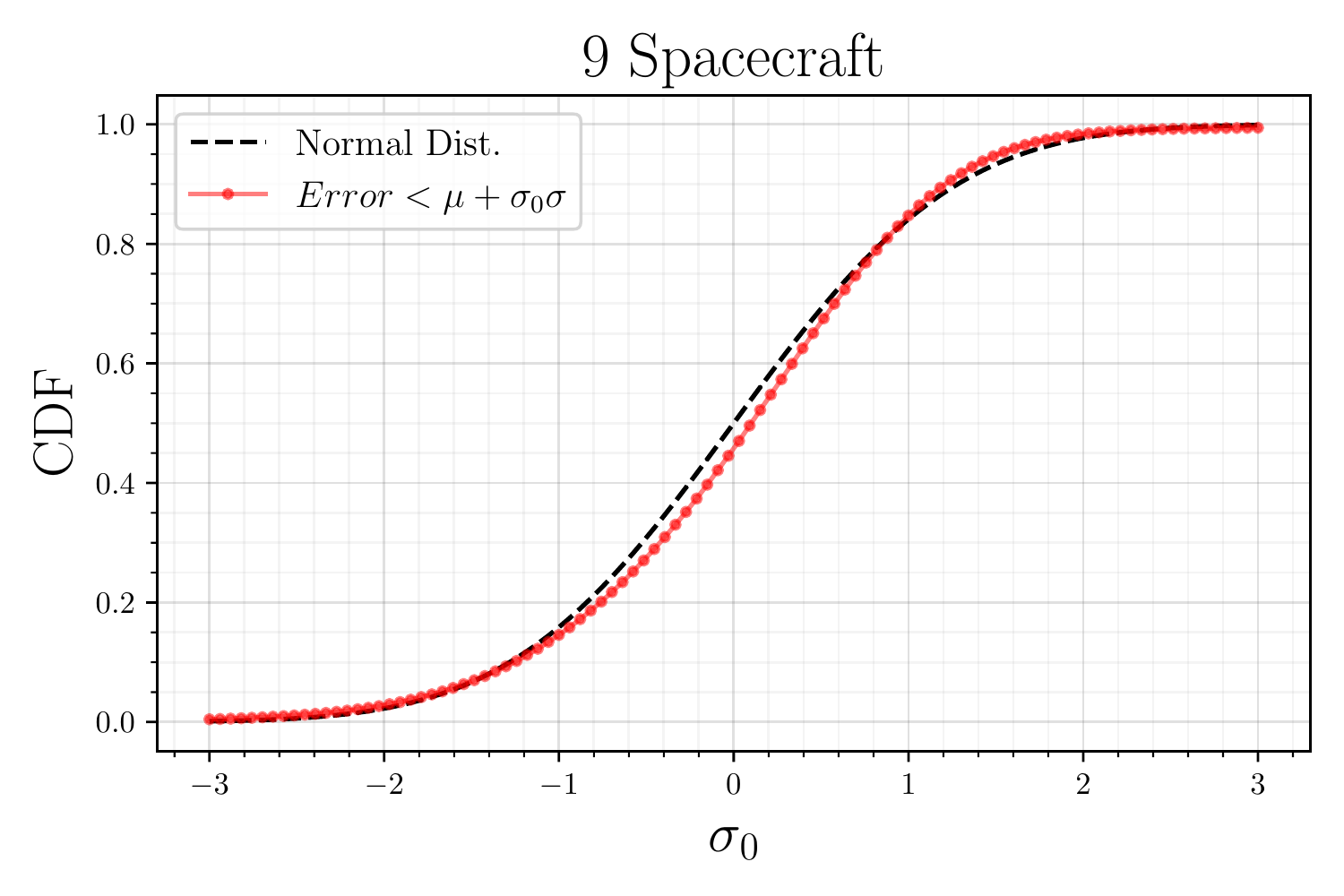}
\end{tabular}
\caption{{\small Using our dataset and learned equations for $\mu$ and $\sigma$, we check what proportion of points fall within the threshold $\mu \pm \sigma_0 \sigma$ for values of $\sigma_0$ in the interval $[-3,3]$.  For ease of visualization, we have taken the $\log_{10}$ of the error value found. Therefore, if our equations perfectly fit the data, we expect that the CDF of a standard normal distribution (shown as black dashes) would be identical. }}
\label{fig:PPC_cdfs} 
\end{figure}

\newpage
\bibliography{References}{}
\bibliographystyle{aasjournal}



\end{document}